\documentclass[twocolumn]{aastex631}
\newcommand{\Msun}{M\ensuremath{_\odot}\,}

\def\M{M\,}
\def\N{NGC\,}

\def\U{UGC\,}

\def\HI{H{\sc i}\,}

\def\kms{$\textrm{km~s$^{-1}$ }$}
\def\nb{\textsc{nbursts}}
\def\green{\textcolor[rgb]{0.00,0.00,0.00}}
\def\red{\textcolor[rgb]{0.00,0.00,0.00}}

\usepackage{newtxtext,newtxmath}
\usepackage{epstopdf}
\begin{document}
\title{A closer look at the extended edge-on low-surface brightness galaxies}
%\author[Saburova et al.]
\author[0000-0002-4342-9312]{Anna S. Saburova}
\affiliation{Sternberg Astronomical Institute, Moscow M.V. Lomonosov State University, Universitetskij pr., 13,  Moscow, 119234, Russia}
\author[0000-0002-1750-2096]{Damir Gasymov}
\affiliation{Sternberg Astronomical Institute, Moscow M.V. Lomonosov State University, Universitetskij pr., 13,  Moscow, 119234, Russia}
\author[0000-0001-8427-0240]{Evgenii V. Rubtsov}
\affiliation{Sternberg Astronomical Institute, Moscow M.V. Lomonosov State University, Universitetskij pr., 13,  Moscow, 119234, Russia}
\author[0000-0002-7924-3253]{Igor V. Chilingarian}
\affiliation{Center for Astrophysics --- Harvard and Smithsonian, 60 Garden Street MS09, Cambridge, MA 02138, USA}
\affiliation{Sternberg Astronomical Institute, Moscow M.V. Lomonosov State University, Universitetskij pr., 13,  Moscow, 119234, Russia}
\author[0000-0002-2516-9000]{Sviatoslav Borisov}
\affiliation{Department of Astronomy, University of Geneva, Chemin Pegasi 51, 1290 Versoix, Switzerland}
\author[0000-0001-7113-8152]{Ivan Gerasimov}
\affiliation{Sternberg Astronomical Institute, Moscow M.V. Lomonosov State University, Universitetskij pr., 13,  Moscow, 119234, Russia}
\author[0000-0002-9609-7980]{Fedor Kolganov}
\affiliation{Faculty of Physics and Earth System Sciences, Leipzig University, Linnestraße 5, 04103 Leipzig, Germany}
\author[0000-0002-1091-5146]{Anastasia V. Kasparova}
\affiliation{Sternberg Astronomical Institute, Moscow M.V. Lomonosov State University, Universitetskij pr., 13,  Moscow, 119234, Russia}
\author[0000-0002-1098-6649]{Roman I. Uklein}
\affiliation{Special Astrophysical Observatory, Russian Academy of Sciences, Nizhniy Arkhyz, Karachai-Cherkessian Republic 357147, Russia}
\author[0000-0003-3104-3372]{Michal B\'ilek}
\affiliation{Coll\`ege de France, 11 place Marcelin Berthelot, 75005 Paris, France}
\affiliation{LERMA, Observatoire de Paris, CNRS, PSL Univ., Sorbonne Univ., 75014
Paris, France}
\affiliation{FZU – Institute of Physics of the Czech Academy of Sciences, Na Slovance
1999/2, Prague 182 21, Czech Republic}
\author[0000-0003-3255-7340]{Kirill A. Grishin}
\affiliation{Sternberg Astronomical Institute, Moscow M.V. Lomonosov State University, Universitetskij pr., 13,  Moscow, 119234, Russia}
\affiliation{Universit\'e Paris Cit\'e, CNRS(/IN2P3), Astroparticule et Cosmologie, F-75013 Paris, France}
\author[0000-0001-9914-4466]{Anatoly Zasov}
\affiliation{Sternberg Astronomical Institute, Moscow M.V. Lomonosov State University, Universitetskij pr., 13,  Moscow, 119234, Russia}
\affiliation{Faculty of Physics, Moscow M.V. Lomonosov State University, Leninskie gory 1,  Moscow, 119991, Russia}
\author[0000-0002-8297-6386]{Mariia Demianenko}
\affiliation{Max-Planck-Institut für Astronomie, Königstuhl 17, 69117 Heidelberg, Germany}
\author[0000-0002-6425-6879]{Ivan Yu.~Katkov}
\affiliation{New York University Abu Dhabi, P.O. Box 129188, Abu Dhabi, UAE}
\affiliation{Center for Astrophysics and Space Science (CASS) 
New York University Abu Dhabi \\
PO Box 129188, Abu Dhabi 
United Arab Emirates}
\affiliation{Sternberg Astronomical Institute, Moscow M.V. Lomonosov State University, Universitetskij pr., 13,  Moscow, 119234, Russia}
\author{Ana Lalovi\' c}
\affiliation{Astronomical Observatory, Volgina 7, 11060 Belgrade, Serbia}
\author{Srdjan Samurovi\'c}
\affiliation{Astronomical Observatory, Volgina 7, 11060 Belgrade, Serbia}

\begin{abstract}
To understand the origin of extended disks of low-surface brightness (LSB) galaxies, we studied in detail 4 such systems with large disks seen edge-on. Two of them are edge-on giant LSB galaxies (gLSBGs) recently identified by our team. The edge-on orientation of these systems boosts their surface brightnesses that provided an opportunity to characterize stellar populations spectroscopically and yielded the first such measurements for edge-on gLSBGs. We collected deep images of one galaxy using the 1.4-m Milankovi\'c Telescope which we combined with the archival  Subaru Hyper Suprime-Cam and DESI Legacy Surveys data available for the three other systems, and measured the structural parameters of the disks. We acquired deep long-slit spectra with the Russian 6-meter telescope and the 10-m Keck~II telescope and estimated stellar population properties in the high- and low-surface brightness regions as well as the gas-phase metallicity distribution. The gas metallicity gradients are shallow to flat in the range between $0$ and $-0.03$~dex per exponential disk scale length, which is consistent with the extrapolation of the gradient -- scale length relation for smaller disk galaxies. Our estimates of stellar velocity dispersion in the LSB disks as well as the relative thickness of the disks indicate the dynamical overheating. Our observations favor mergers as the essential stage in the formation scenario for massive LSB galaxies.
\end{abstract}

\section{Introduction}
\label{sec:intro}

\begin{deluxetable*}{lcccc}
    \tablecaption{Observed properties of galaxies sample. Index $\mathrm{z_0/h}$ means correction of magnitude from edge-on orientation to face-on. The first part of the table shows the general properties of the sample. $\mathrm{M_{g}}$ is absolute magnitude in DECaLS g-band ($\dagger$ marks magnitude in SDSS g-band). The second part shows dynamical properties of galaxies, $\mathrm{V_{rot}}$ was obtained from SCORPIO-2 spectra and used in Sec.~\ref{subsubsec:spectral_analysis}, dynamical mass $\mathrm{M_{dyn}}$ was calculated within $4h$ distance, $\mathrm{M_{sph}}$ is an estimation of spheroid mass based on the disk relative thickness according to \citep{Khoperskovetal2010} and stellar mass estimate $\mathrm{M_{\star}}$ was described in Sec.~\ref{dis:met}. The third part is described in Sec.~\ref{sec:gas-phase_met} and in Fig.~\ref{fig:abundgradient}. The fourth part is devoted to the surface photometry of the galaxies, all parameters are described in Sec.~\ref{subsubsec:phot_analysis}. Mass-to-light ratio (M/L) is given in $(\mathrm{M_\odot/L_\odot})$ terms for parameters of stellar populations presented in the Fig.~\ref{fig:spectra} for disk and bulge regions. Last part shows $\chi^2$ and p-value for W1 and W2 bands (see Sec.~\ref{sec:check_AGN}). 
    \label{tab:mainproperties} \label{variability} \label{tab_photometry} \label{tab:BFMM}} %}
    \tablehead{
 \colhead{Name (ID)} & \colhead{UGC 8197 (i)} & \colhead{LEDA 1044131 (ii)} & \colhead{FGC 150 (iii)} & \colhead{2MASX J02300754-0240451 (iv)}\\ [-0.5cm]
         }
         \startdata 
         RA (J2000)  & 13:05:58.07 & 02:23:07.58 & 01:20:00.70 & 02:30:07.54\\
         DEC (J2000) & 58:48:39.3  & -05:29:03.6 & 13:03:12.0 & -02:40:45.1 \\
         z (Helio)   & 0.028203 & 0.069335 & 0.043041 & 0.058675 \\
         Distance, Mpc   & 125  & 306 & 190 & 259 \\ 
         $\mathrm{M_{g}}$ & $-19.3$ & $-20.0$ & $-19.1^{\dagger}$ & $-18.9$ \\
         $\mathrm{R_{27.7, z_0/h}}$,~arcsec & 32.5 & 18.2 & 25.5 & 21.5 \\
         $\mathrm{R_{27.7, z_0/h}}$,~kpc & 18.5 & 24.4 & 21.9 & 24.8 \\ 
         \hline 
         $\mathrm{V_{rot}},~$\kms$ $ & 149 & 195 & 211 & 178 \\ 
         $\mathrm{M_{dyn}},~10^{11}\mathrm{M}_\odot$ & 0.94 $\pm$ 0.03 & 1.92 $\pm$ 0.04 & 5.49 $\pm$ 0.40 & 2.08 $\pm$ 0.07 \\
         $\mathrm{M_{\star}},~10^{11}\mathrm{M}_\odot$ & 0.46 $\pm$ 0.11 & 0.87 $\pm$ 0.2 & 0.04 $\pm$ 0.01 & 0.12 $\pm$ 0.03 \\ 
         $\mathrm{M_{sph}},~10^{11}\mathrm{M}_\odot$ & 0.33 $\pm$ 0.01 & --- & 4.05 $\pm$ 0.31 & 1.36 $\pm$ 0.05 \\
         \hline
         $\rm 12 + \log(O/H)$ & $8.621 \pm 0.016$ & $8.626 \pm 0.074$ & $8.444 \pm 0.054$ & $8.507 \pm 0.077$ \\
         O/H grad., $\rm dex\ kpc^{-1}$ & $-0.020 \pm 0.003$ & $-0.011 \pm 0.008$ & $-0.001 \pm 0.004$ & $-0.011 \pm 0.006$ \\ 
         \hline % \hline
         Data source & DECAM-$z$ & HSC-$i$ (unw.) & Serb.-L(g) & HSC-$i$ \\
         $\mathrm{R_{knee}}$,~arcsec & 30 & 16 & 22 & 18 \\
         $h$,~arcsec  & 8.0 $\pm$ 0.2 & 4.1 $\pm$ 0.1 / $8.6 \pm 0.6^{\dagger}$ & 15.4 $\pm$ 1.1 & 6.1 $\pm$ 0.2 \\
         $\mathrm{z_0}$,~arcsec  & 1.76 $\pm$ 0.03 & --- & 1.51 $\pm$ 0.02 & 0.78 $\pm$ 0.01 \\
         $h$,~kpc  & 4.5 $\pm$ 0.1 & 5.5 $\pm$ 0.1 / $11.5 \pm 0.8^{\dagger}$ & 13.2 $\pm$ 1.0 & 7.0 $\pm$ 0.2 \\
         $\mathrm{z_0}$,~kpc  & 1.00 $\pm$ 0.02 & --- & 1.30 $\pm$ 0.02 & 0.89 $\pm$ 0.01 \\
         $\mathrm{\mu_0}$  & 20.3 $\pm$ 0.2 & 20.3 $\pm$ 0.3 & 23.3 $\pm$ 0.0 & 20.8 $\pm$ 0.0 \\
         $\mathrm{\mu_{0,z_0/h}}$  & 22.0 $\pm$ 0.1 & --- & 25.8 $\pm$ 0.2 & 23.0 $\pm$ 0.1 \\
         $(M/L)_{\textrm{bulge}}$  & 2.73 & 2.61 & 0.75 & 1.33 \\
         $(M/L)_{\textrm{disk}}$  & 1.30 & 0.51 & 0.42 & 0.39 \\ \hline
         $\chi^2_{W1}$ & 64.72 $\pm$ 16.09 & 15.14  $\pm$  5.29 & 4.09  $\pm$  1.19 & 474.11  $\pm$  38.08 \\
         p-value$_{W1}$ & 0.0 & 0.0 & $ 5 \cdot 10^{-9} $ & 0.0 \\
         $\chi^2_{W2}$ & 8.47 $\pm$ 2.37 & 2.05  $\pm$  0.62 & 1.34  $\pm$  0.43 & 30.5  $\pm$  5.25 \\
         p-value$_{W2}$ & 0.0 & 0.002 & 0.088 & 0.0 \\ 
         \enddata
   \end{deluxetable*}
   
To understand the evolution and formation of disk galaxies, it is essential to study all their subclasses, including those with extreme properties. The general principles governing ``normal'' systems should also apply to these extreme cases. That is why our picture of the evolution of galaxies will not be complete without understanding the extreme cases. Historically, research on disk galaxies was primarily focused on high-surface brightness systems.  Freeman's seminal paper in \citeyear{Freeman1970} postulated a universal central surface brightness for galactic thin disks. Almost two decades later, \citet{Bothun1987} discovered galaxies with much dimmer central surface brightnesses (fainter than 22 mag./arcsec$^2$ in the B-band), primarily in sparse environments. As instrumentation and detectors evolved over the years, it became evident that the number of low-surface brightness (LSB) disks was significantly underestimated, revealing the diversity of LSB galaxies.

Among LSB galaxies, those with giant disks (gLSBGs) deserve special attention. These galaxies have extremely extended LSB disks with radii up to 130~kpc \citep{Boissier2016}.  They have the largest disks in the Universe, and they are the largest concentrations of baryons (M$_{\textrm{b}} \approx 10^{11}$~\Msun) with substantial angular momentum. Their dynamical masses reach up to $10^{12}$~\Msun ~within the disk radius \citep{saburovaetal2021}. The assembly of very massive pressure-supported galaxies (giant ellipticals, cluster dominants) is explained by multiple major and minor mergers. However, the same scenario cannot be directly applied to gLSBGs, because to preserve the angular momentum in the end-product, the majority of infalling galaxies must have their orbital spins aligned, which is hard to reproduce in the $\Lambda$CDM cosmology \citep{Pawlowski2024}. 
Rings or disks of satellites were indeed found around M31 \citep{ibata2013}, CenA \citep{2018Sci...359..534M}, and NGC~4490 \citep{Karachentsev2024}, but even if one takes the combined baryonic mass in all satellites in the `plane’ of these systems, it will fall order of magnitude short of the typical mass of a gLSB disk. Thus the question of the formation of giant low surface brightness galaxies is still actively debated \citep{Kasparova2014, Galaz2015, Hagen2016, Boissier2016,Saburovaetal2018, Saburova2019, saburovaetal2021}. 

With the advent of the deep wide-field photometric surveys such as Hyper Suprime-Cam \citep[HSC,][]{2018PASJ...70S...1M} Subaru Strategic \green{Program} \citep{hsc2019} and the Dark Energy Spectroscopic Instrument (DESI) Legacy Surveys \citep{Dey2019}, our team conducted a search for gLSBGs by visual inspection of multi-color images \citep{saburova2023}. This was necessary because the training set for an automated search, e.g. using machine learning had been too small. Our efforts resulted in a large sample of gLSBGs and smaller LSB galaxies with similar morphology. Notably, we discovered the first edge-on gLSBGs, which became the subjects of this work. In this study, we focus on four edge-on galaxies characterized by extended LSB disks, with their central regions dominated by high surface brightness (HSB) disks. Among these, FGC~150 is a \emph{bona fide} gLSBG, while LEDA~1044131, also identified as a gLSBG by \citet{saburova2023}, presents a more complex case because of the disk asymmetry or lopsidedness which however shows regular rotation with the velocity amplitude of 150~\kms and can be traced in the GALEX UV image up to the radius of 60~kpc.  The remaining two galaxies have large disks and may represent intermediate cases between gLSBGs and smaller `classical' LSB galaxies.

The distinction between gLSBGs and intermediate-size LSB galaxies is rather arbitrary. We adopt a threshold of four disk radial scale lengths ($4h$) of at least 50~kpc as the criterion  (see \citealt{saburovaetal2021}). At the same time, \citet{Du2023} used the `diffuseness index' criterion $\mu_{B}$(0) + 5log (h) $>$ 27.0 where $\mu_{B}$(0) is the disk B-band central surface brightness, \citet{Zhu2023} relied on the \HI disk size radius. Observations demonstrate that gLSBGs do not form a distinct population but rather occupy the high-mass and large-size end of the distribution functions for disk systems \citep{saburova2023}. They also do not deviate by their position on the angular momentum versus baryonic mass relation \citep{Mancera2021}. A key question is whether gLSBGs are a particular instance of extended ultraviolet (XUV) disks found in normal spiral galaxies  \citep{GildePaz2005, Thilker2005}. At least two gLSBGs, Malin~1 and UGC~1382 were classified as XUV disks in the past \citep{Hagen2016, Boissier2016}. Dwarf LSB galaxies might also exhibit properties similar to gLSBGs. For example, the low-mass LSB galaxy Ark~18, located in the Eridanus void \citep{Egorova2021}, has an extended blue disk surrounding a high surface brightness component, similar to some gLSBGs. This suggests that gLSBGs might be extreme cases of a more common phenomenon occurring in normal galaxies. That is why here we study smaller but still extended disk sizes in addition to gLSBGs.

\begin{deluxetable}{lcccc}
\tablecaption{Observing log. All cases except the one marked by $^*$ (DEIMOS) correspond to SCORPIO-2. \label{log}}
\tablehead{
\colhead{Galaxy ID} &\colhead{Slit PA} & \colhead{Date} & \colhead{Exp.time}& \colhead{Seeing}\\ [-0.25cm]
\colhead{}&  \colhead{(deg)}  & \colhead{} &    \colhead{ (s)} &        \colhead{(arcsec)}
}
\decimalcolnumbers
\startdata 
(i)                &   67  &08.06.2021 &   10801   &   2.5\\
(ii)            &   105 &30.11.2022 &   3600    &   2.1\\
(ii)$^*$        &   91.2&25.10.2022 &   4800    &   0.7\\
(iii)                 &   159 &17.08.2020 &   8400    &   3.0\\
(iv) &   336 &07.11.2021 &   9900    &   1.5\\
\enddata
\end{deluxetable}

The edge-on orientation of these systems is beneficial because it enhances their surface brightness making possible the assessment of stellar population properties of LSB disks using optical spectroscopy not only in emission but also in absorption lines. The goal of this paper is to determine the conditions for LSB disk formation and select the most likely evolutionary scenario. To shed light on the origin of LSB disks and help to locate missing reservoirs of accreted gas, we focused on the stellar population properties, internal dynamics, and gas metallicity of four of these edge-on LSB galaxies.

The main properties of these galaxies are summarized in the first block of Table~\ref{tab:mainproperties}, which includes the coordinates listed in NASA/IPAC Extragalactic Database (NED)\footnote{\url{https://ned.ipac.caltech.edu/}}, redshifts obtained in the current paper and the corresponding distances, and the {\it g}-band absolute magnitudes calculated from DESI Legacy Survey data (a combination of Dark Energy Camera Legacy Survey, DECaLS DR9 and Beijing--Arizona Sky Survey, BASS, \citealp{2017PASP..129f4101Z}) for all galaxies except FGC~150, for which we used Sloan Digital Sky Survey (SDSS) DR8 \citep{Aihara2011} measurements. The last two lines of the block give the radius of the deprojected {\it g}-band isophote 27.7~mag~arcsec$^{-2}$ in arcsec and kpc. This isophote level is chosen the same as in \citep{saburova2023} to facilitate the comparison. The photometric systems of DECaLS, BASS, SDSS, and HSC-SSP are different at a few per~cent level \citep{2023PASP..135h4102T}, however, the dominating sources of uncertainties of integrated and surface photometry at such low surface brightnesses are the systematics of the background subtraction, and they significantly exceed the differences of photometric systems.

The paper is organized as follows. In Sect.~\ref{sec:obs} we give the details of the spectroscopic and photometric observations and the data reduction and analysis. Particularly in Sect.~\ref{subsubsec:inclination} we discuss the effect of the deviation from the edge-on position and the dust on the estimate of the photometric parameters. Sect.~\ref{sec:obs_results} is devoted to the discussion of the main results of the analysis -- the surface photometry, the properties of the stellar population, the kinematics profiles, and the oxygen abundance of gas and emission line excitation mechanisms. In Sect.~\ref{dis} we explore the potential formation scenarios for these galaxies. Additionally, we examine their positions on the Tully-Fisher relation, the relationship between radial metallicity gradients and disk size, and the correlation between stellar mass and gas metallicity.  %We summarise the main results of the paper in Sect.~\ref{sec:summary}.

\section{observations and data reduction}
\label{sec:obs}
\subsection{Long-slit spectral observations}
\label{subsec:long-slit_obs}
We performed long-slit spectral observations of all four galaxies with the multi-mode focal reducer SCORPIO-2 \citep{AfanasievMoiseev2011} mounted in the primary focus of the 6-m telescope BTA of SAO RAS (proposals 2020-2-456, 2021-1--527, 2021-2-595, P.I.: A.~Saburova). We used the VPHG1200@540 grism covering the 3600–7070~\AA\ wavelength range with $\sim5.2$~\AA\ spectral resolution (FWHM) and the 0.36~arcsec pixel$^{-1}$ scale. The slit width was 1~arcsec for all cases except LEDA~1044131, for which we used a narrower 0.69~arcsec-wide slit oriented along the principal plane of the HSB edge-on disk. The data were processed using the \textsc{idl} pipeline. The data reduction has been described in detail, e.g. in \citet{Zasov2015}, and includes bias subtraction, flat correction, wavelength calibration, and sky subtraction, which was performed using a technique, described in \cite{2011ASPC..442..143K}.

HSB and LSB disks in LEDA~1044131 are misaligned and we observed this galaxy along the LSB disk using DEIMOS \citep{2003SPIE.4841.1657F} operated on the Keck-II 10~m telescope (proposal 2022B-N149, P.I.: I.~Chilingarian). We used a 0.7~arcsec long slit and the 900ZD grating. The spectrum covers the 5000-8400~\AA\ wavelength range with the spectral resolving power $\text{R}\sim 3600$ (1.5--2.4~\AA\ FWHM) and a pixel scale of 0.12~arcsec~pix$^{-1}$. The data reduction was performed using a dedicated \textsc{idl} data reduction pipeline for DEIMOS built upon a generic long-slit/multi-slit data reduction toolkit originally developed for the TDS double-beam spectrograph operated at the 2.5~m telescope of the Moscow State University \citep{2020AstL...46..836P}. The substantial difference with the SCORPIO-2 data reduction was the use of atmospheric airglow lines for wavelength calibration instead of an internal arc lamp. Unlike SCORPIO-2 the DEIMOS's instrumental line spread function does not vary along the slit. Therefore the sky background subtraction was performed by averaging out the outer parts of the slit, which do not contain flux from the galaxy, and then subtracting the sky vector from each row of the frame.

We provide the observing log in Table~\ref{log}. For both DEIMOS and SCORPIO-2 observations, except the case of 2MASX~J02300754-0240451, the instrumental line-spread function (LSF) was determined from fitting the twilight sky spectra observed with the same setup fitted against a high-resolution solar template spectrum from \citep{1984sfat.book.....K}. For the observations of 2MASX~J02300754-0240451, the LSF was determined by fitting emission lines in the arc spectrum with Gauss-Hermite profiles, due to the lack of twilight observations for that night. All spectra were corrected for atmospheric absorption using a technique described in detail in \cite{2023ApJS..266...11B}.

\subsubsection{Analysis of spectral data}
\label{subsubsec:spectral_analysis}

Primary reduction yielded spatially resolved spectra and LSF estimates. Further analysis of each spectrum involved several sequential steps.  In these steps we used the {\sc idl} software package \nb{} full spectrum fitting technique \citep{Chilingarian2007a, Chilingarian2007b} with XSL simple stellar population models \citep{Verro2022}, simple stellar population models based on the empirical X-Shooter Spectral Library ($3500-24800$~\AA, R~$\sim 10000$) with the  \citet{Kroupa2001} initial mass function and PARSEC/COLIBRI isochrones \citep{Bressan2012, Marigo2013}. The \nb{} also fits emission line profiles with Gaussian-Hermite functions concurrently with stellar populations, using a bounded linear inversion to constrain emission line ratios  (e.g. for [N{\sc ii}] and [O{\sc iii}] doublets). Line fluxes, their uncertainties, and kinematics for absorption and one or several sets of emission lines are determined in the same minimization loop.

One of the primary objectives of this paper was to estimate the properties of the stellar population within the LSB disk region. To fit the absorption lines in this low-brightness area, we employed extensive binning along the slit to achieve reasonable signal-to-noise ratios. To obtain unbiased values for velocity dispersion and stellar population parameters, we corrected the data for galaxy rotation. Below we provide detailed information on the binning process and the correction for the rotation.

\begin{figure*}
    \centering
    \includegraphics[width=0.4\hsize]{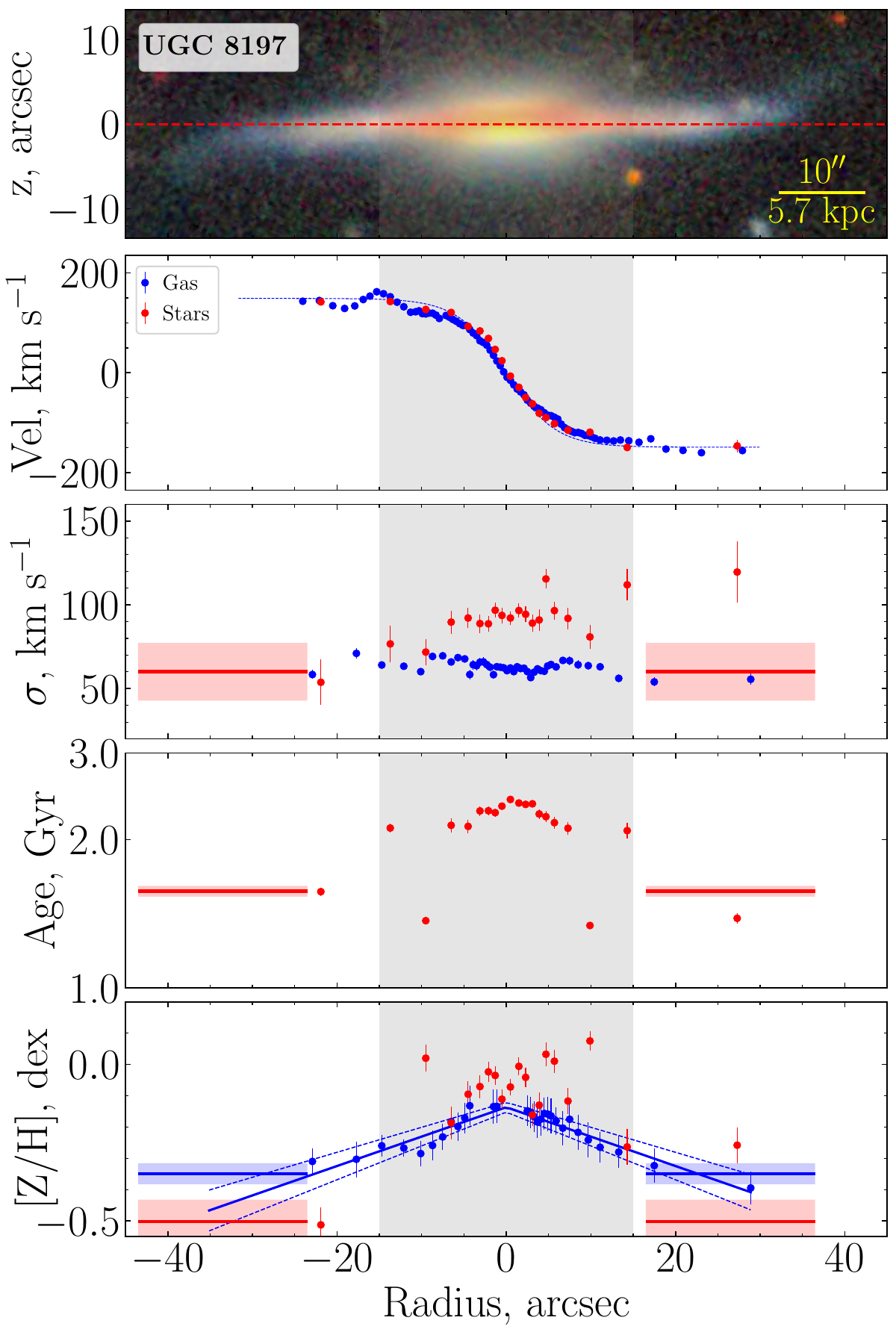}
    \includegraphics[width=0.4\hsize]{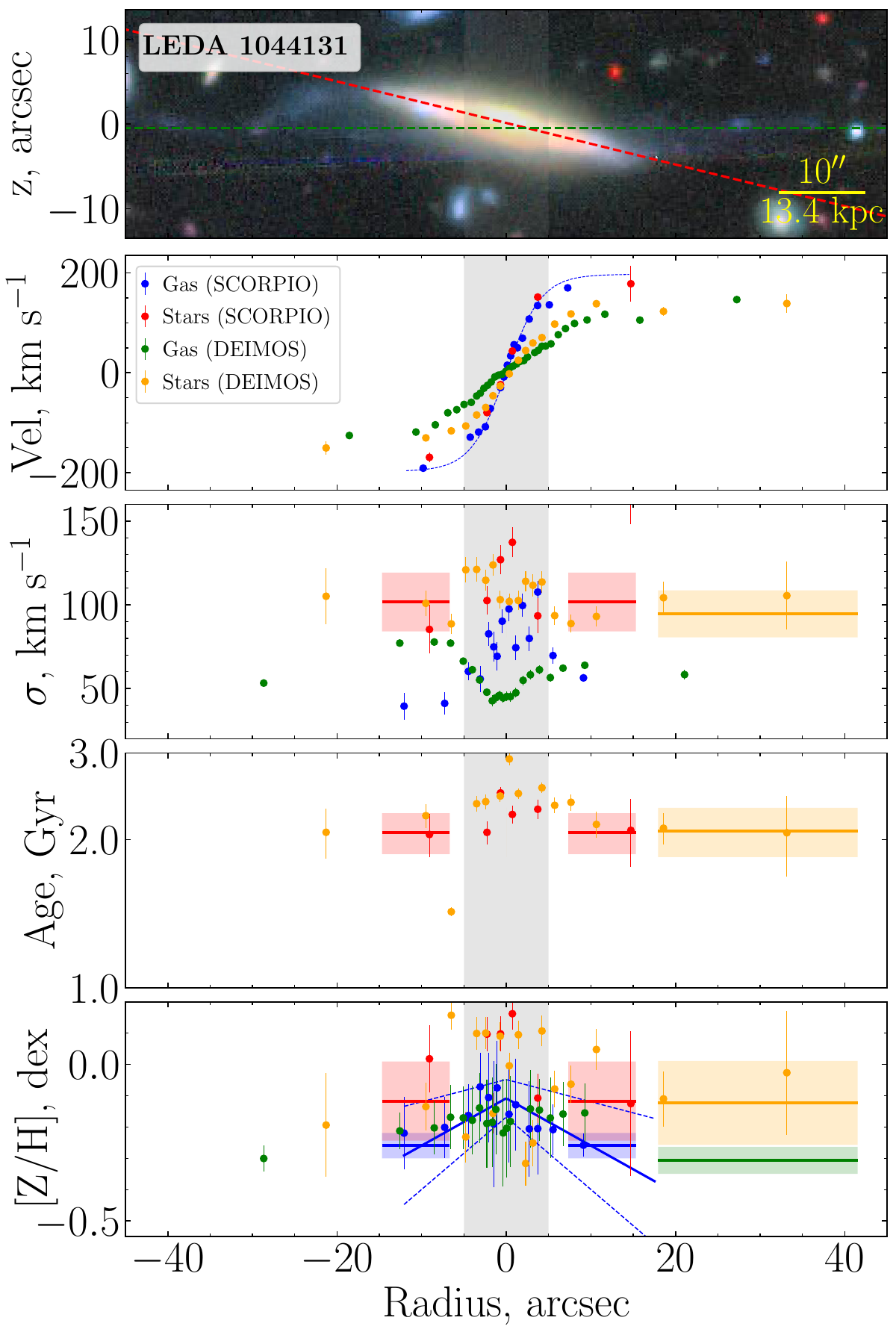}
    \includegraphics[width=0.4\hsize]{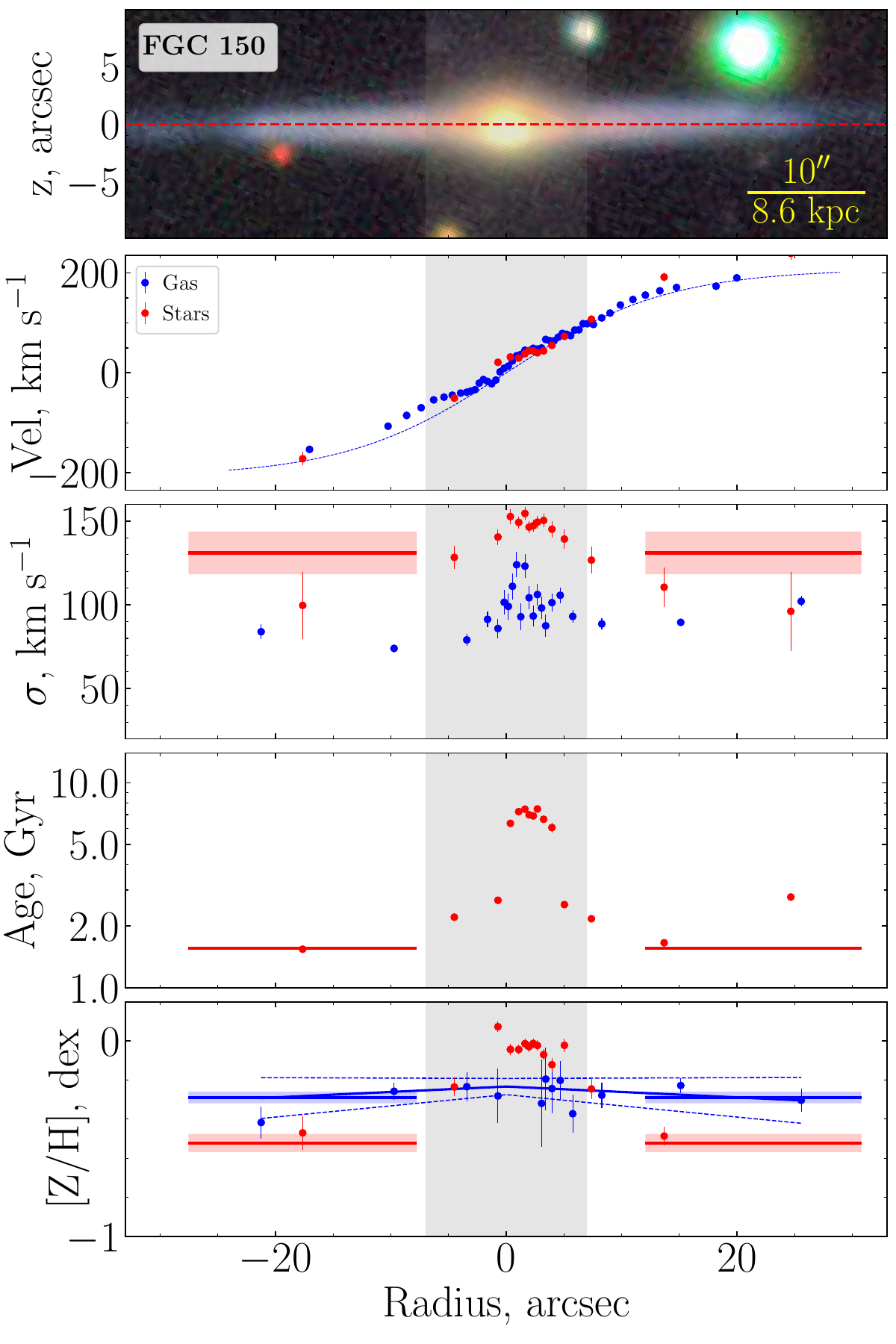}
    \includegraphics[width=0.4\hsize]{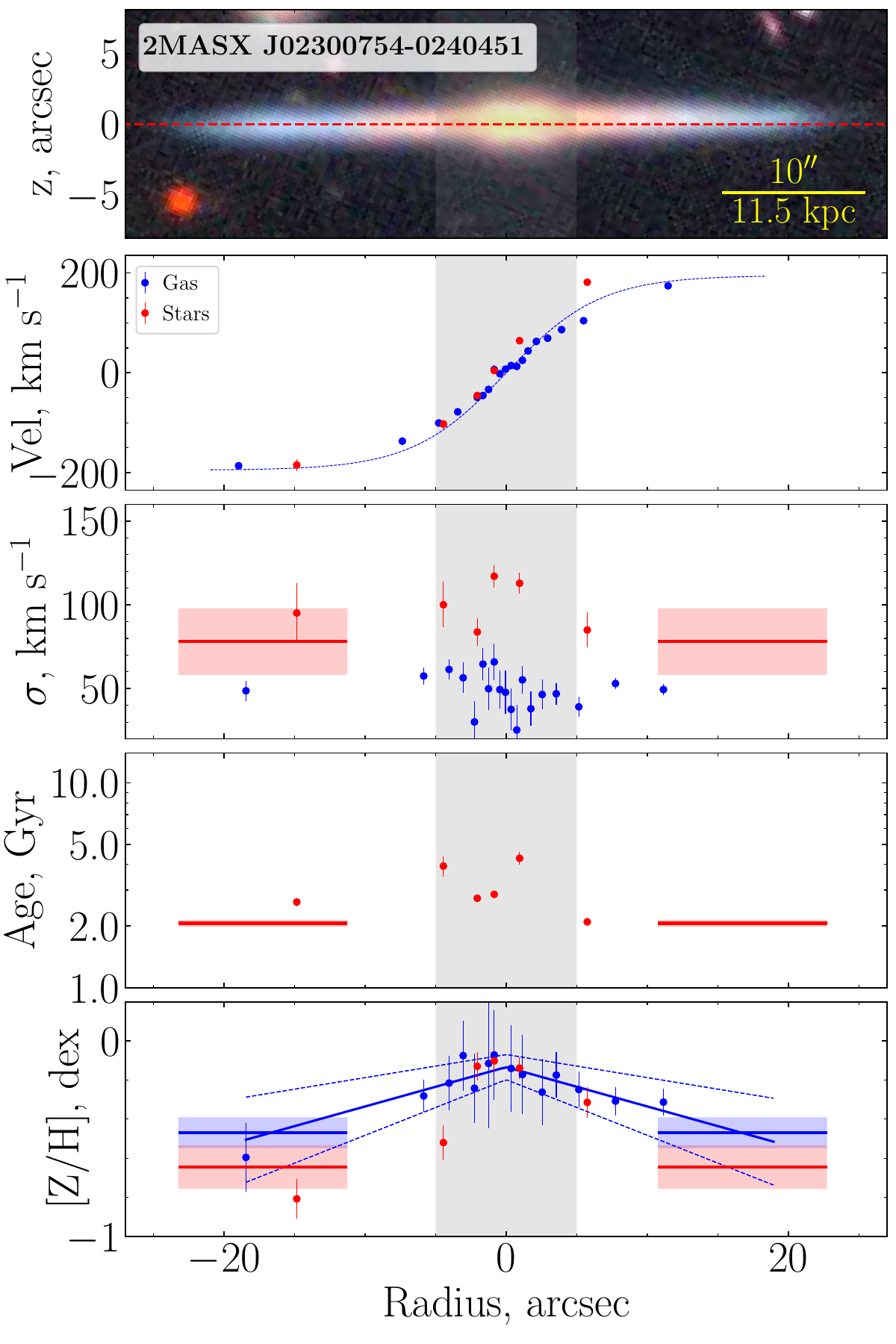}
    \caption{Results of the analysis of SCORPIO-2 long-slit spectra for each galaxy. Top to bottom panels show the image with slit location (red line), line-of-sight velocities, velocity dispersions, SSP-equivalent ages, and metallicities with relative gas-phase metallicities (corrected to $12+\log$(O/H)$_\odot = 8.69$). Blue symbols correspond to ionized gas and red to stellar parameters. For LEDA~1044131 we also show the results of the analysis of a DEIMOS long-slit spectrum (green slit on the image). The grey area is defined by photometric analysis (Fig.~\ref{fig:rad_profs}) and corresponds to the parameters of bulges or bars because they dominate in the integrated light. LSB disk spectra corrected for disk kinematics were integrated in large bins; colored lines show the results of the analysis in these bins.}
    \label{fig:spectra}
\end{figure*}

In the first step, we applied one-dimensional adaptive binning to the spectrum along the slit within a narrow wavelength range (accounting for the spread due to the galaxy's rotation) near H$\mathrm{\alpha}$. This allowed us to determine radial velocities from bright emission lines at various distances from the galaxy's center. The derived rotation curve $v(R)$ was approximated by the function:
\begin{equation}
    v(R)  =  2 v_{\textrm{rot}} \tanh((R - R_c) / h) + v_{\textrm{sys}}
\end{equation}
where $R_c$ is the center position; $h$ is the disk scale length; $v_{sys}$ is the systematic velocity; $v_{rot}$ is the maximum rotation velocity. Then the entire spectrum was corrected pixel-by-pixel for the rotation of the galaxy (Fig.~\ref{fig:spectra} top panels). Further steps were performed on the corrected spectra.

In the second step, adaptive binning was again applied at the continuum level near H$\mathrm{\beta}$. This enabled us to get optimal signal-to-noise ratio (SNR) in the stellar continuum for a more reliable determination of the properties of stellar populations (SSP age T~(Gyr) and metallicity [Fe/H]~(dex)), but in a smaller number of bins than in the first step (Fig.~\ref{fig:spectra} bottom panels).

In the third step, adaptive binning was once again applied in the close vicinity of H$\mathrm{\alpha}$ to obtain a larger number of bins with optimal SNR in the strong emission lines to refine the velocity dispersions from the first step with the stellar populations determined in the second step (Fig.~\ref{fig:spectra} middle panels).

In the final step, we determined the properties of stellar populations and emission lines in LSB disks as a single bin (the resulting spectra are shown in Fig.~\ref{fig:lsb} for red and yellow regions in Fig.\ref{fig:spectra}).  In Fig.~\ref{fig:spectra} we also show the colour-composite images of the galaxies (top panel) with the overplotted position of the slit (horizontal line) and the results of gas-phase metallicity estimates described below (bottom panel).

\subsubsection{Gas-phase metallicity measurements}
\label{sec:gas-phase_met}
\begin{figure*}
    \centering
    \includegraphics[width=0.8\hsize]{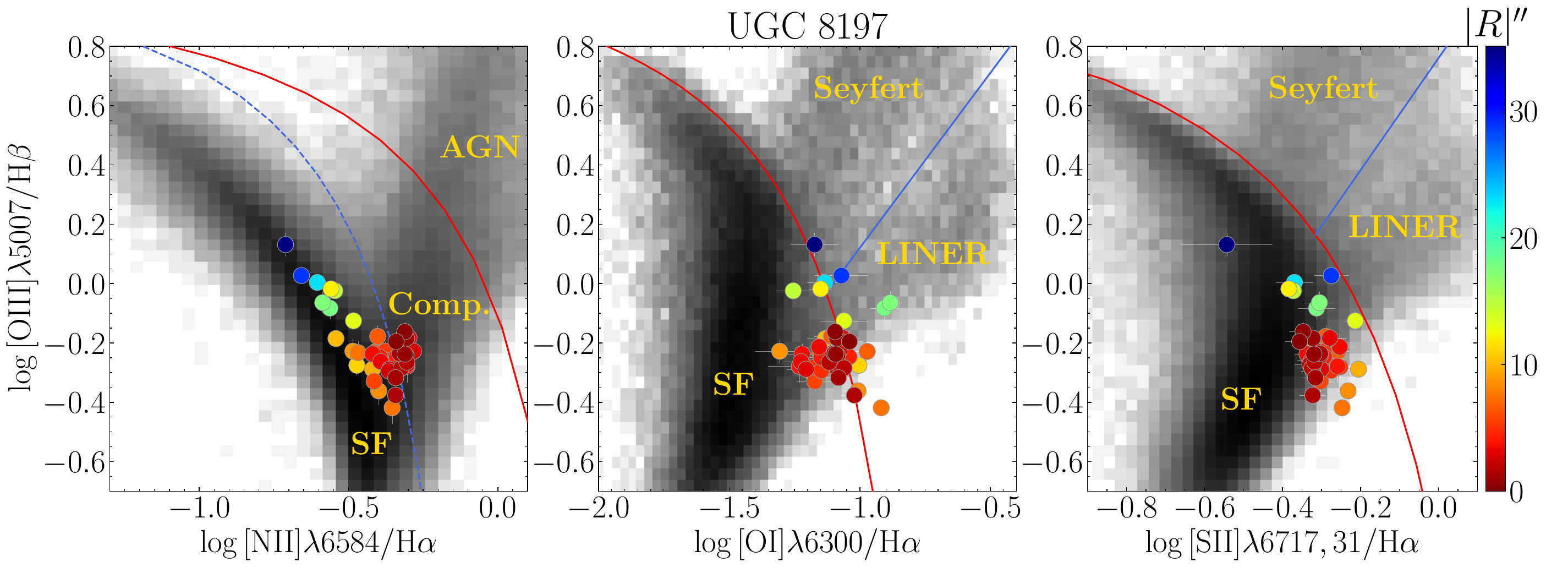}
    \includegraphics[width=0.8\hsize]{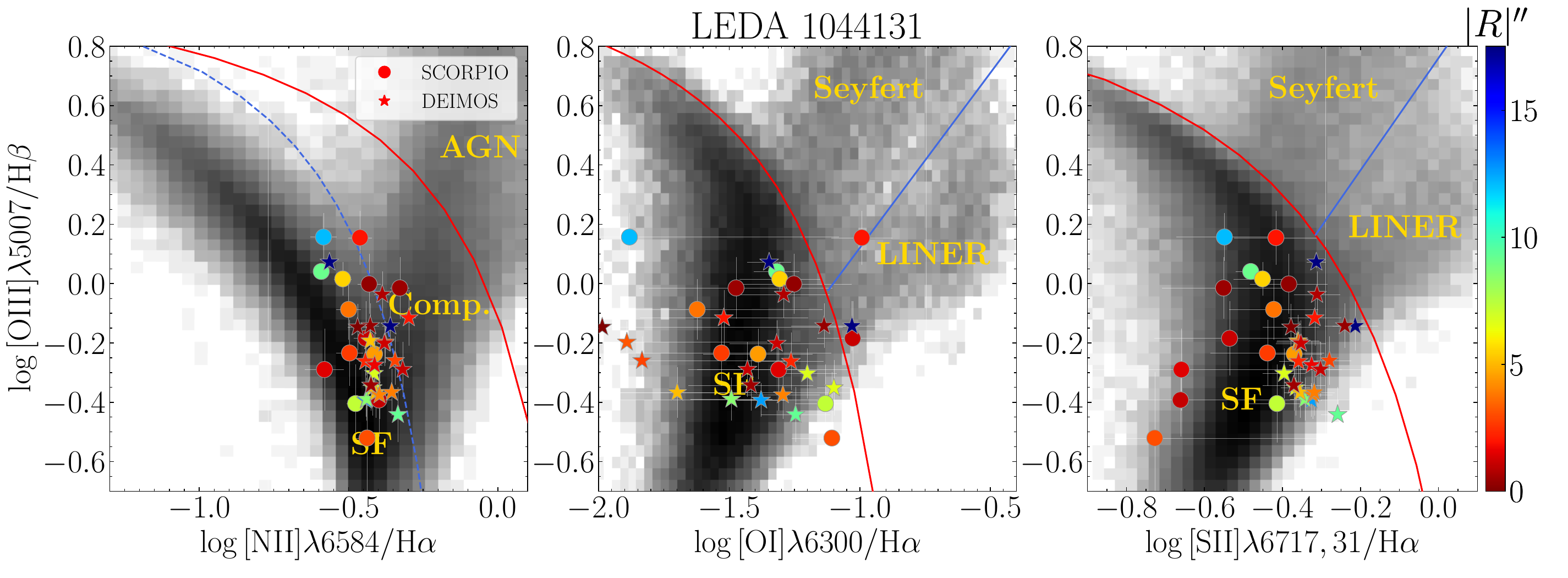}
    \includegraphics[width=0.8\hsize]{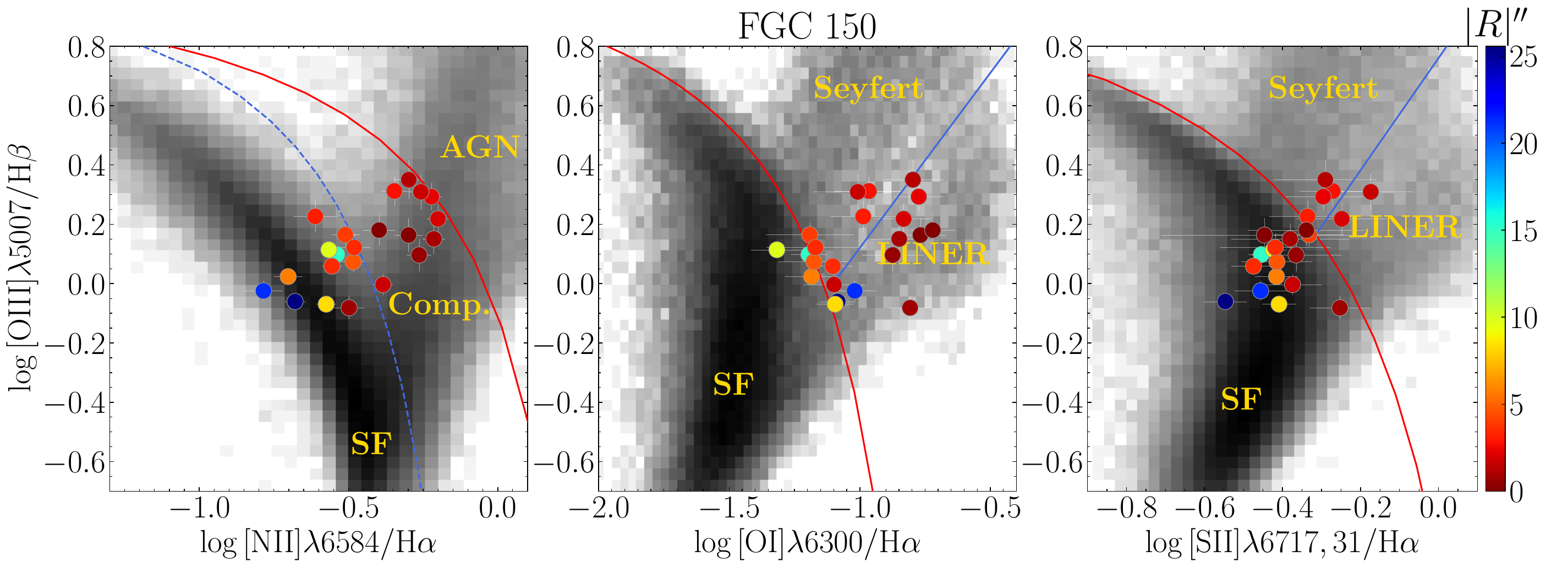}
    \includegraphics[width=0.8\hsize]{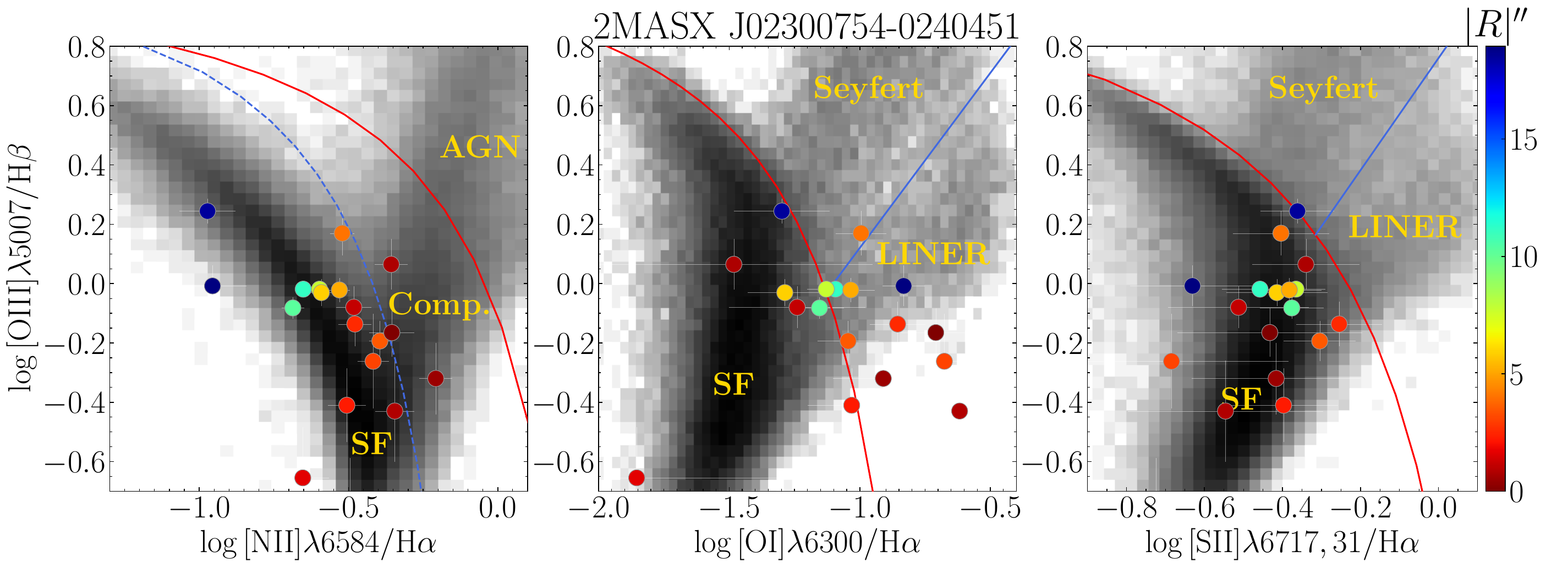}

    \caption{Resolved BPT diagrams along slits. Symbols are color-coded corresponding to their galactocentric distance. The blue dashed line in the BPT-BPT-NII plot represents the demarcation line introduced by \citet{Kauffmann2003}, which distinguishes between the star-forming and the so-called composite regions. The additional demarcation lines are taken from \citet{Kewley2001}. They separate composite and AGN regions on the BPT-NII diagram and categorize star-forming, Seyfert/LINER regions in the BPT-OI and BPT-SII diagrams. Density plots of emission line measurements from the RCSED database
    \citep[\url{http://rcsed.sai.msu.ru},][]{Chilingarian2017ApJS..228...14C} are shown in grey.}
\label{fig:bpt}
\end{figure*}

As a result of the simultaneous spectral fitting described above, we obtained emission line fluxes, which were used in the subsequent analysis of ionization mechanisms and ionized gas metallicities. \citet{BPT} first empirically demonstrated that ionization can be distinguished by several mechanisms using a few easily measured line ratios. \citet{Veilleux1987} refined this method to use line ratios that minimize the impact of dust reddening correction. In Fig.~\ref{fig:bpt} we present BPT diagrams for each galaxy in our sample with demarcation curves from \citet{Kewley2001} and \citet{Kauffmann2003}.
We excluded regions with line flux ratios indicating composite or shock gas excitation from the gas metallicity analysis. Finally, we checked other spectra for poor flux measurements and removed them from further analysis.

\begin{figure}
    \includegraphics[width = \linewidth]{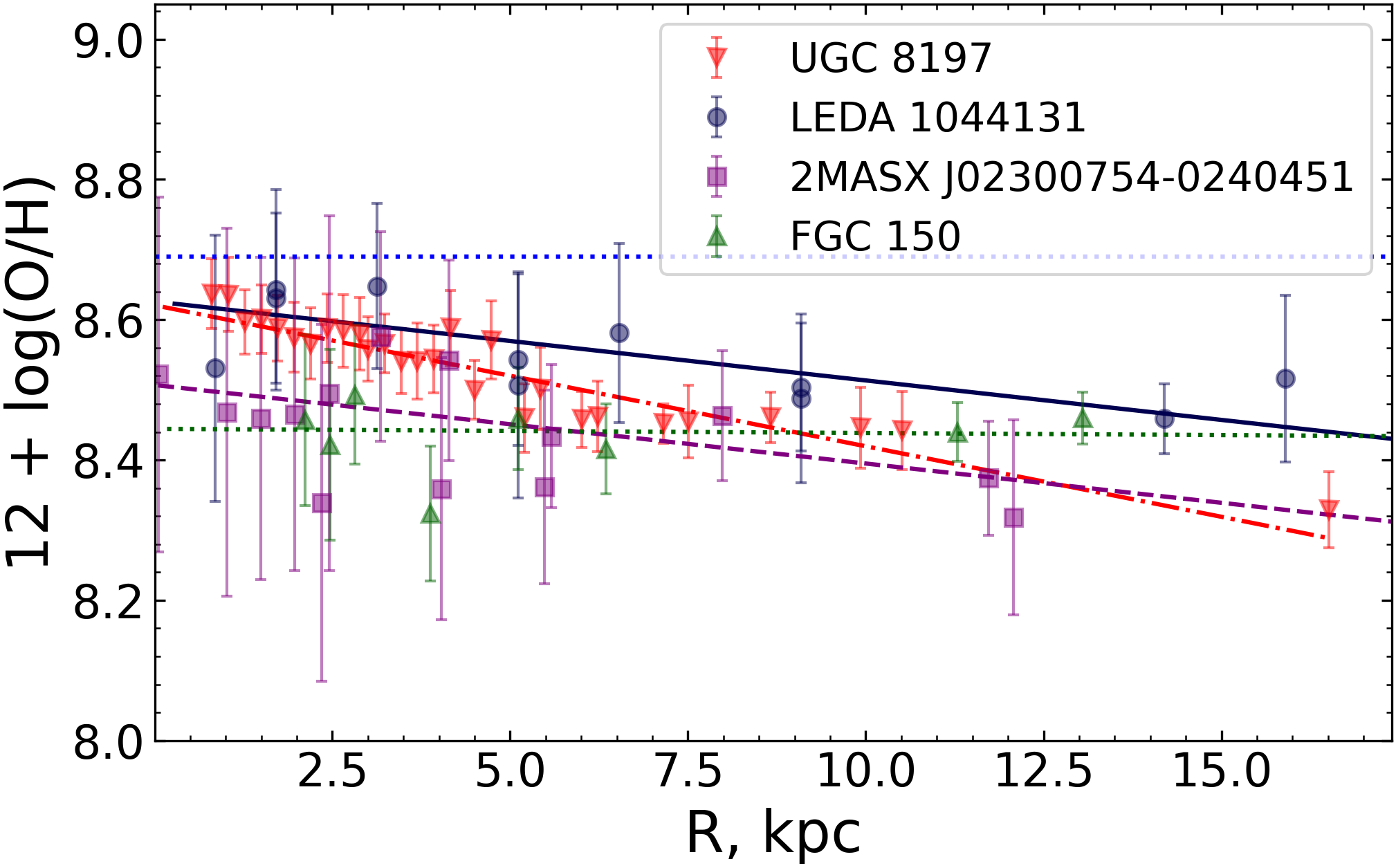}
    \caption{ The radial oxygen abundance 12 + log(O/H)  profiles for each galaxy determined with the S-calibration \citep{Pilyugin16}.
    In dark blue, red, green, and violet colors, we plot individual metallicity measurements at particular radial distances (symbols) and their linear best-fit approximation in different lines for UGC~8197 (dash dotted), LEDA~1044131 (solid), FGC~150 (dotted) and 2MASX~J02300754-0240451 (dashed) respectively. The blue dotted line reflects solar metallicity.}
    \label{fig:abundgradient}
\end{figure}

To measure gas metallicity, we first corrected emission line fluxes for dust attenuation in star-forming regions. We used \citet{Fitzpatrick1999} extinction curve with $R_V = 3.1$ accessible via \textsc{PyNeb} package \citep{Luridiana2015} with $\rm H\alpha/\rm H \beta$ line ratio and their theoretical ratio 2.86 for case-B recombination and temperature $10~000$~K \citep{Osterbrock2006}. For our sample, we derived gas metallicity with the S2 method from \citet{Pilyugin16}. The radial distribution of metallicity is shown in Fig.~\ref{fig:abundgradient} for each galaxy. We present metallicity gradients in Table~\ref{tab:BFMM} (the third block) in the units of $\mathrm{dex\ kpc^{-1}}$. Below, we compare these measurements with disk scale lengths in Sect.~\ref{dis:met}.

\begin{center}

\end{center}

\subsection{Surface photometry}
\label{subsec:surface_photometry}
\subsubsection{Observations with the Milankovi\'c telescope}
\label{subsubsec:milankovic_obs}
We performed deep photometric observations of FGC~150 with 1.4-m Milankovi\'c telescope equipped with Andor~iKon-L~936~CCD camera in the optical unfiltered `white' band  (L-band) with a field of view of 13.3 $\times$ 13.3~arcmin, which proved to be useful for photometric studies of LSB features \citep{2021POBeo.100..169V}. The observations were performed on 01.10.2021 and 18.10.2022. In total, we acquired 105 images with a duration of 200~sec each, which corresponds to the total integration time of 5.83~hours. 
To avoid large-scale background variations in the resulting co-add, the larger dithering strategy is necessary \citep[e.g.,][]{Duc2015}. Following it, we used the random offsets of up to 3.6~arcmin, centered on the galaxy.

The data reduction pipeline will be described in detail in B\'ilek et al., \textit{in prep.} In brief, following the basic photometric and astrometric calibration, we carefully model and subtract the background. It is a two-iteration process based on the original Milankovi\'c Telescope deep-imaging pipeline by \citet{muller2019}. The first iteration consists of the following steps:
\begin{enumerate}
    \item  A mask is created for each image from the objects detected by the \textit{sep} source extractor.
    \item The unmasked regions of each image are robustly fitted by a first degree polynomial and the image is divided by the polynomial.
    \item  The unmasked parts of all normalized images are robustly averaged into a master sky-flat.
    \item For each frame, the master sky-flat is multiplied by the respective polynomial and subtracted from the frame.
    \item All background-subtracted images are co-added.
\end{enumerate}
The resulting image already shows most of the final LSB features. Iteration two is the same as the first with three differences:
\begin{enumerate}
    \item A master mask is generated from the output of the first iteration to hide virtually all objects, including the faint diffuse emission. It is created in part by \textit{sep} again and in part by manual drawing. In the algorithm for the first iteration the mask from its Step 1. is unified with the master mask. 
    \item The same as in the Step 2. of the first iteration, but now a third-order polynomial is used.
    \item When making the final co-add frame, image weights deduced from the images themselves are applied.
\end{enumerate}

\begin{figure}
\centering
\includegraphics[width=\linewidth]{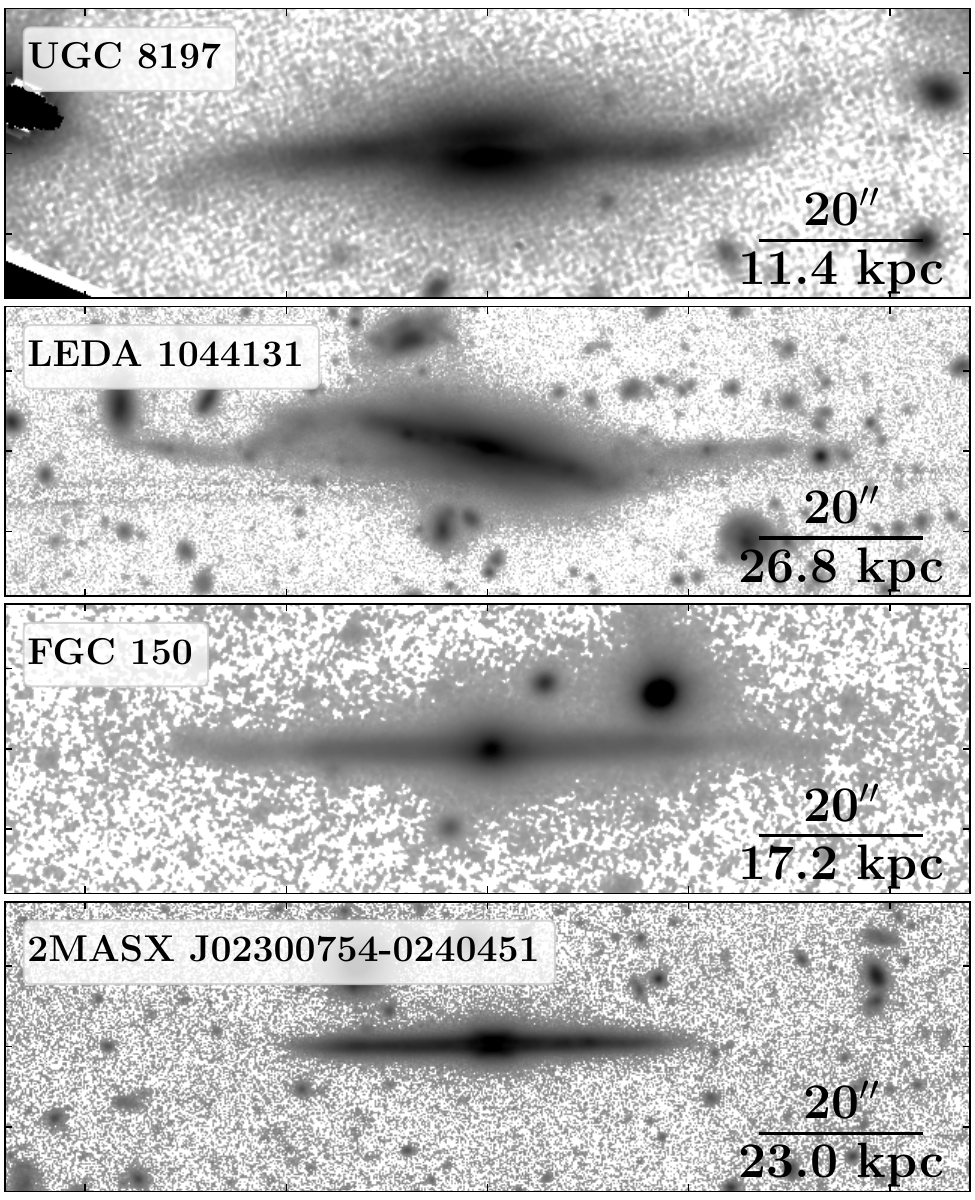}
\caption{Deep images of the galaxies, a `white light' $L$-band image for FGC~150 obtained with Milankovi\'c telescope, and co-added $g+r$ images from HSC~SSP for LEDA~1044131 and 2MASX~J02300754-0240451, and from BASS for UGC~8197.}
\label{fig_deep_images}
\end{figure}

The resulting deep image is shown in the third panel of Fig.~\ref{fig_deep_images}.

\subsubsection{Analysis of photometric observations}
\label{subsubsec:phot_analysis}
To proceed with surface photometry we combined the data from Milankovi\'c telescope with the publicly available archival images from DESI Legacy Surveys (UGC~8197) and HSC~SSP~DR2 (for LEDA~1044131 and 2MASX~J02300754-0240451). To estimate structural parameters of our edge-on galaxies we performed the analysis described in detail in \citet{Katkov2019}. % for FGC~150 and 

LEDA~1044131 has a prominent warp. We proceeded with the analysis by unwarping the galaxy's image as follows: First, we adopted the positional angle from the literature to define the main disk plane. Next, we identified the location of the warped disk on the image by measuring the maximum intensity across the disk. We then fitted the position of the warped part of the galaxy disk, as defined by these measurements, with a 2- or 3-degree polynomial to mitigate the impact of noise on the measurements of the disk's maximum intensity. This polynomial was unique for each side of the galaxy. Finally, we projected these warps onto the main plane, using the determined polynomials as references for the observed disk of the galaxy.

\begin{figure*}
    \centering
    \includegraphics[width = \linewidth]{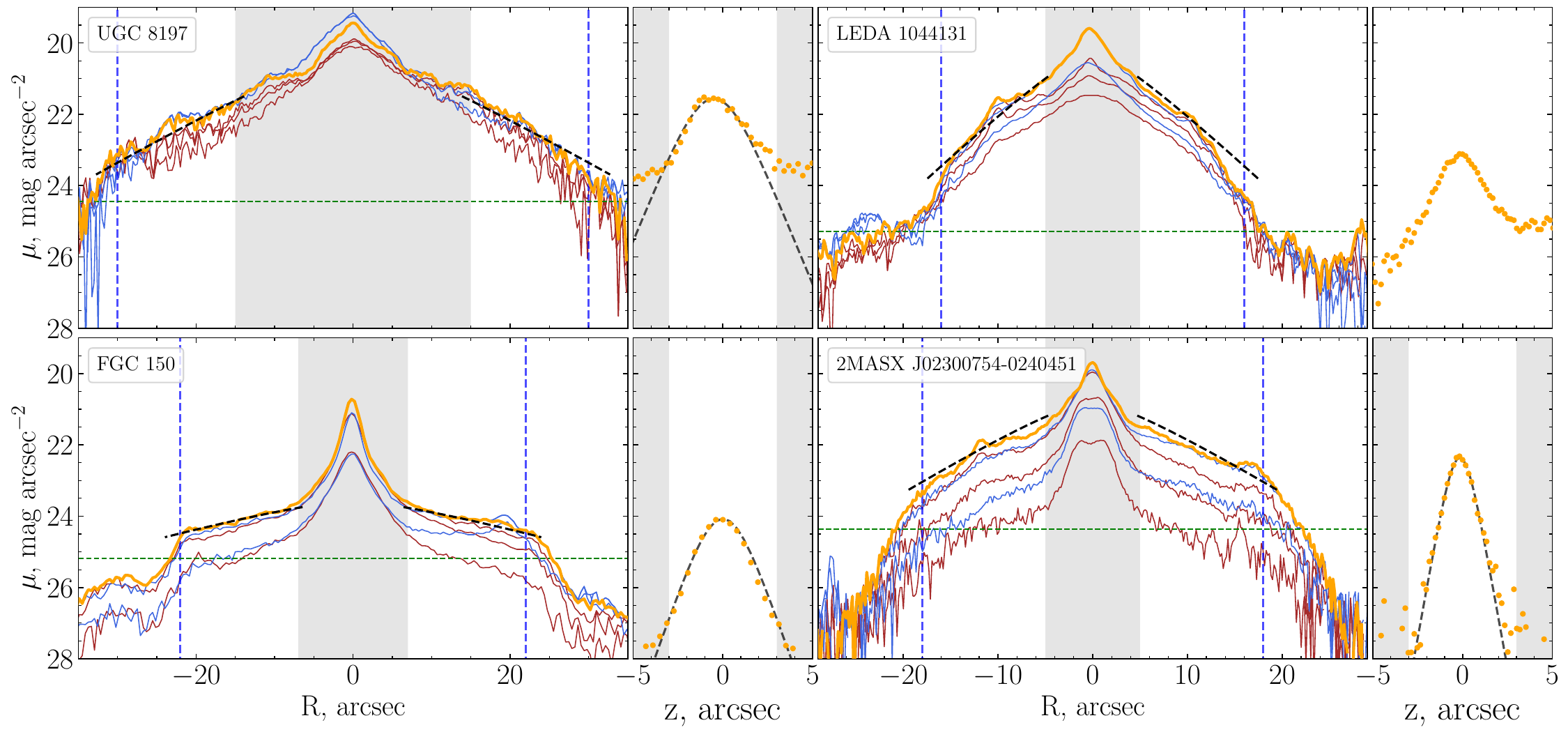}
    \caption{The radial and \red{vertical} photometric profiles \red{on the right side of each galaxy (see Fig.~\ref{fig_deep_images})}. The radial profiles was taken at different $z$ distances relative to the main planes of the galaxies (from $-2$ to $2$ arcsec with a $0.5$~arcsec step). The vertical profiles were taken at $R\sim -15 .. -16$~arcsec for all galaxies. Solid orange lines are profiles at the center plane ($z=0$), blue lines show profiles at negative $z$, red is for positive $z$, \red{and black dashed lines show models (Eq.~\ref{eq:rad_prof} for radial profiles or Eq.~\ref{eq:vert_prof} for vertical) with parameters from Table~\ref{tab:mainproperties}}. Grey shaded regions are excluded from analysis, the blue dashed lines indicate $\pm R_\textrm{knee}$. All disks were fitted between the shaded region ($R_\textrm{bulge}$) and $R_\textrm{knee}$. For LEDA~1044131 we also obtained the radial scale of the outer disk by fittting the region between $R_\textrm{knee}$ and $R=45$ arcsec. Green dashed lines are $27.7$ isophotes in (SDSS-g)-band, which were recalculated in images bands (from Tab.~\ref{tab_photometry}) and deprojected from edge-on orientation to face-on. $R_{27.7, z_0/h}$ are obtained from intersection between central profiles and $27.7$ isophotes.
    }
    \label{fig:rad_profs}
\end{figure*}

After the unwrapping, all the images were processed using the approach described in \citet{Katkov2019}. From the images (see the data sources in Table~\ref{tab_photometry}), we calculated radial profiles parallel to the central cut of the galaxy within the range $\Delta z \sim \pm 2$~arcsec, and perpendicular to this cut (parallel to the minor axis) within the range between $R_\textrm{bulge}$ and $R_\textrm{knee}$. All radial profiles exhibit a complex ``knee'' structure characterized by two distinct exponential segments (see Fig.~\ref{fig:rad_profs}). \red{The presence of a ``knee'' or break in the surface brightness profile raises an intriguing question. Most of the considered galaxies exhibit type II profiles, where the outer part of the surface brightness profile is steeper than the inner part. The exception is LEDA~1044131, which shows a more complex, asymmetric profile. According to \citet{Pohlen2006}, type II profiles are found in 66 per cent of late-type galaxies. These profiles are more common in disc-dominated galaxies located in the field \citep{2018AJ....156..118S}. However, there are no significant differences in the morphological asymmetries of galaxies with different profile types, indicating that environmental disturbances or recent satellite accretion do not significantly impact the formation of surface brightness profile breaks \citep{2020ApJ...897...79T}. Evidence suggests that type II breaks occur in regions where the age of stellar populations transitions from young to old \citep{Yoachim2012, 2022MNRAS.509..261P}. This is supported by the presence of a minimum in the colour and mass-to-light ratio radial profiles at the break radius, while the mass profiles do not show a corresponding break \citep{2008ApJ...683L.103B}.}

To determine the radial disk scales of all galaxies, only the inner section from $R_\textrm{bulge}$ to $R_\textrm{knee}$ was used, as this segment is dominated by the LSB disk.\footnote{LEDA 1044131 shows more complex behaviour of the surface brightness profile, so for this galaxy we also obtained the scale length for the outermost region between $R_\textrm{knee}$ and R=45 arcsec.}  The photometric profiles were described by the following function
\begin{equation}
    \label{eq:rad_prof}
    I_R(R, z) \propto \dfrac{R}{h(z)} K_1 \bigg( \dfrac{R}{h(z)} \bigg),
\end{equation}
where $h(z)$ is the exponential scale length at a given $z$, and $K_1$ is the modified Bessel function of the second kind and first order. 
We estimated the FWHM seeing quality for all images using stars in the field and we used a model that convolves the profiles (Eq.~\ref{eq:vert_prof} and Eq.~\ref{eq:rad_prof}) with a Gaussian PSF. The effect of seeing is minimal for the radial decomposition because the central zone is masked, but it is significant for determining the accurate value of the vertical scale $z_0$.

\begin{figure}
    \centering
    \includegraphics[width = \linewidth]{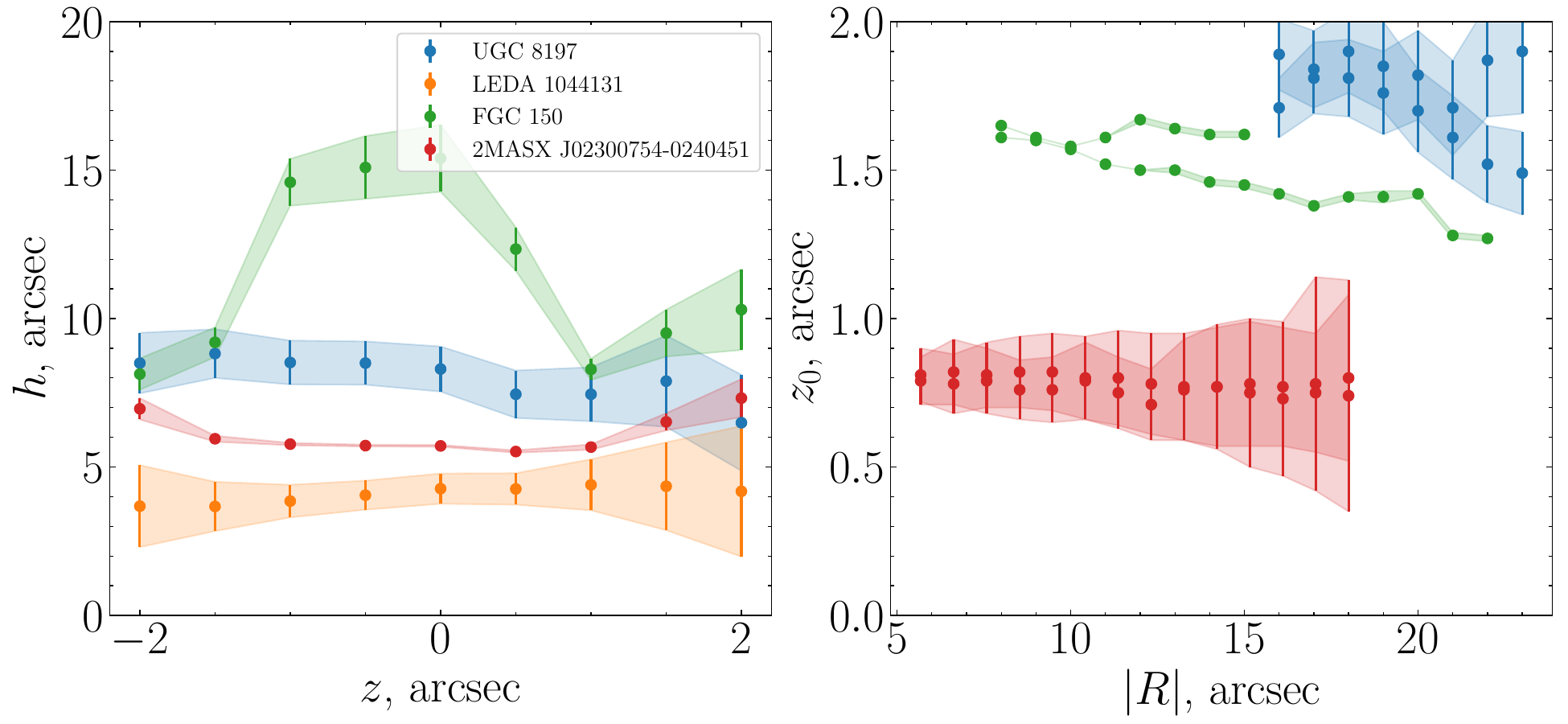}
    \caption{The scale lengths $h$ and heights $z_0$ for LSB disks along vertical distance $z$ and $R$, respectively. All galaxies (with the exception of  $h(z)$ for FGC~150) do not show trends and we used mean values in Table~\ref{tab_photometry}.}
    \label{fig:h_z0_profs}
\end{figure}

Fig.~\ref{fig:h_z0_profs} shows that the radial scale $h$ of all galaxies, except FGC~150, varies slightly within the chosen $z$ limits; thus, the mean value of $h(z)$ is presented in Tab.~\ref{tab_photometry}. The galaxy FGC~150 exhibits a symmetric decrease in $h(z)$. Both the image and the Balmer decrement H$\alpha$/H$\beta$ indicate that the absorption in this galaxy is minimal, suggesting that this behavior of $h(z)$ is not due to the presence of a dust in the disk (see Fig.\ref{fig:radial_scale_dust}). Therefore, for this galaxy, we used the radial scale from the central cut ($z = 0$).

The vertical profiles of the galaxies were analyzed similarly. We estimated the parameters of the vertical profiles for a given galactocentric distance $R$ using a multi-component model \citep{1942ApJ....95..329S}: 
\begin{equation}
    \label{eq:vert_prof}
    I_z(R, z) \propto \sum_{i} \mu_i (R) \textrm{sech}^2{\bigg( \dfrac{z}{z_{0i} (R)} \bigg)},
\end{equation}
where $\mu_i$ and $z_{0i}$ denote the mid-plane surface brightness and the scale height of each disk component. 

All galaxies are accurately described by a single-component model. Within the examined range of radial distances from $R_\textrm{bulge}$ to $R_\textrm{knee}$, no trends in $z_0$(R) are observed. Therefore, the mean value of $z_0$ is provided in Table~\ref{tab_photometry}.

\subsubsection{The impact of the lower inclination and dust on the analysis}
\label{subsubsec:inclination}
It is evident from the images that at least two galaxies, UGC~8197 and LEDA~1044131, are oriented not strictly edge-on and have an inclination angle slightly less than $i=90^\mathrm{\circ}$. Deviation from the edge-on orientation leads to the overestimation of disk thickness \citep[see][]{Kasparova2020}. To numerically evaluate this effect, we analysed model images of dust-free inclined disks with different axial ratios. Following the prescription for the vertical density profile of an isothermal disk in the equilibrium state by \citet{vanderKruit1981}, we created the model of an exponential disk with the 3D luminosity density given by
\begin{equation}
    I(R, z) = I_0e^{-R/h}\mathrm{sech}^2\left(\frac{z}{z_0}\right),
\end{equation}
where $I_0$ is the central luminosity density, $h$ is the disk scale length, and $z_0$ is its scale height. To simulate 2D images of a galaxy, we integrated the luminosity density along the line-of-sight in each pixel from different view points. For each image, we fitted the vertical profile with the $\mathrm{sech}^2$ law to estimate how the thickness estimate is biased.

\begin{figure}%[h!]
\centering
\includegraphics[width=1\linewidth]{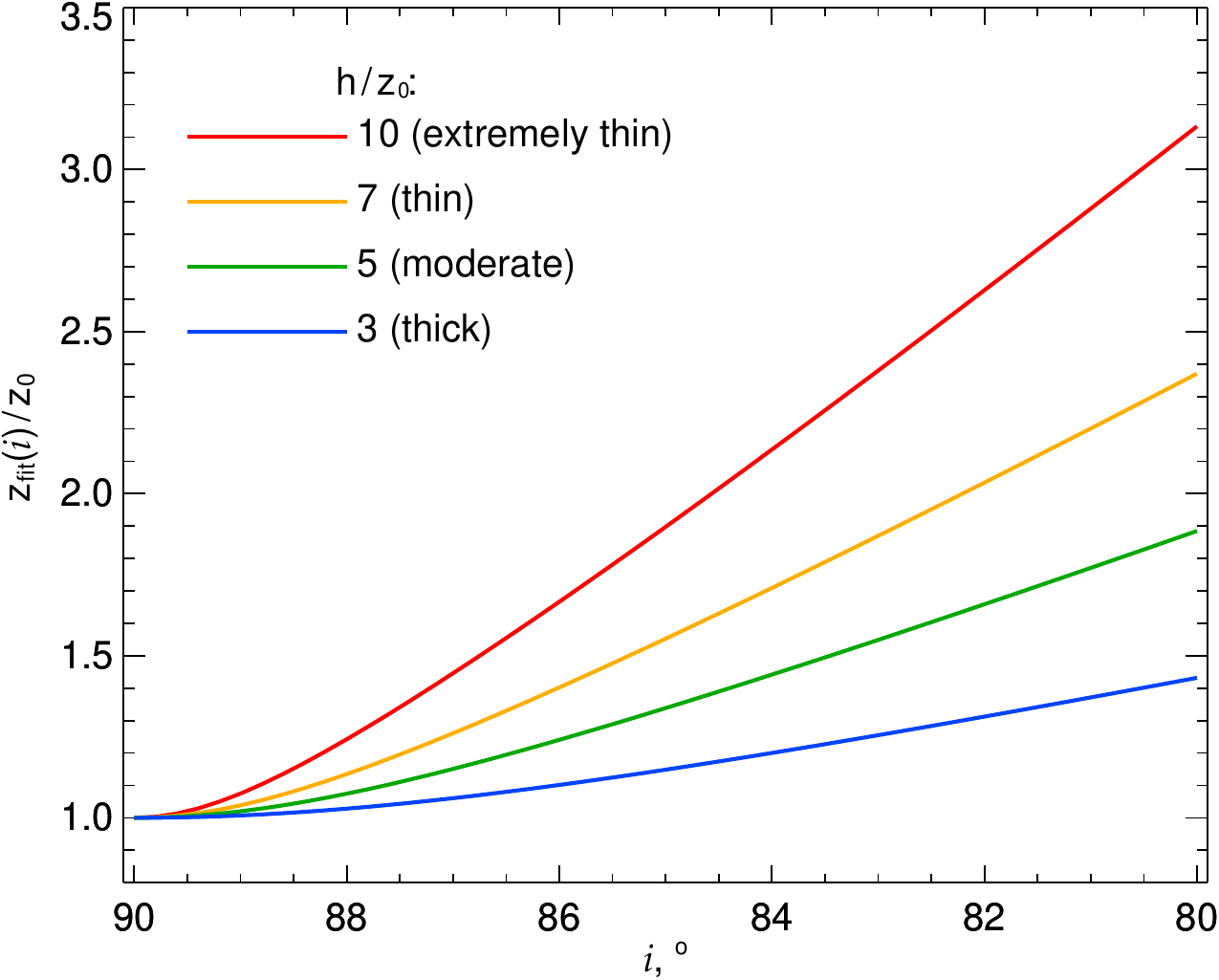}
\caption{Ratio of the best-fit scale height to its real value (when $i=90^\mathrm{\circ}$) as a function of inclination for four different $h/z_0$ ratios.}
\label{fig:z0_inclination}
\end{figure}

Fig.~\ref{fig:z0_inclination} shows how deviation from the edge-on orientation affects the computed thickness. As one can see from the figure, at a given angle, the effect increases with the major-to-minor axes ratio. For extremely thin galaxies with an axial ratio of 10, even a slight deviation of 5$^\mathrm{\circ}$ ($i$=85$^\mathrm{\circ}$) from the edge-on orientation can lead to an overestimation of their thickness by nearly a factor of 2. However, even at this inclination, the vertical profile of a galaxy is very similar to the $\mathrm{sech}^2$-like shape. On top of this, real-life measurements have some noise that makes it even harder to tell whether a galaxy is observed strictly edge-on.

\begin{figure}
\centering
\hspace*{0.75cm}\includegraphics[width=0.91\linewidth]{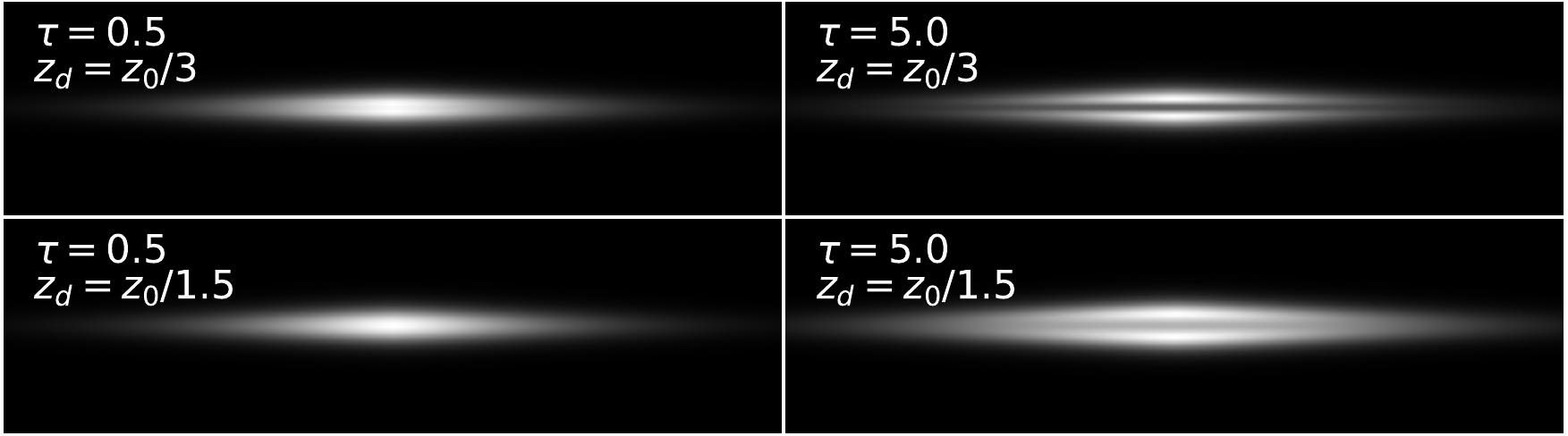}

\includegraphics[width=\linewidth]{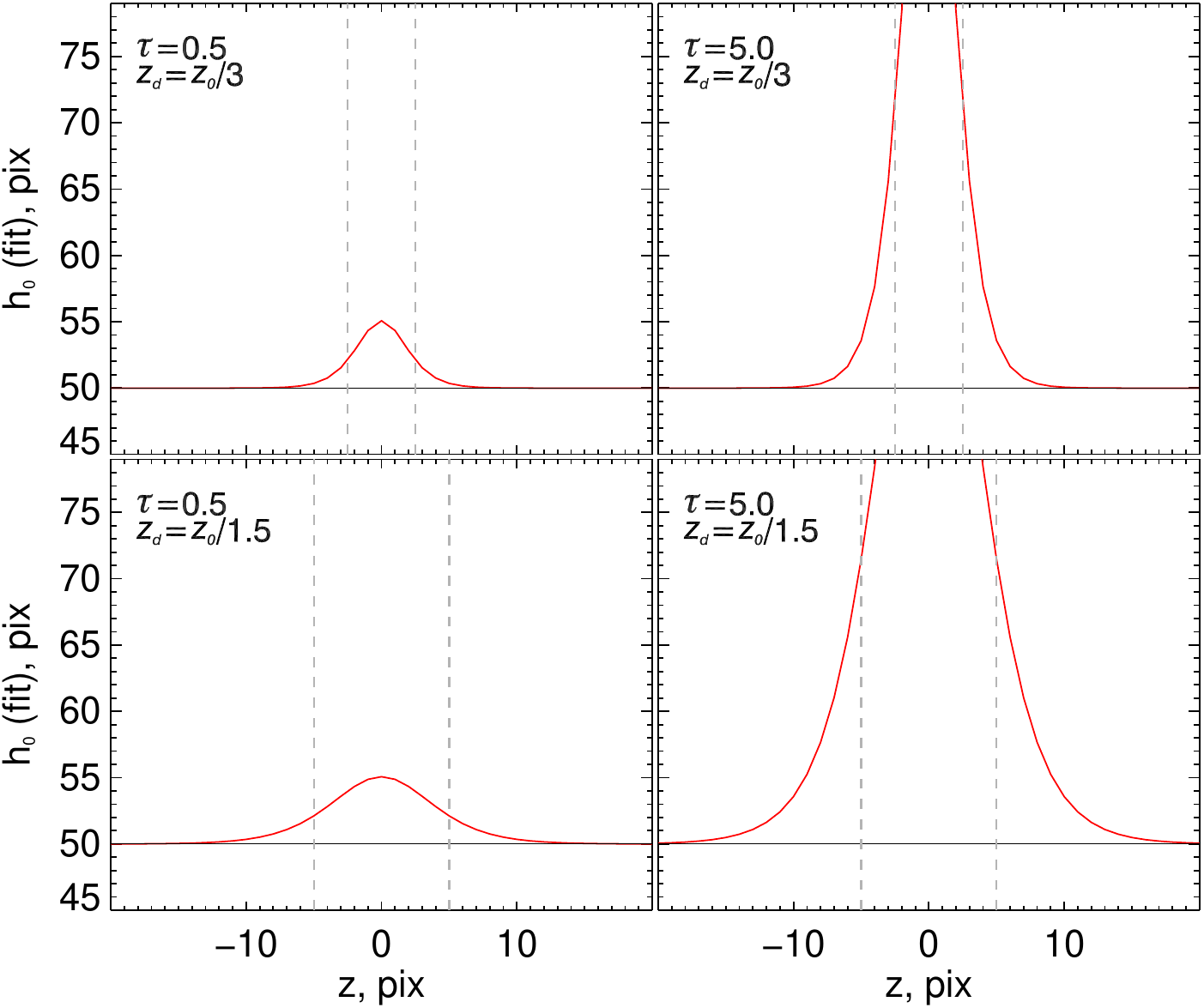}
\caption{\textbf{Top:} Model images of galaxies with different absorption coefficients $\tau$ and dust disk thicknesses $z_d$. \textbf{Bottom:} Disk scale length estimated from radial profiles at different heights $z$ (in pixels) above/below the mid-plane for the four models with different $\tau$ and dust disk thicknesses $z_d$ (red curves). The horizontal black curves show the real value of the scale length. Grey dashed lines correspond to $z_d$ of the models. The total semi-height of each model is 60 pixels.}
\label{fig:radial_scale_dust} \label{fig:models_dust}
\end{figure}

Three of our galaxies (UGC~8197, LEDA~1044131, and 2MASX~J02300754-0240451) have a prominent dust lane, which usually helps to visually analyse the inclination of a galaxy with respect to the edge-on position. On the other hand, this can make it difficult to estimate structural parameters. To evaluate the influence of dust, we added absorption to the radiative transfer in our model and simulated images of edge-on galaxies. In our model, the dust distribution follows the same equation as stars (Fig.~\ref{fig:z0_inclination}) but with a smaller vertical scale $z_d$. We then visually determined the mid-plane region to be masked and fitted the vertical profile with the $\mathrm{sech}^2$ law. We tested this approach with two different values of $z_d$: thin dust lane with $z_0$/3 and thick dust lane with $z_0$/1.5. For each value of $z_d$, we also made tests with $\tau_\mathrm{eo}$=0.5 and $\tau_\mathrm{eo}$=5, where $\tau_\mathrm{eo}$ is the total optical depth through the center of the edge-on disk. Fig.~\ref{fig:models_dust} demonstrates the model images for all four cases. Our modeling has shown that the presence of dust leads to the overestimation of disk scale length. This effect has earlier been found in \citep{Savchenko2023}. However, masking the attenuated region around the mid-plane helps to minimize this effect as well. Also Fig.~\ref{fig:radial_scale_dust} shows how the determined scale length depends on the height above/below the mid-plane where the radial profile is built. In addition, our models allowed us to estimate the inclinations for these three galaxies as $i\approx88.5^\mathrm{\circ}$ (UGC~8197), $i\approx85^\mathrm{\circ}$ (LEDA~1044131), and $i\approx89.5^\mathrm{\circ}$ (2MASX~J02300754-0240451).

We conducted a blind test in which our well-experienced team members responsible for surface photometry did not know the values of the structural parameters of the models. In all four cases, their approach showed a very good agreement with the actual values of $z_0$ and $h$. We thus conclude that our approach is robust and might be applied to real observations.

\section{Properties of individual galaxies}
\label{sec:obs_results}

\subsection{LEDA~1044131}

LEDA~1044131 is characterised by the ``normally'' looking inner part surrounded by the LSB structures of complex morphology demonstrating a strong warp. Such structures can be a result of gravitational interaction, e.g. with the in-falling satellites  \citep[see, e.g.][]{1989MNRAS.237..785O}. However alternative explanations of the warp formation were proposed, e.g. the misalignment between the stellar and gaseous disks \citep{2010MNRAS.408..783R} or between the disk and the halo \citep{1999ApJ...513L.107D}, bending modes or waves in the disks \citep{2004AA...425...67R}. The closest massive neighbor to LEDA~1044131 with a known redshift, LEDA~1043075, is located in 546~kpc in the projection of the SW direction (the velocity difference is about 900~$\textrm{km~s$^{-1}$}$) which could hardly lead to the formation of warped LSB structures. At the same time, LEDA~1044131 is surrounded by clumpy small objects with unknown redshifts which could be the satellite candidates or the remnants of the intruder galaxy. 

The gaseous kinematics of LEDA~1044131 is quite complex (see Fig. \ref{fig:spectra}). 
The emission lines show splitting in the region of the LSB disk where one component of the line shows slower rotation than another. It is likely related to the warp, we see the parts of the galaxy with different orientations in the line-of-sight. The stellar velocity dispersion remains almost constant with the value of 100~\kms, which can indicate the dynamical overheating of the disk (that means the stellar velocity dispersion is higher than needed for the disk to be gravitationally stable, see also the discussion). The radial distribution of the SSP-equivalent stellar age is also surprisingly flat with the mean value of around 2~Gyr. It can mean that the whole disk is forming stars at a rate that is independent of radius.The stellar and gaseous metallicities are solar in the center and become sub-solar in the periphery. From the BPT diagram it is evident that several disk regions are above the `maximum starburst line' (these regions were excluded from the metallicity estimation.

\subsection{FGC~150} 
FGC~150 was included in the flat galaxies catalogue with a major-to-minor axis ratio $a/b>7$ \citep{Karachentsev1993}. Our photometric analysis confirms this conclusion (see Table~\ref{tab_photometry}). Even though the inner part of the disk looks flat, the outermost regions show bending which is well visible in the deep image (see Fig.~\ref{fig_deep_images}, third panel). What is also visible from the image are two clumpy structures, near both sides of the disk which could be possible satellites that could give rise to the disk warp.

Stars show a flat radial gradient of the velocity in the inner part of the galaxy and a sharp decrease of the age from 10~Gyr in the center to 2~Gyr in the LSB disk where the stellar metallicity also decreases abruptly. At the same time the gaseous metallicity profile is very flat. The inner areas of the galaxy lie in the LINER region on the BPT diagram indicating possible presence of AGN.

\subsection{UGC~8197}
UGC~8197 is the least massive and extended galaxy of the sample with a disk size comparable to that of the Milky Way (see Table~\ref{tab_photometry}), however it demonstrates similar morphology as the more extended systems of our sample. 
As it is evident from the HSC image it has a prominent warp which can be a result of the gravitational interaction. 
The galaxy LEDA~214068 located North of UGC~8197 is a background galaxy ($\Delta v \sim 30000$~\kms) and could not lead to the formation of warp. The slit crosses the outer region of the bright clump seen NE from the galaxy which also appears to be a background source according to the spectral data. The rotation velocity reaches the plateau at the distance of 10~arcsec (5.7~kpc) from the center. Ionized gas shows an almost constant velocity dispersion radial profile. 
The radial profile of stellar velocity dispersion rises to the periphery on one side of the disk.  BPT diagram (Fig.~\ref{fig:bpt}) shows that most of the regions of the galaxy occupy the photoionization domain below the `maximum starburst line' which makes possible the ionised gas metallicity estimates. 
The central region of UGC~8197 possesses young stars with near-solar metallicity.  The gas has the solar value of metallicity in the inner region. At the same time, the LSB disk is characterised by a young metal-poor stellar population and a slightly sub-solar value of gas metallicity. It can indicate the ongoing star formation in the LSB disk. The presence of enriched gas evidences against the recent accretion of gas from the cosmic filament to this system. 

\subsection{2MASX~J02300754-0240451}
2MASX~J02300754-0240451 is the galaxy with a very extended `regular' disk  (see Table~\ref{tab_photometry}).  Its disk also does not show any significant warping at the periphery. No neighbour massive galaxies are observed within 500~kpc. Both ionised gas and stars show a steep increase in velocity in the inner region where the X-shaped bulge which is most likely a bar
is situated. The ionised gas has a change of the velocity gradient and the minimum of the velocity dispersion which is expected for bar regions \citep{chung2004,saburova2017}.  The bulge of  2MASX~J02300754-0240451 has quite low age of around 5~Gyr which decreases slightly to the periphery.  Most of the regions of 2MASX~J02300754-0240451 belong to the photoionization domain which allowed us to estimate the oxygen abundance. The ionised gas metallicity changes from nearly solar to slightly sub-solar in the outer regions. It argues against the fresh accretion of metal-poor gas from the filament or a dwarf gas-rich satellite.

\subsection{Tentative Active Galactic Nuclei}
\label{sec:check_AGN}
The BPT diagrams (Fig.~\ref{fig:bpt}) demonstrate that the innermost regions of three out of four galaxies lie in the locus of LINERs, which can be an indication of the presence of an AGN \citep{Heckman1980}. The findings of \citep{Ramya2011, Subramanian2016, saburovaetal2021} show us that gLSBGs tend to have too low central black hole masses for their stellar velocity dispersion. Therefore these AGNs can be powered by black holes in a low-mass SMBH regime.

The recent studies of low-mass AGNs variability \citep{2022ApJ...936..104W, Cleland_2022} use light curves derived from mid-infrared (MIR) unWISE (Wide-field Infrared Survey Explorer, WISE + NEOWISE) images \citep{Meisner_2023}, as well as optical Zwicky Transient Facility Forced Photometry \citep{masci2023new} light curves of galaxy centers to prove the activity of central black hole accretion processes. The AGN orientation can be arbitrary relative to its host galaxy inclination, however, \citet{2021A&A...650A..75G} found an observational lack of AGN Type 1 detections in spiral galaxies, because of host galaxy dust contribution. To check the evidence of AGN activity, we propose to use MIR variability, because the contribution of the host galaxy on the long timescales is expected to be constant, and we do not expect to find optical variability. 

We downloaded time-resolved unWISE coadds (unTimely catalog) using the \texttt{unTimely} Python package \citep{Meisner_2023}. Each coadded image includes more than 12 images taken during each WISE sky-pass, with the standard cadence of unTimely light curves being 6 months. We chose the aperture photometry radius as 3 arcseconds for each source because a bigger aperture radius leads to capturing more than one source for 2MASX J02300754-0240451, FGC 150. 
We downloaded ZTF Forced Photometry light curves of these objects and standard stars for calibration. We note that the ZTF light curves of LEDA~1044131 and 2MASX~J02300754-0240451 have huge gaps and poor coverage of dates, hereby they are out of our consideration. To calibrate ZTF Forced Photometry light curves we used a post-processing algorithm outlined in \citet{Demianenko2024} and standard stars correction (Demianenko et al. in prep.).

Because these time-domain surveys were not designed to check the variability of faint AGN on short time scales and do not have sufficient cadence, we do not pretend to deliver timescales of the variability ($\tau$) from stochastic processes (e.g. damped random walk or Gaussian Processes models). We applied the simplest approach: check forced photometry light curves to test the presence of variability. 
We used $\chi^2$ statistics, which are widely used to calculate the $p$-value, and we compared the $p$-value with a certain significance level to conclude the presence of variability \citep{Sokolovsky2017}. Our null hypothesis was that all observation points in the light curve are independent random variables $ x ~  \sim \mathcal{N}(\mu,\,\sigma^{2}) $, which form the constant. Then we used the weighted mean as a robust estimator of the model being a constant source obtained by the method of likelihood maximization.
The number of observations affects the $\chi^2$ calculation, therefore we applied the bootstrap technique to calculate the error $\sigma_{chi^2}$ of $\chi^2$ estimation.

For each photometric bandpass, we calculated the $p$-value of $\chi^2$ distribution with the number of degrees of freedom calculated as the number of observations in a given passband minus one. Since even after all calibration steps we expect to have systematic errors in light curves, we use reduced $\chi^2+3\sigma_{\chi^2} > 2$ as additional strict variability criteria.
Therefore, the variability confirmation criterion is given as the following conditions:
\begin{equation}
    \text{p-value} < 0.05,~
    \chi^2+3\sigma_{\chi^2} > 2.
\end{equation}
 
The Table~\ref{tab:mainproperties} lists reduced $\chi^2$ and $p$-value for $W1$ (3.4~$\mu m$) and $W2$ (4.6~$\mu m$)
~photometric bands respectively. We found strong variability in $W1$ passband in all sources, $W2$ variability in UGC 8197 and 2MASX J02300754-0240451. We did not find optical variability as expected.
This is consistent with AGN behavior in mid-IR  \citep{2019ApJ...886...33L}, where we expect to see the presence of variability in $W1$, a decrease in the variability amplitude in $W2$ compared to $W1$, and the RMS of $W1$ $\lesssim$0.1~mag. 

We found our galaxies as extended radio sources at  3GHz in the Quick Look catalog of the Very Large Array Sky Survey (VLASS) \citep{2020RNAAS...4..175G} without the signature of point sources at their centers. However, the absence of strong radio emission does not imply the absence of a low-mass AGN (as does an X-ray non-detection).  
We did not find any archival X-ray data deep enough to validate our AGN candidates. As a summary, we propose centers of these galaxies as potential AGNs, with deep X-ray follow-up observations required to confirm their nature.

\section{Discussion}\label{dis}
The primary objective of this paper is to explore the formation mechanisms of LSB galaxies with extended disks. We do not advocate for a single formation scenario for all the galaxies studied, particularly considering the range of scenarios proposed by \citet{saburovaetal2021} for gLSBGs. However, given the similar morphology of these systems, it is plausible to hypothesize that they share a common formation process. 

Several formation scenarios could be responsible for a building-up of the extended disks \citet{saburovaetal2021}, we shortly discuss them below. \\

{\bf Scenario a:} 
major merger with fine-tuned orbital parameters can lead to the formation of extended disks \citep{Saburova2018, Zhuetal2018}. {\it What is expected in the scenario?} \begin{itemize}
    \item Flat or shallow radial gradients of gas metallicity and stellar age/metallicity.
    \item The stellar disk should be dynamically overheated having the velocity dispersion ($\sigma_*$) significantly above the critical value needed for the gravitational stability of the disk.
    \item The collisional excitation of gas could be seen in the BPT diagnostic diagrams.
\end{itemize}  
{\bf Scenario b:} the formation of gLSBG can occur via minor mergers with gas-rich dwarf galaxies \citep{Penarrubia2006}.
{\it What is expected in the scenario?} \begin{itemize}
\item A significant radial gas-phase metallicity gradient should be observed because the metallicity of dwarf satellites is expected to be low.
\item The remnants of disrupted satellites should be seen in the velocity and gas metallicity profiles.

\item Stellar velocity dispersion $\sigma_*$ should be close to critical. Thus the disk should not be significantly dynamically overheated which is observed in most of the spiral and some of the lenticular galaxies \citep{Zasov2011}.
\end{itemize}  

\
{\bf Scenario c:} unusual properties of giant disks could be explained by the unusually large radial scale of a dark matter (DM) halo, a sparse dark halo with a shallow potential well \citet{Kasparova2014}. {\it What is expected in the scenario?} 
\begin{itemize}
\item Large dark matter halo radial scale and the low central density as reconstructed from the rotation curve modeling.
\item The stellar velocity dispersion is close to that expected for a marginally gravitationally stable disk.
\item The presence of age and metallicity radial gradients in stellar population which is indicative of the ``standard'' inside-out formation.
\end{itemize}  
{\bf Scenario d:} a giant disk is formed by accretion of metal-poor gas from a cosmic filament on-to a pre-existing elliptical or spiral HSB galaxy formed in a ``standard way'' by accretion and mergers \citet{Saburova2019}. 

{\it What is expected in the scenario?} 
\begin{itemize} \item Sharp gas metallicity gradient should be observed because the disk is built-up from pristine metal-poor material.

\item Kinematically decoupling such as counter-rotation should be common ($\sim$50\% of cases).
\item Inward non-circular motions should be seen in the gLSB disk.
\item The old stellar disks should be gravitationally stable at the same extent as in the case of the absence of a newly formed gaseous extension.
\end{itemize}

Let's examine the proposed scenarios using the data obtained in this study.

\subsection{Metallicity gradients}\label{dis:met}

The metallicity radial gradients could be used as a clue to a formation scenario of a galaxy. The considered galaxies have a low slope of the gas metallicity (see Table \ref{tab:mainproperties}). The different slopes support distinct scenarios of disk formation. In this work, we also test the merger scenario. For galaxies formed in this way, simulations predict a flat abundance gradient caused by the tidal torques transferring the matter to the centers of galaxies and mixing the gas in the peripheral areas of the disks \citep{rupke2010}. On the other hand, \citet{BresolinKennicutt2015} showed a tight relation between metallicity gradient and B-band scale length, which they speculated as an occurrence of local mass surface density --- ($\Sigma_M-Z$; \citealt{Rosales-Ortega2012}) or local stellar surface brightness ($\Sigma_L-Z$; \citealt{Pilyugin2014b}) metallicity relation. Thus to figure out if the obtained radial gradients of metallicity are indeed low and can speak in favour of the merger scenario the most informative is not the sole value of the gradient but its correspondence to the disk size and possible deviation from this relation. 

\begin{figure}
    \centering
    \includegraphics[width = \linewidth]{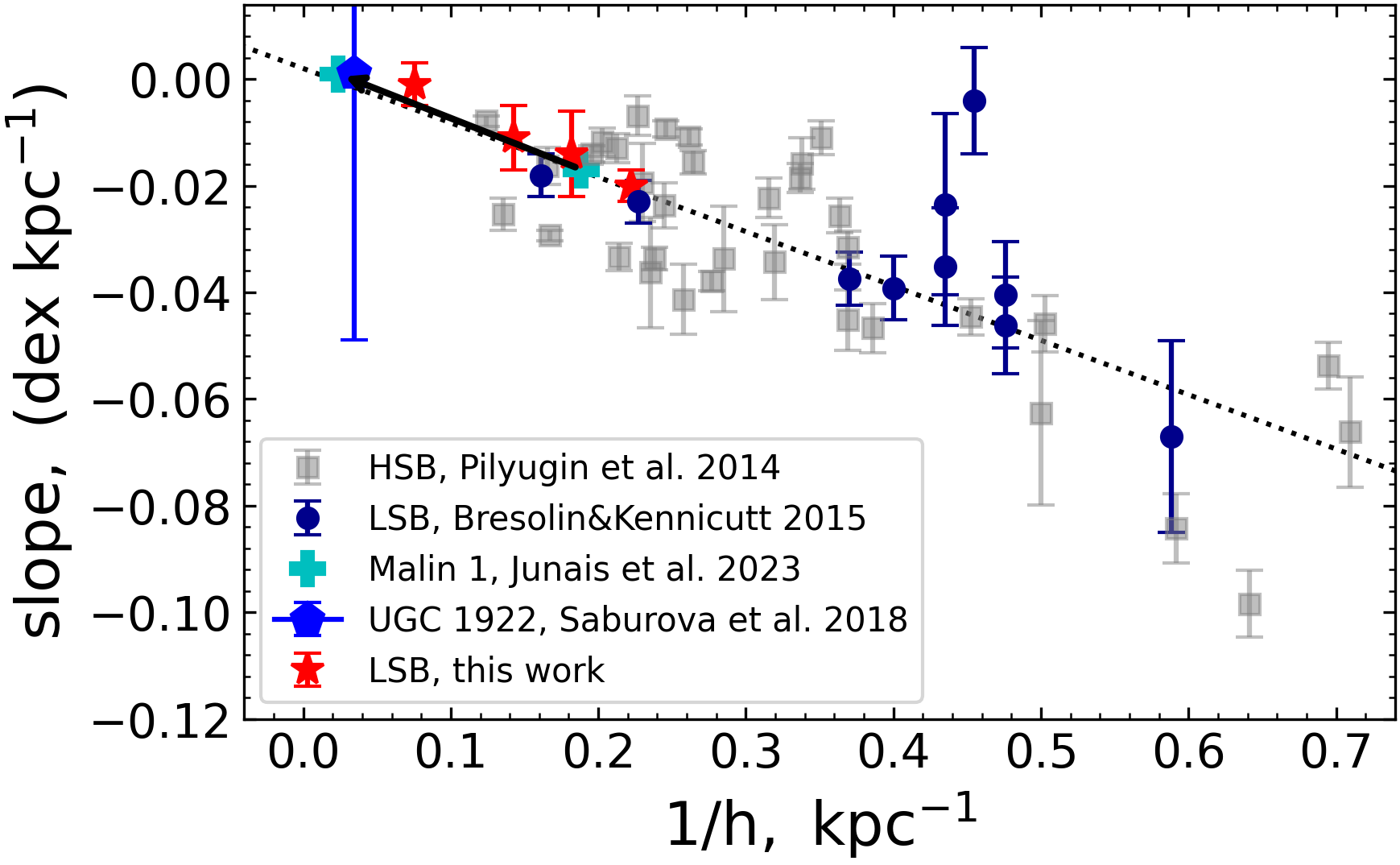}
    \caption{The relation of metallicity radial gradient and the disk scale length for different samples. By grey squares, we present the sample of HSB galaxies (slopes from \citealt{Pilyugin2014a}, and scale lengths from \citealt{Pilyugin2014b}). The \citet{BresolinKennicutt2015}'s LSB sample is plotted as dark blue circles, while ours is shown in red stars. We additionally overlaid two galaxies available in the literature: UGC~1922 \citep{Saburova2018} and Malin~1 \citep{Junais2024} as blue pentagon and cyan plus, respectively. Note that Malin~1 has two disks. The black dashed line is eq.~1 adopted from \citet{BresolinKennicutt2015}.}
    \label{fig:HZR}
\end{figure}

{In Fig.~\ref{fig:HZR}, we demonstrate the comparison of the radial ionized gas metallicity gradient 
%MB of stars or gas?
with the exponential disk radial scale length. Note that these scale lengths were not obtained in the B band (here we neglect the difference between the scale estimates in different bands). The line corresponds to the relation described by Eq.~1 in \citet{BresolinKennicutt2015}. The position of the late-type LSB galaxies from \citet{BresolinKennicutt2015} is shown by dark blue squares, while the HSB sample \citep{Pilyugin2014a, Pilyugin2014b} is overlaid as grey squares. The galaxies of our sample are shown by red stars. We additionally plotted values for 2
%MB probably: gLSBGs
gLSBGs from literature: UGC~1922 \citep{Saburova2018} and Malin~1 \citep{Junais2024} as blue pentagon and cyan plus, respectively. As one can see from Fig.~\ref{fig:HZR} all galaxies of our sample have the radial gradients of metallicity that agree well with that expected from their disk sizes according to the data on other LSB and HSB galaxies. It might testify against significant radial mixing of metals in the galaxies of our sample which would lead to the deviation of the galaxies towards the top area of the diagram. We note that observations of oxygen abundance radial profile flattening is more difficult for galaxies with huge disks.}

\begin{figure}
    \centering
    \includegraphics[width = \linewidth]{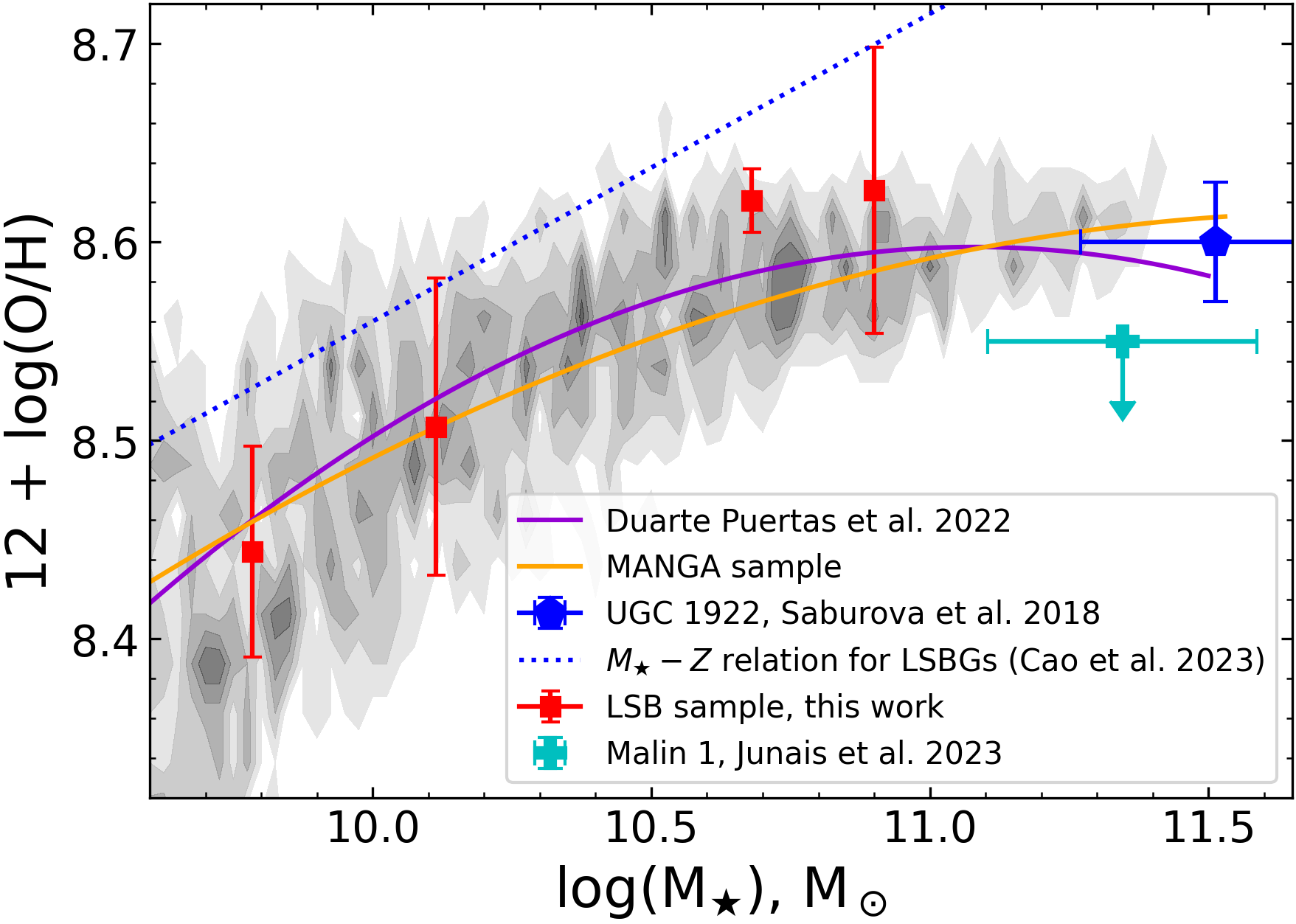}
    \caption{Mass-metallicity relation. The LSB sample analysed in this work is shown in red squares (data are summarized in Table~\ref{tab:mainproperties}). Malin~1 and UGC~1922 from the literature are superimposed with cyan plus and blue pentagon, respectively (\citealt{Junais2024} provided the O3N2-calibrated metallicity measurements for Malin 1, while \citealt{Saburova2018} is the reference for the oxygen abundance of UGC~1922. Masses for both galaxies are taken from \citealt{saburovaetal2021} In addition, we overlay the relation obtained for local Universe (z = 0) from \citet{DuartePuertas2022} (violet), which uses empirical calibrations, that agree well with the MANGA sample metallicities determined with the S2 calibration \citet{Pilyugin16} (superimposed background 2D source density distribution and its  2-nd degree polynomial mean approximation shown by the orange line).}
    \label{fig:MZR}
\end{figure}

The good agreement of the radial metallicity gradient with that expected for the observed disk scale lengths for normal spiral galaxies speaks against the \emph{recent} major merger scenario. The BPT diagram does not show the collisional excitation of gas which makes it even less possible.

The gas metallicity data are also useful to consider the possibility of another scenario -- the formation of the extended disks containing the emission regions from the metal-poor gas of the cosmic filaments.  Previously we already noted that the gas-phase metallicity is only slightly sub-solar in the outer regions of the considered galaxies which does not get in line with that scenario.

In Fig.~\ref{fig:MZR}, we also compare the central gas metallicities (which are close to the mean within effective radius due to the relatively low radial gradients) with the stellar masses. We derived the stellar mass estimates from the DECaLS and SDSS integrated magnitudes in the bands used for the photometric analysis, combined with the photometric parameters of the disks and the mass-to-light ratios of the bulge and disk regions obtained from the spectral fitting (see Table~\ref{tab_photometry}). For the comparison, we also show the relations found by \citet{ DuartePuertas2022,2023ApJ...948...96C}. As one can see from Fig.~\ref{fig:MZR}, the positions of our galaxies agree well with that found by \citet{DuartePuertas2022} for other galaxies, which means that the gaseous metallicities are not too low for the expected stellar mass which could be expected in the recent accretion from the cosmic filament scenario. At the same time, the position of our systems deviates from that of the blue edge-on LSB galaxies that demonstrate inefficient oxygen abundance enhancement \citep{2023ApJ...948...96C}, which can indicate the different evolution history of these types of LSB galaxies.

Another argument against the metal-poor gas in-fall is that the metallicity gradients are not lower than expected from the observed disk sizes.

\subsection{On the gravitational stability of the disks}

 As previously noted, the difference between the stellar velocity dispersion and its threshold value for the gravitational stability of the disk serves as a crucial diagnostic criterion that can illuminate the formation scenario. A thin isothermal stellar disk is locally stable when the radial stellar velocity dispersion,  $\sigma_r$,  exceeds the critical value:
\begin{equation}
(\sigma_{r})_{\rm crit}=Q_{T,\rm crit}\cdot 3.36G \Sigma_d / \varkappa,
\end{equation}
where $\varkappa$ is the epicyclic frequency, $\Sigma_d$ is the surface density of the disk, and $Q_{T,\rm crit}$ is the Toomre's stability parameter which is equal to unity for small radial perturbations of a thin disk. 

In more realistic cases,  $Q_{T, \rm crit}$  for a marginally stable disk may differ from unity due to non-radial perturbations, finite disk thickness, or the presence of multiple disk components \citep[see discussion in][]{2017MNRAS.472..166G}. Additionally, the Toomre expression is derived under the epicyclic approximation, which is valid when the radial velocity dispersion is small compared to the circular rotation velocity. Numerical simulations bypass these restrictions, demonstrating that for more realistic models, the Toomre parameter typically ranges between 2 and 3 for stellar or two-component stellar+gas disks, depending on disk parameters and initial conditions  \citep[see, e.g.][and references therein]{Khoperskov2003,  Inoue2016, 2017MNRAS.472..166G, 2017PhyU...60....3Z}.  
Current estimates of the Toomre parameter for several stellar-gaseous disks of LSB  galaxies almost universally demonstrate that their disks are gravitationally stable at all radii. The minimum values of the stability parameter $(Q_T)_{min}$ are close to 3, with the minimum $Q_T$ reached at some distance from the center, in the region corresponding to R/h$\sim$2–4 \citep{2017MNRAS.472..166G, 2023MNRAS.526...29A}.  For super-thin LSBs observed edge-on, the data are contradictory. For instance, for FGC~2366, $(Q_T)_{min}\sim 3$ \citep{2023MNRAS.526...29A}, while for UGC~7321, $Q_T$ is significantly higher ($Q_T>>1$) \citep{2014MNRAS.439..929G}.

For the galaxies considered in this study, only the upper limit $(Q_T)_{up}$ can be estimated for a given distance from the center, as there is no data on their gas density, whereas the contribution of \HI to the rotation curves of gLSB galaxies can be comparable to that of their stellar disks \citep[see f.e.][]{Pickering1997}.

We estimated the disk surface density from the photometric data presented in Sect. \ref{subsubsec:phot_analysis} and the mass-to-light ratios obtained from the spectral fitting (see Table \ref{tab:mainproperties}). In this estimation, we took into account the extinction calculated from the spectra using the Balmer decrement. 
Formally calculated $(Q_T)_{up}$ values for the corresponding R at the radial distances of the LSB bins (see Fig. \ref{fig:rad_profs}) are the following: 20 (R=10.2~kpc), 135 (R=12.6~kpc), 9.5 (R=9.2~kpc), 70 (R=16.6~kpc) for LEDA~1044131, FGC~150, UGC~8197 and 2MASX~J02300754-0240451. Despite a rough estimate of the values, they clearly indicate a strong dynamic overheating of the disks. Note that these estimates were done under the assumption that the observed stellar rotation velocities are close to circular ones and that the stellar velocity dispersion is isotropic in the first approximation. In the stellar disks of galaxies, $\sigma_r > \sigma_\phi$  (for galaxies with a flat rotation curve in the epicyclic approximation $\sigma_r/\sigma_\phi \approx \sqrt{2} >1$), but this inequality only strengthens the conclusion about the large reserve of stability of disks. Even a comparable amount of cool gas with the velocity dispersion $\approx10$~km/s would not change this result. 

Such overheating of the disks would mean that the observed star formation there is not related to the large-scale instability. The reason for the high $Q_T$ values is the very low surface density of the stellar disks for their high angular rotation speed due to a massive dark halo. Only a multiple excess of a gas surface density over the stellar one could alter this conclusion, bringing the disk to a marginally stable state with $Q_T < 3$. 

One can formally estimate the scale height of the stellar disk for the velocity dispersion of stars obtained from our observations.  If disk thickness is determined by its gravity, its half-thickness $z_0 = \sigma_z^2/\pi G\Sigma_*$. Assuming that the vertical velocity dispersion $\sigma_z$, as in "normal" galaxies, does not differ much from the azimuthal one, which for edge-on galaxies is close to the observed line-of-sight dispersion, it is easy to see that the resulting scale heights for our galaxies turn out to be unrealistically large, comparable to the galaxy radius, which is incompatible with the existence of a disk. However, stellar disks of the considered galaxies are not too thick: the estimates of $z_0$ from photometry give the values of $z_0$ close to 1~kpc (see Table \ref{tab:mainproperties}).  This indicates the presence of some additional force holding stars near the disk plane, which significantly exceeds the force of self-gravity.

A massive dark halo can provide an additional force compressing the disk, in addition to disk self-gravity. The acceleration component perpendicular to the disk plane associated with halo gravity grows linearly with distance $z \ll R$ from the disk plane (as in the case of a homogeneous disc), and is equal to $g_z(R,z) = M_h(R)/(R^2+z^2)\cdot(z/R)$, where $M_h(R) = V_{rot}^2R/G$ is the halo mass within a sphere of radius $R$. Simple estimates of the expected scaleheight  of low density equilibrium stellar disks of our galaxies allow us to conclude that the observed values of $z_0$ (about 1~kpc) require the vertical dispersion of stellar velocities to be significantly smaller than the circular rotation velocity in the dominant field of spherical halo. This is consistent with very close values of the measured stellar and gas rotation velocities (see Fig. \ref{fig:spectra}).  However, the estimates of stellar velocity dispersion in the outer disk regions are averaged over a large interval of $R$, and therefore cannot be used to estimate the vertical scale profile of the disks directly.

Below we consider if the high values of the stellar velocity dispersion can be a result of a projection effect. Since our galaxies are edge-on, we see the regions along a line-of-sight at different radial distances. In the absence of a significant amount of dust, it can lead to the artificial increase of the observed stellar velocity dispersion. In order to assess the change of the LOSVD parameters in comparison to the intrinsic galaxy kinematics we calculated a series of simple luminosity-weighted LOSVD models for a set of given radial velocity dispersion and rotational velocities at large distances from the galaxy centers. For the modeling we used several general assumptions about the kinematics in the LSB disk: (i) a flat rotation curve in the outer region of the disk, (ii) a Gaussian distribution of the radial and azimuthal velocity components at each point of the disk, (iii) a fixed ratio of the velocity dispersion components $q = \sigma_{r}/\sigma_{\phi}$. The parameters of the model are: rotational velocity $V_{\text{rot}}$, radial velocity dispersion $\sigma_{r}$, and  $q$, which was assumed to be equal to $1/\sqrt{2}$\, as in the case of the epicyclic approximation for the flat rotation curve. Finally, the LOSVD is determined as a luminosity-weighted sum of projected azimuthal and radial velocity distributions onto the line of sight. The model is essentially a simplified version of one discussed in \citet{Zasov_Khoperskov2003}. We performed the calculations of the LOSVD for the grid of realistic rotational velocities: $V_{\text{LOS}} < V_{\text{rot}} < V_{\text{LOS}}+60$~km/s, and radial velocity dispersions: $50 < \sigma_r < 170$~km/s, where $V_{\text{LOS}}$ is the measured line-of-sight rotational velocity for each galaxy, taken at the distance of 4 exponential scale lengths from the center. These simulations showed that the real velocity of rotation is on average about 10\% higher than the directly measured from the line-of-sight data, and $\sigma_{\text{LOS}}$ is overestimated by 5-10\% with respect to the azimuthal velocity dispersion. Thus the projection effect does not lead to the significant overestimation of the Toomre parameter.

\subsection{On the dark matter halos}
\label{sec:dark_matter_halo}
Another question that needs to be answered is related to the properties of the dark matter halos in considered galaxies. The edge-on position despite having its advantage has also a shortcoming that hampers the reliable estimate of the inner part of the rotation curve from the long-slit spectral data. The projection effects together with the internal absorption make the line-of-sight velocity radial profiles of little use to reconstruct the mass distribution of the edge-on galaxies \citep{Zasov_Khoperskov2003}. Thus we did not attempt to get the structural parameters of the dark matter halo, however, it is still possible to calculate the mass fraction of the dark halo. 
According to the photometric mass estimates, the dark matter halo dominates by mass in four considered galaxies (see Table \ref{tab:mainproperties}).

Another method for calculating the mass fraction of the dark matter halo involves using numerical N-body simulations, which reveal a correlation between the ratio of the disk's vertical and radial scales and the mass ratio of the spherical subsystem to the disk. Specifically, as the relative thickness of the disk decreases, the contribution of the spherical subsystem to the total mass increases \citep{Khoperskovetal2010}. The spherical subsystem encompasses both the bulge and the dark matter halo. In scenarios where disks are dynamically overheated, as in our case, this approach provides the lowest estimate for the spheroidal mass. However even for these estimates the masses of the spheroidal components make a bulk contribution to the total dynamical masses (see Table \ref{tab:mainproperties}, second block) which confirms the previous conclusion.

\begin{figure}
\centering
\includegraphics[width=1.0\hsize, trim={1.5cm 1.5cm 1.5cm 1.5cm}]{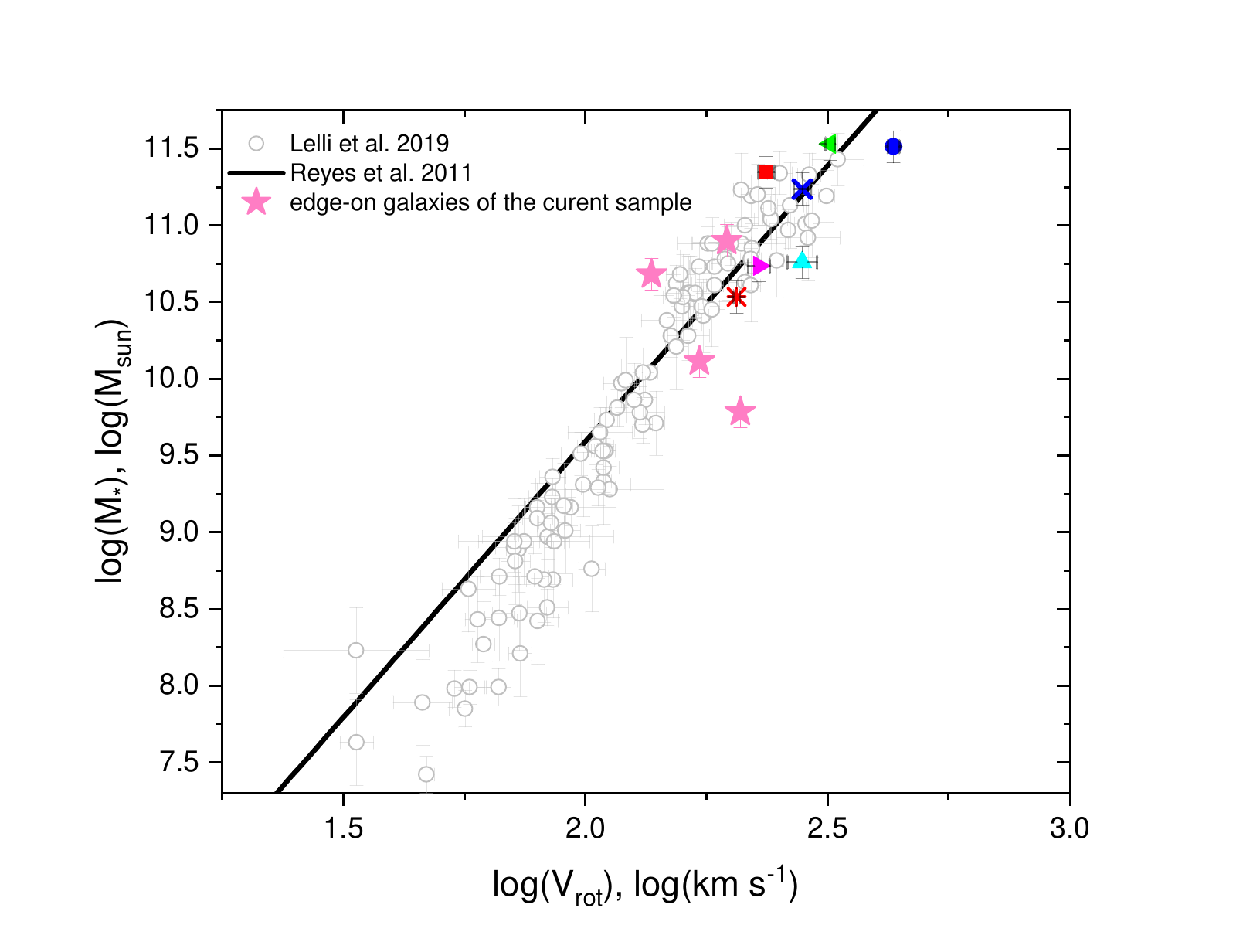}
\caption{The Tully-Fisher relation connecting the stellar mass and rotation velocity. Coloured symbols correspond to LSB systems. Stars show the position of edge-on galaxies considered in the current paper, and other coloured symbols demonstrate gLSBGs from \citet{saburovaetal2021}. The line gives the relation found by \citet{Reyes2011}. Grey open circles correspond to the systems from SPARC database \citep{Lelli2016sparc}. }

\label{TF}
\end{figure}

An indirect conclusion on the ratio between baryon and dark mass can be drawn from the Tully--Fisher \citeyear{1977A&A....54..661T} relation, specifically, its baryonic variant that relates the mass of baryons with the rotation velocity \citep[see, e.g.,][]{Ponomareva2018,Lellietal2019}. However, because the neutral gas mass measurements normally used to construct the TF relation are not available for the galaxies considered here, we use the stellar TF relation that relates the stellar mass with the rotation velocity at the plateau of the rotation curves. The stellar TF relation is shown in Fig.~\ref{TF}. Pink stars denote the positions of our four galaxies, filled symbols correspond to the 7 gLSBGs from \citet{saburovaetal2021}. Open circles correspond to the systems from the Spitzer photometry and accurate rotation curves database (SPARC) \citep{Lelli2016sparc}, the line gives the relation found by \citet{Reyes2011}. Fig.~\ref{TF} shows that three out of four galaxies lie within spread on the top part of the baryonic TF relation. However, FGC~150 seems to lack stellar mass for its rotation velocity. It can either indicate the higher dark matter fraction in this system or a high fraction of neutral gas which would put this galaxy in line with other disk galaxies on the baryonic TF relation. 

\section{Conclusions}
\red{The galaxies studied in this paper represent gLSBGs and an intermediate ``chain'' between gLSBGs and moderately-sized low-surface brightness galaxies (LSBGs). The stellar and dynamical masses of these galaxies are comparable to those of well-studied gLSBGs  \citep{saburovaetal2021}. They exhibit flat radial ionized gas metallicity gradients, similar to other gLSBGs (see Fig. \ref{fig:HZR}). The stellar population of the bulges, however, appears to be younger than that observed in other gLSBGs.  In the sample from \citet{saburova2023}, only 8~per~cent exhibit a truncated surface brightness profile (type II), as we see in the galaxies presented here. However, most galaxies from \citet{saburova2023} are not observed edge-on. According to \citep{saburova2023}, approximately 13,000 gLSBGs are expected to be found across the entire sky up to a redshift of $z=0.1$, indicating that these galaxies are relatively rare. The published gLSBG samples contains only a few dozen of these systems, and the number varies depending on the criteria used to define gLSBGs.}

\red{Based on the observational data collected, we conclude that gLSBGs do not constitute a unique class of objects but represent the high-mass, low-surface density, large-disk end of the population of disky galaxies. However, the question remains as to which circumstances are favorable for the formation of these giant disks.} Thus this paper aims to clarify the evolutionary paths of low surface brightness (LSB) galaxies with extended disks. Let's summarize the arguments supporting the scenarios proposed at the beginning of Section \ref{dis}. According to our data, the most promising scenarios involve mergers. However, the observational data do not distinguish between mergers with different mass ratios. We observe flat ionized gas metallicity radial profiles in the studied galaxies. Thus most probably the disks were not formed from the gas-rich metal-poor dwarf galaxies. At the same time the metallicity gradient values align with those expected for the disk scale length of normal spiral galaxies, suggesting that the formation paths of our systems are not particularly extreme or exotic. Although disk galaxies can form through major mergers, this is likely not the primary formation channel for these systems, as major mergers tend to destroy the disk structure \citep{Rodriguez-Gomez2017,Kulier2020}. Thus, we give less credibility to the major merger scenario, though we do not discard it completely. The disks of our systems look thin and regular (perhaps, except LEDA~1044131), and appear to be virialized, suggesting that they have completed at least 3--4 full rotation periods, which exceeds 2~Gyr for all our objects. The stellar velocity dispersion and relative disk thickness support the presence of dynamically overheated disks, which is consistent with a merger scenario, as are the flat metallicity profiles. Additionally, deep images of the galaxies reveal possible remnants of disrupted satellites in the form of clumps, which, however, do not have a spectroscopic confirmation.

Because we do not observe extremely low metallicity at the outer regions of these galaxies and the central value of the gas metallicity agrees with that expected for a given stellar mass for normal spiral galaxies, we conclude that the formation of LSB disks via gas accretion from intergalactic filaments is quite improbable. We cannot rule out the possibility that the dark matter halos of the observed galaxies possess peculiar properties, such as a high radial scale and low central density. It is possible that we are observing end-products of mergers combined with these specific dark matter halo characteristics. However, the mergers look essential in the formation scenario of LSB disks.

\red{To advance the study of gLSBGs, we need \HI and optical spectral observations of a larger sample of these systems. This will allow us to examine their properties in greater detail, including the parameters of their dark matter halos, and provide further insights into their nature. Additionally, deep X-ray observations of the central regions of gLSBGs can be used to assess the activity of their nuclei.}

\subsection*{Acknowledgement} We thank the anonymous referee for valuable comments that helped to improve the paper. {AZ, IG, DG, AK, ER research on the analysis and fitting of the long-slit spectra, ionised gas metallicity estimates, and interpretation of the kinematic and metallicity profiles and evaluation of the photometric parameters was supported by The Russian Science Foundation (RSCF) grants No.~23-12-00146. IC's research is supported by the Telescope Data Center at Smithsonian Astrophysical Observatory.  A.~L. and S.~S. acknowledge support by the Astronomical Station Vidojevica and funding from the Ministry of Science, Technological Development and Innovation of the Republic of Serbia, contract No.~451-03-66/2024-03/200002, and by the EC through project BELISSIMA (call FP7-REGPOT-2010-5, No.~256772). MD is grateful to Frank Kiwy for valuable explanations about the unTimely catalog explorer tool.

Observations with the SAO RAS telescopes are supported by the Ministry of Science and Higher Education of the Russian Federation. The renovation of telescope equipment is currently provided within the national project ''Science and universities.''  This work was supported by a NASA Keck PI Data Award (project 2022B-N149), administered by the NASA Exoplanet Science Institute. Data presented herein were obtained at the W. M. Keck Observatory from telescope time allocated to the National Aeronautics and Space Administration through the agency's scientific partnership with the California Institute of Technology and the University of California. The Observatory was made possible by the generous financial support of the W. M. Keck Foundation. The authors wish to recognize and acknowledge the very significant cultural role and reverence that the summit of Maunakea has always had within the indigenous Hawaiian community. We are most fortunate to have the opportunity to conduct observations from this mountain.

The Legacy Surveys consist of three individual and complementary projects: the Dark Energy Camera Legacy Survey (DECaLS; Proposal ID 2014B-0404; PIs: David Schlegel and Arjun Dey), the Beijing-Arizona Sky Survey (BASS; NOAO Prop. ID 2015A-0801; PIs: Zhou Xu and Xiaohui Fan), and the Mayall z-band Legacy Survey (MzLS; Prop. ID 2016A-0453; PI: Arjun Dey). DECaLS, BASS and MzLS together include data obtained, respectively, at the Blanco telescope, Cerro Tololo Inter-American Observatory, NSF’s NOIRLab; the Bok telescope, Steward Observatory, University of Arizona; and the Mayall telescope, Kitt Peak National Observatory, NOIRLab. Pipeline processing and analyses of the data were supported by NOIRLab and the Lawrence Berkeley National Laboratory (LBNL). The Legacy Surveys project is honored to be permitted to conduct astronomical research on Iolkam Du’ag (Kitt Peak), a mountain with particular significance to the Tohono O’odham Nation.

The Hyper Suprime-Cam (HSC) collaboration includes the astronomical communities of Japan and Taiwan, and Princeton University. The HSC instrumentation and software were developed by the National Astronomical Observatory of Japan (NAOJ), the Kavli Institute for the Physics and Mathematics of the Universe (Kavli IPMU), the University of Tokyo, the High Energy Accelerator Research Organization (KEK), the Academia Sinica Institute for Astronomy and Astrophysics in Taiwan (ASIAA), and Princeton University. Funding was contributed by the FIRST program from Japanese Cabinet Office, the Ministry of Education, Culture, Sports, Science and Technology (MEXT), the Japan Society for the Promotion of Science (JSPS), Japan Science and Technology Agency (JST), the Toray Science Foundation, NAOJ, Kavli IPMU, KEK, ASIAA, and Princeton University. 

This work made use of Astropy\footnote{http://www.astropy.org}, a community-developed core Python package and an ecosystem of tools and resources for astronomy \citep{astropy:2013, astropy:2018, astropy:2022}, particularly Photutils \citep{larry_bradley_2023_7946442}, and in addition Python package SciPy \citep{2020SciPy-NMeth}.} This paper makes use of software developed for the Large Synoptic Survey Telescope. We thank the LSST Project for making their code available as free software at  http://dm.lsst.org

\bibliography{LSB.bib}{}

\begin{thebibliography}{}
\expandafter\ifx\csname natexlab\endcsname\relax\def\natexlab#1{#1}\fi
\providecommand{\url}[1]{\href{#1}{#1}}
\providecommand{\dodoi}[1]{doi:~\href{http://doi.org/#1}{\nolinkurl{#1}}}
\providecommand{\doeprint}[1]{\href{http://ascl.net/#1}{\nolinkurl{http://ascl.net/#1}}}
\providecommand{\doarXiv}[1]{\href{https://arxiv.org/abs/#1}{\nolinkurl{https://arxiv.org/abs/#1}}}

\bibitem[{{Aditya} {et~al.}(2023){Aditya}, {Banerjee}, {Kamphuis}, {Mosenkov},
  {Makarov}, \& {Borisov}}]{2023MNRAS.526...29A}
{Aditya}, K., {Banerjee}, A., {Kamphuis}, P., {et~al.} 2023, \mnras, 526, 29,
  \dodoi{10.1093/mnras/stad2599}

\bibitem[{{Afanasiev} \& {Moiseev}(2011)}]{AfanasievMoiseev2011}
{Afanasiev}, V.~L., \& {Moiseev}, A.~V. 2011, Baltic Astronomy, 20, 363.
\newblock \doarXiv{1106.2020}

\bibitem[{{Aihara} {et~al.}(2011){Aihara}, {Allende Prieto}, {An}, {Anderson},
  {Aubourg}, {Balbinot}, {Beers}, {Berlind}, {Bickerton}, {Bizyaev}, {Blanton},
  {Bochanski}, {Bolton}, {Bovy}, {Brandt}, {Brinkmann}, {Brown}, {Brownstein},
  {Busca}, {Campbell}, {Carr}, {Chen}, {Chiappini}, {Comparat}, {Connolly},
  {Cortes}, {Croft}, {Cuesta}, {da Costa}, {Davenport}, {Dawson}, {Dhital},
  {Ealet}, {Ebelke}, {Edmondson}, {Eisenstein}, {Escoffier}, {Esposito},
  {Evans}, {Fan}, {Femen{\'\i}a Castell{\'a}}, {Font-Ribera}, {Frinchaboy},
  {Ge}, {Gillespie}, {Gilmore}, {Gonz{\'a}lez Hern{\'a}ndez}, {Gott}, {Gould},
  {Grebel}, {Gunn}, {Hamilton}, {Harding}, {Harris}, {Hawley}, {Hearty}, {Ho},
  {Hogg}, {Holtzman}, {Honscheid}, {Inada}, {Ivans}, {Jiang}, {Johnson},
  {Jordan}, {Jordan}, {Kazin}, {Kirkby}, {Klaene}, {Knapp}, {Kneib},
  {Kochanek}, {Koesterke}, {Kollmeier}, {Kron}, {Lampeitl}, {Lang}, {Le Goff},
  {Lee}, {Lin}, {Long}, {Loomis}, {Lucatello}, {Lundgren}, {Lupton}, {Ma},
  {MacDonald}, {Mahadevan}, {Maia}, {Makler}, {Malanushenko}, {Malanushenko},
  {Mandelbaum}, {Maraston}, {Margala}, {Masters}, {McBride}, {McGehee},
  {McGreer}, {M{\'e}nard}, {Miralda-Escud{\'e}}, {Morrison}, {Mullally},
  {Muna}, {Munn}, {Murayama}, {Myers}, {Naugle}, {Neto}, {Nguyen}, {Nichol},
  {O'Connell}, {Ogando}, {Olmstead}, {Oravetz}, {Padmanabhan},
  {Palanque-Delabrouille}, {Pan}, {Pandey}, {P{\^a}ris}, {Percival},
  {Petitjean}, {Pfaffenberger}, {Pforr}, {Phleps}, {Pichon}, {Pieri}, {Prada},
  {Price-Whelan}, {Raddick}, {Ramos}, {Reyl{\'e}}, {Rich}, {Richards}, {Rix},
  {Robin}, {Rocha-Pinto}, {Rockosi}, {Roe}, {Rollinde}, {Ross}, {Ross},
  {Rossetto}, {S{\'a}nchez}, {Sayres}, {Schlegel}, {Schlesinger}, {Schmidt},
  {Schneider}, {Sheldon}, {Shu}, {Simmerer}, {Simmons}, {Sivarani}, {Snedden},
  {Sobeck}, {Steinmetz}, {Strauss}, {Szalay}, {Tanaka}, {Thakar}, {Thomas},
  {Tinker}, {Tofflemire}, {Tojeiro}, {Tremonti}, {Vandenberg}, {Vargas
  Maga{\~n}a}, {Verde}, {Vogt}, {Wake}, {Wang}, {Weaver}, {Weinberg}, {White},
  {White}, {Yanny}, {Yasuda}, {Yeche}, \& {Zehavi}}]{Aihara2011}
{Aihara}, H., {Allende Prieto}, C., {An}, D., {et~al.} 2011, \apjs, 193, 29,
  \dodoi{10.1088/0067-0049/193/2/29}

\bibitem[{{Aihara} {et~al.}(2019){Aihara}, {AlSayyad}, {Ando}, {Armstrong},
  {Bosch}, {Egami}, {Furusawa}, {Furusawa}, {Goulding}, {Harikane}, {Hikage},
  {Ho}, {Hsieh}, {Huang}, {Ikeda}, {Imanishi}, {Ito}, {Iwata}, {Jaelani},
  {Kakuma}, {Kawana}, {Kikuta}, {Kobayashi}, {Koike}, {Komiyama}, {Li},
  {Liang}, {Lin}, {Luo}, {Lupton}, {Lust}, {MacArthur}, {Matsuoka}, {Mineo},
  {Miyatake}, {Miyazaki}, {More}, {Murata}, {Namiki}, {Nishizawa}, {Oguri},
  {Okabe}, {Okamoto}, {Okura}, {Ono}, {Onodera}, {Onoue}, {Osato}, {Ouchi},
  {Shibuya}, {Strauss}, {Sugiyama}, {Suto}, {Takada}, {Takagi}, {Takata},
  {Takita}, {Tanaka}, {Terai}, {Toba}, {Uchiyama}, {Utsumi}, {Wang}, {Wang}, \&
  {Yamada}}]{hsc2019}
{Aihara}, H., {AlSayyad}, Y., {Ando}, M., {et~al.} 2019, \pasj, 71, 114,
  \dodoi{10.1093/pasj/psz103}

\bibitem[{{Astropy Collaboration} {et~al.}(2013){Astropy Collaboration},
  {Robitaille}, {Tollerud}, {Greenfield}, {Droettboom}, {Bray}, {Aldcroft},
  {Davis}, {Ginsburg}, {Price-Whelan}, {Kerzendorf}, {Conley}, {Crighton},
  {Barbary}, {Muna}, {Ferguson}, {Grollier}, {Parikh}, {Nair}, {Unther},
  {Deil}, {Woillez}, {Conseil}, {Kramer}, {Turner}, {Singer}, {Fox}, {Weaver},
  {Zabalza}, {Edwards}, {Azalee Bostroem}, {Burke}, {Casey}, {Crawford},
  {Dencheva}, {Ely}, {Jenness}, {Labrie}, {Lim}, {Pierfederici}, {Pontzen},
  {Ptak}, {Refsdal}, {Servillat}, \& {Streicher}}]{astropy:2013}
{Astropy Collaboration}, {Robitaille}, T.~P., {Tollerud}, E.~J., {et~al.} 2013,
  \aap, 558, A33, \dodoi{10.1051/0004-6361/201322068}

\bibitem[{{Astropy Collaboration} {et~al.}(2018){Astropy Collaboration},
  {Price-Whelan}, {Sip{\H{o}}cz}, {G{\"u}nther}, {Lim}, {Crawford}, {Conseil},
  {Shupe}, {Craig}, {Dencheva}, {Ginsburg}, {Vand erPlas}, {Bradley},
  {P{\'e}rez-Su{\'a}rez}, {de Val-Borro}, {Aldcroft}, {Cruz}, {Robitaille},
  {Tollerud}, {Ardelean}, {Babej}, {Bach}, {Bachetti}, {Bakanov}, {Bamford},
  {Barentsen}, {Barmby}, {Baumbach}, {Berry}, {Biscani}, {Boquien}, {Bostroem},
  {Bouma}, {Brammer}, {Bray}, {Breytenbach}, {Buddelmeijer}, {Burke},
  {Calderone}, {Cano Rodr{\'\i}guez}, {Cara}, {Cardoso}, {Cheedella}, {Copin},
  {Corrales}, {Crichton}, {D'Avella}, {Deil}, {Depagne}, {Dietrich}, {Donath},
  {Droettboom}, {Earl}, {Erben}, {Fabbro}, {Ferreira}, {Finethy}, {Fox},
  {Garrison}, {Gibbons}, {Goldstein}, {Gommers}, {Greco}, {Greenfield},
  {Groener}, {Grollier}, {Hagen}, {Hirst}, {Homeier}, {Horton}, {Hosseinzadeh},
  {Hu}, {Hunkeler}, {Ivezi{\'c}}, {Jain}, {Jenness}, {Kanarek}, {Kendrew},
  {Kern}, {Kerzendorf}, {Khvalko}, {King}, {Kirkby}, {Kulkarni}, {Kumar},
  {Lee}, {Lenz}, {Littlefair}, {Ma}, {Macleod}, {Mastropietro}, {McCully},
  {Montagnac}, {Morris}, {Mueller}, {Mumford}, {Muna}, {Murphy}, {Nelson},
  {Nguyen}, {Ninan}, {N{\"o}the}, {Ogaz}, {Oh}, {Parejko}, {Parley}, {Pascual},
  {Patil}, {Patil}, {Plunkett}, {Prochaska}, {Rastogi}, {Reddy Janga},
  {Sabater}, {Sakurikar}, {Seifert}, {Sherbert}, {Sherwood-Taylor}, {Shih},
  {Sick}, {Silbiger}, {Singanamalla}, {Singer}, {Sladen}, {Sooley},
  {Sornarajah}, {Streicher}, {Teuben}, {Thomas}, {Tremblay}, {Turner},
  {Terr{\'o}n}, {van Kerkwijk}, {de la Vega}, {Watkins}, {Weaver}, {Whitmore},
  {Woillez}, {Zabalza}, \& {Astropy Contributors}}]{astropy:2018}
{Astropy Collaboration}, {Price-Whelan}, A.~M., {Sip{\H{o}}cz}, B.~M., {et~al.}
  2018, \aj, 156, 123, \dodoi{10.3847/1538-3881/aabc4f}

\bibitem[{{Astropy Collaboration} {et~al.}(2022){Astropy Collaboration},
  {Price-Whelan}, {Lim}, {Earl}, {Starkman}, {Bradley}, {Shupe}, {Patil},
  {Corrales}, {Brasseur}, {N{"o}the}, {Donath}, {Tollerud}, {Morris},
  {Ginsburg}, {Vaher}, {Weaver}, {Tocknell}, {Jamieson}, {van Kerkwijk},
  {Robitaille}, {Merry}, {Bachetti}, {G{"u}nther}, {Aldcroft},
  {Alvarado-Montes}, {Archibald}, {B{'o}di}, {Bapat}, {Barentsen}, {Baz{'a}n},
  {Biswas}, {Boquien}, {Burke}, {Cara}, {Cara}, {Conroy}, {Conseil}, {Craig},
  {Cross}, {Cruz}, {D'Eugenio}, {Dencheva}, {Devillepoix}, {Dietrich},
  {Eigenbrot}, {Erben}, {Ferreira}, {Foreman-Mackey}, {Fox}, {Freij}, {Garg},
  {Geda}, {Glattly}, {Gondhalekar}, {Gordon}, {Grant}, {Greenfield}, {Groener},
  {Guest}, {Gurovich}, {Handberg}, {Hart}, {Hatfield-Dodds}, {Homeier},
  {Hosseinzadeh}, {Jenness}, {Jones}, {Joseph}, {Kalmbach}, {Karamehmetoglu},
  {Ka{l}uszy{'n}ski}, {Kelley}, {Kern}, {Kerzendorf}, {Koch}, {Kulumani},
  {Lee}, {Ly}, {Ma}, {MacBride}, {Maljaars}, {Muna}, {Murphy}, {Norman},
  {O'Steen}, {Oman}, {Pacifici}, {Pascual}, {Pascual-Granado}, {Patil},
  {Perren}, {Pickering}, {Rastogi}, {Roulston}, {Ryan}, {Rykoff}, {Sabater},
  {Sakurikar}, {Salgado}, {Sanghi}, {Saunders}, {Savchenko}, {Schwardt},
  {Seifert-Eckert}, {Shih}, {Jain}, {Shukla}, {Sick}, {Simpson},
  {Singanamalla}, {Singer}, {Singhal}, {Sinha}, {Sip{H{o}}cz}, {Spitler},
  {Stansby}, {Streicher}, {{{S}}umak}, {Swinbank}, {Taranu}, {Tewary},
  {Tremblay}, {Val-Borro}, {Van Kooten}, {Vasovi{'c}}, {Verma}, {de Miranda
  Cardoso}, {Williams}, {Wilson}, {Winkel}, {Wood-Vasey}, {Xue}, {Yoachim},
  {Zhang}, {Zonca}, \& {Astropy Project Contributors}}]{astropy:2022}
{Astropy Collaboration}, {Price-Whelan}, A.~M., {Lim}, P.~L., {et~al.} 2022,
  \apj, 935, 167, \dodoi{10.3847/1538-4357/ac7c74}

\bibitem[{{Bakos} {et~al.}(2008){Bakos}, {Trujillo}, \&
  {Pohlen}}]{2008ApJ...683L.103B}
{Bakos}, J., {Trujillo}, I., \& {Pohlen}, M. 2008, \apjl, 683, L103,
  \dodoi{10.1086/591671}

\bibitem[{{Baldwin} {et~al.}(1981){Baldwin}, {Phillips}, \& {Terlevich}}]{BPT}
{Baldwin}, J.~A., {Phillips}, M.~M., \& {Terlevich}, R. 1981, \pasp, 93, 5,
  \dodoi{10.1086/130766}

\bibitem[{{Boissier} {et~al.}(2016){Boissier}, {Boselli}, {Ferrarese},
  {C{\^o}t{\'e}}, {Roehlly}, {Gwyn}, {Cuillandre}, {Roediger}, {Koda},
  {Mu{\~n}os Mateos}, {Gil de Paz}, \& {Madore}}]{Boissier2016}
{Boissier}, S., {Boselli}, A., {Ferrarese}, L., {et~al.} 2016, \aap, 593, A126,
  \dodoi{10.1051/0004-6361/201629226}

\bibitem[{{Borisov} {et~al.}(2023){Borisov}, {Chilingarian}, {Rubtsov},
  {Ledoux}, {Melo}, {Grishin}, {Katkov}, {Goradzhanov}, {Afanasiev},
  {Kasparova}, \& {Saburova}}]{2023ApJS..266...11B}
{Borisov}, S.~B., {Chilingarian}, I.~V., {Rubtsov}, E.~V., {et~al.} 2023,
  \apjs, 266, 11, \dodoi{10.3847/1538-4365/acc321}

\bibitem[{{Bothun} {et~al.}(1987){Bothun}, {Impey}, {Malin}, \&
  {Mould}}]{Bothun1987}
{Bothun}, G.~D., {Impey}, C.~D., {Malin}, D.~F., \& {Mould}, J.~R. 1987, \aj,
  94, 23, \dodoi{10.1086/114443}

\bibitem[{Bradley {et~al.}(2023)Bradley, Sip{\H o}cz, Robitaille, Tollerud,
  Vin{\'{\i}}cius, Deil, Barbary, Wilson, Busko, Donath, G{\"u}nther, Cara,
  Lim, Me{\ss}linger, Conseil, Bostroem, Droettboom, Bray, Bratholm, Barentsen,
  Craig, Rathi, Pascual, Perren, Georgiev, de~Val-Borro, Kerzendorf, Bach,
  Quint, \& Souchereau}]{larry_bradley_2023_7946442}
Bradley, L., Sip{\H o}cz, B., Robitaille, T., {et~al.} 2023, astropy/photutils:
  1.8.0, 1.8.0,  Zenodo, \dodoi{10.5281/zenodo.7946442}

\bibitem[{{Bresolin} \& {Kennicutt}(2015)}]{BresolinKennicutt2015}
{Bresolin}, F., \& {Kennicutt}, R.~C. 2015, \mnras, 454, 3664,
  \dodoi{10.1093/mnras/stv2245}

\bibitem[{{Bressan} {et~al.}(2012){Bressan}, {Marigo}, {Girardi}, {Salasnich},
  {Dal Cero}, {Rubele}, \& {Nanni}}]{Bressan2012}
{Bressan}, A., {Marigo}, P., {Girardi}, L., {et~al.} 2012, \mnras, 427, 127,
  \dodoi{10.1111/j.1365-2966.2012.21948.x}

\bibitem[{{Cao} {et~al.}(2023){Cao}, {Wu}, {Galaz}, {Kalari}, {Cheng}, {Li}, \&
  {Wang}}]{2023ApJ...948...96C}
{Cao}, T.-w., {Wu}, H., {Galaz}, G., {et~al.} 2023, \apj, 948, 96,
  \dodoi{10.3847/1538-4357/acc864}

\bibitem[{{Chilingarian} {et~al.}(2007{\natexlab{a}}){Chilingarian},
  {Prugniel}, {Sil'Chenko}, \& {Koleva}}]{Chilingarian2007a}
{Chilingarian}, I., {Prugniel}, P., {Sil'Chenko}, O., \& {Koleva}, M.
  2007{\natexlab{a}}, in IAU Symposium, Vol. 241, IAU Symposium, ed.
  A.~{Vazdekis} \& R.~{Peletier}, 175--176, \dodoi{10.1017/S1743921307007752}

\bibitem[{{Chilingarian} {et~al.}(2007{\natexlab{b}}){Chilingarian},
  {Prugniel}, {Sil'Chenko}, \& {Afanasiev}}]{Chilingarian2007b}
{Chilingarian}, I.~V., {Prugniel}, P., {Sil'Chenko}, O.~K., \& {Afanasiev},
  V.~L. 2007{\natexlab{b}}, \mnras, 376, 1033,
  \dodoi{10.1111/j.1365-2966.2007.11549.x}

\bibitem[{{Chilingarian} {et~al.}(2017){Chilingarian}, {Zolotukhin}, {Katkov},
  {Melchior}, {Rubtsov}, \& {Grishin}}]{Chilingarian2017ApJS..228...14C}
{Chilingarian}, I.~V., {Zolotukhin}, I.~Y., {Katkov}, I.~Y., {et~al.} 2017,
  \apjs, 228, 14, \dodoi{10.3847/1538-4365/228/2/14}

\bibitem[{{Chung} \& {Bureau}(2004)}]{chung2004}
{Chung}, A., \& {Bureau}, M. 2004, \aj, 127, 3192, \dodoi{10.1086/420988}

\bibitem[{Cleland \& McGee(2022)}]{Cleland_2022}
Cleland, C., \& McGee, S.~L. 2022, Monthly Notices of the Royal Astronomical
  Society, 515, 5905–5913, \dodoi{10.1093/mnras/stac2188}

\bibitem[{{Debattista} \& {Sellwood}(1999)}]{1999ApJ...513L.107D}
{Debattista}, V.~P., \& {Sellwood}, J.~A. 1999, \apjl, 513, L107,
  \dodoi{10.1086/311913}

\bibitem[{{Demianenko} {et~al.}(2024){Demianenko}, {Grishin}, {Toptun},
  {Chilingarian}, {Katkov}, {Goradzhanov}, \& {Kuzmin}}]{Demianenko2024}
{Demianenko}, M., {Grishin}, K., {Toptun}, V., {et~al.} 2024, in Astronomical
  Society of the Pacific Conference Series, Vol. 535, Astronomical Society of
  the Pacific Conference Series, ed. B.~V. {Hugo}, R.~{Van Rooyen}, \& O.~M.
  {Smirnov}, 283, \dodoi{10.48550/arXiv.2201.03712}

\bibitem[{{Dey} {et~al.}(2019){Dey}, {Schlegel}, {Lang}, {Blum}, {Burleigh},
  {Fan}, {Findlay}, {Finkbeiner}, {Herrera}, {Juneau}, {Landriau}, {Levi},
  {McGreer}, {Meisner}, {Myers}, {Moustakas}, {Nugent}, {Patej}, {Schlafly},
  {Walker}, {Valdes}, {Weaver}, {Y{\`e}che}, {Zou}, {Zhou}, {Abareshi},
  {Abbott}, {Abolfathi}, {Aguilera}, {Alam}, {Allen}, {Alvarez}, {Annis},
  {Ansarinejad}, {Aubert}, {Beechert}, {Bell}, {BenZvi}, {Beutler}, {Bielby},
  {Bolton}, {Brice{\~n}o}, {Buckley-Geer}, {Butler}, {Calamida}, {Carlberg},
  {Carter}, {Casas}, {Castander}, {Choi}, {Comparat}, {Cukanovaite}, {Delubac},
  {DeVries}, {Dey}, {Dhungana}, {Dickinson}, {Ding}, {Donaldson}, {Duan},
  {Duckworth}, {Eftekharzadeh}, {Eisenstein}, {Etourneau}, {Fagrelius},
  {Farihi}, {Fitzpatrick}, {Font-Ribera}, {Fulmer}, {G{\"a}nsicke},
  {Gaztanaga}, {George}, {Gerdes}, {Gontcho}, {Gorgoni}, {Green}, {Guy},
  {Harmer}, {Hernand ez}, {Honscheid}, {Huang}, {James}, {Jannuzi}, {Jiang},
  {Joyce}, {Karcher}, {Karkar}, {Kehoe}, {Kneib}, {Kueter-Young}, {Lan},
  {Lauer}, {Le Guillou}, {Le Van Suu}, {Lee}, {Lesser}, {Perreault Levasseur},
  {Li}, {Mann}, {Marshall}, {Mart{\'\i}nez-V{\'a}zquez}, {Martini}, {du Mas des
  Bourboux}, {McManus}, {Meier}, {M{\'e}nard}, {Metcalfe},
  {Mu{\~n}oz-Guti{\'e}rrez}, {Najita}, {Napier}, {Narayan}, {Newman}, {Nie},
  {Nord}, {Norman}, {Olsen}, {Paat}, {Palanque-Delabrouille}, {Peng},
  {Poppett}, {Poremba}, {Prakash}, {Rabinowitz}, {Raichoor}, {Rezaie},
  {Robertson}, {Roe}, {Ross}, {Ross}, {Rudnick}, {Safonova}, {Saha},
  {S{\'a}nchez}, {Savary}, {Schweiker}, {Scott}, {Seo}, {Shan}, {Silva},
  {Slepian}, {Soto}, {Sprayberry}, {Staten}, {Stillman}, {Stupak}, {Summers},
  {Sien Tie}, {Tirado}, {Vargas-Maga{\~n}a}, {Vivas}, {Wechsler}, {Williams},
  {Yang}, {Yang}, {Yapici}, {Zaritsky}, {Zenteno}, {Zhang}, {Zhang}, {Zhou}, \&
  {Zhou}}]{Dey2019}
{Dey}, A., {Schlegel}, D.~J., {Lang}, D., {et~al.} 2019, \aj, 157, 168,
  \dodoi{10.3847/1538-3881/ab089d}

\bibitem[{{Du} {et~al.}(2023){Du}, {Cheng}, {Du}, {Du}, \& {Wu}}]{Du2023}
{Du}, W., {Cheng}, C., {Du}, P., {Du}, L., \& {Wu}, H. 2023, \apj, 959, 105,
  \dodoi{10.3847/1538-4357/ad05bd}

\bibitem[{{Duarte Puertas} {et~al.}(2022){Duarte Puertas}, {Vilchez},
  {Iglesias-P{\'a}ramo}, {Moll{\'a}}, {P{\'e}rez-Montero}, {Kehrig},
  {Pilyugin}, \& {Zinchenko}}]{DuartePuertas2022}
{Duarte Puertas}, S., {Vilchez}, J.~M., {Iglesias-P{\'a}ramo}, J., {et~al.}
  2022, \aap, 666, A186, \dodoi{10.1051/0004-6361/202141571}

\bibitem[{{Duc} {et~al.}(2015){Duc}, {Cuillandre}, {Karabal}, {Cappellari},
  {Alatalo}, {Blitz}, {Bournaud}, {Bureau}, {Crocker}, {Davies}, {Davis}, {de
  Zeeuw}, {Emsellem}, {Khochfar}, {Krajnovi{\'c}}, {Kuntschner}, {McDermid},
  {Michel-Dansac}, {Morganti}, {Naab}, {Oosterloo}, {Paudel}, {Sarzi}, {Scott},
  {Serra}, {Weijmans}, \& {Young}}]{Duc2015}
{Duc}, P.-A., {Cuillandre}, J.-C., {Karabal}, E., {et~al.} 2015, \mnras, 446,
  120, \dodoi{10.1093/mnras/stu2019}

\bibitem[{{Egorova} {et~al.}(2021){Egorova}, {Egorov}, {Moiseev}, {Saburova},
  {Grishin}, \& {Chilingarian}}]{Egorova2021}
{Egorova}, E.~S., {Egorov}, O.~V., {Moiseev}, A.~V., {et~al.} 2021, arXiv
  e-prints, arXiv:2103.00211.
\newblock \doarXiv{2103.00211}

\bibitem[{{Faber} {et~al.}(2003){Faber}, {Phillips}, {Kibrick}, {Alcott},
  {Allen}, {Burrous}, {Cantrall}, {Clarke}, {Coil}, {Cowley}, {Davis}, {Deich},
  {Dietsch}, {Gilmore}, {Harper}, {Hilyard}, {Lewis}, {McVeigh}, {Newman},
  {Osborne}, {Schiavon}, {Stover}, {Tucker}, {Wallace}, {Wei}, {Wirth}, \&
  {Wright}}]{2003SPIE.4841.1657F}
{Faber}, S.~M., {Phillips}, A.~C., {Kibrick}, R.~I., {et~al.} 2003, in Society
  of Photo-Optical Instrumentation Engineers (SPIE) Conference Series, Vol.
  4841, Instrument Design and Performance for Optical/Infrared Ground-based
  Telescopes, ed. M.~{Iye} \& A.~F.~M. {Moorwood}, 1657--1669,
  \dodoi{10.1117/12.460346}

\bibitem[{{Fitzpatrick}(1999)}]{Fitzpatrick1999}
{Fitzpatrick}, E.~L. 1999, \pasp, 111, 63, \dodoi{10.1086/316293}

\bibitem[{{Freeman}(1970)}]{Freeman1970}
{Freeman}, K.~C. 1970, \apj, 160, 811, \dodoi{10.1086/150474}

\bibitem[{{Galaz} {et~al.}(2015){Galaz}, {Milovic}, {Suc}, {Busta}, {Lizana},
  {Infante}, \& {Royo}}]{Galaz2015}
{Galaz}, G., {Milovic}, C., {Suc}, V., {et~al.} 2015, \apjl, 815, L29,
  \dodoi{10.1088/2041-8205/815/2/L29}

\bibitem[{{Garg} \& {Banerjee}(2017)}]{2017MNRAS.472..166G}
{Garg}, P., \& {Banerjee}, A. 2017, \mnras, 472, 166,
  \dodoi{10.1093/mnras/stx1918}

\bibitem[{{Ghosh} \& {Jog}(2014)}]{2014MNRAS.439..929G}
{Ghosh}, S., \& {Jog}, C.~J. 2014, \mnras, 439, 929,
  \dodoi{10.1093/mnras/stu013}

\bibitem[{{Gil de Paz} {et~al.}(2005){Gil de Paz}, {Madore}, {Boissier},
  {Swaters}, {Popescu}, {Tuffs}, {Sheth}, {Kennicutt}, {Bianchi}, {Thilker}, \&
  {Martin}}]{GildePaz2005}
{Gil de Paz}, A., {Madore}, B.~F., {Boissier}, S., {et~al.} 2005, \apjl, 627,
  L29, \dodoi{10.1086/432054}

\bibitem[{{Gkini} {et~al.}(2021){Gkini}, {Plionis}, {Chira}, \&
  {Koulouridis}}]{2021A&A...650A..75G}
{Gkini}, A., {Plionis}, M., {Chira}, M., \& {Koulouridis}, E. 2021, \aap, 650,
  A75, \dodoi{10.1051/0004-6361/202140278}

\bibitem[{{Gordon} {et~al.}(2020){Gordon}, {Boyce}, {O'Dea}, {Rudnick},
  {Andernach}, {Vantyghem}, {Baum}, {Bui}, \&
  {Dionyssiou}}]{2020RNAAS...4..175G}
{Gordon}, Y.~A., {Boyce}, M.~M., {O'Dea}, C.~P., {et~al.} 2020, Research Notes
  of the American Astronomical Society, 4, 175,
  \dodoi{10.3847/2515-5172/abbe23}

\bibitem[{{Hagen} {et~al.}(2016){Hagen}, {Seibert}, {Hagen}, {Nyland}, {Neill},
  {Treyer}, {Young}, {Rich}, \& {Madore}}]{Hagen2016}
{Hagen}, L.~M.~Z., {Seibert}, M., {Hagen}, A., {et~al.} 2016, \apj, 826, 210,
  \dodoi{10.3847/0004-637X/826/2/210}

\bibitem[{{Heckman}(1980)}]{Heckman1980}
{Heckman}, T.~M. 1980, \aap, 87, 152

\bibitem[{{Ibata} {et~al.}(2013){Ibata}, {Lewis}, {Conn}, {Irwin},
  {McConnachie}, {Chapman}, {Collins}, {Fardal}, {Ferguson}, {Ibata}, {Mackey},
  {Martin}, {Navarro}, {Rich}, {Valls-Gabaud}, \& {Widrow}}]{ibata2013}
{Ibata}, R.~A., {Lewis}, G.~F., {Conn}, A.~R., {et~al.} 2013, \nat, 493, 62,
  \dodoi{10.1038/nature11717}

\bibitem[{{Inoue} {et~al.}(2016){Inoue}, {Dekel}, {Mandelker}, {Ceverino},
  {Bournaud}, \& {Primack}}]{Inoue2016}
{Inoue}, S., {Dekel}, A., {Mandelker}, N., {et~al.} 2016, \mnras, 456, 2052,
  \dodoi{10.1093/mnras/stv2793}

\bibitem[{{Junais} {et~al.}(2024){Junais}, {Weilbacher}, {Epinat}, {Boissier},
  {Galaz}, {Johnston}, {Puzia}, {Amram}, \& {Ma{\l}ek}}]{Junais2024}
{Junais}, {Weilbacher}, P.~M., {Epinat}, B., {et~al.} 2024, \aap, 681, A100,
  \dodoi{10.1051/0004-6361/202347669}

\bibitem[{{Karachentsev} {et~al.}(1993){Karachentsev}, {Karachentseva}, \&
  {Parnovskij}}]{Karachentsev1993}
{Karachentsev}, I.~D., {Karachentseva}, V.~E., \& {Parnovskij}, S.~L. 1993,
  Astronomische Nachrichten, 314, 97, \dodoi{10.1002/asna.2113140302}

\bibitem[{{Karachentsev} \& {Kroupa}(2024)}]{Karachentsev2024}
{Karachentsev}, I.~D., \& {Kroupa}, P. 2024, \mnras, 528, 2805,
  \dodoi{10.1093/mnras/stae184}

\bibitem[{{Kasparova} {et~al.}(2020){Kasparova}, {Katkov}, \&
  {Chilingarian}}]{Kasparova2020}
{Kasparova}, A.~V., {Katkov}, I.~Y., \& {Chilingarian}, I.~V. 2020, \mnras,
  493, 5464, \dodoi{10.1093/mnras/staa611}

\bibitem[{{Kasparova} {et~al.}(2014){Kasparova}, {Saburova}, {Katkov},
  {Chilingarian}, \& {Bizyaev}}]{Kasparova2014}
{Kasparova}, A.~V., {Saburova}, A.~S., {Katkov}, I.~Y., {Chilingarian}, I.~V.,
  \& {Bizyaev}, D.~V. 2014, \mnras, 437, 3072, \dodoi{10.1093/mnras/stt1982}

\bibitem[{{Katkov} \& {Chilingarian}(2011)}]{2011ASPC..442..143K}
{Katkov}, I.~Y., \& {Chilingarian}, I.~V. 2011, in Astronomical Society of the
  Pacific Conference Series, Vol. 442, Astronomical Data Analysis Software and
  Systems XX, ed. I.~N. {Evans}, A.~{Accomazzi}, D.~J. {Mink}, \& A.~H. {Rots},
  143, \dodoi{10.48550/arXiv.1012.4125}

\bibitem[{{Katkov} {et~al.}(2019){Katkov}, {Kniazev}, {Kasparova}, \&
  {Sil'chenko}}]{Katkov2019}
{Katkov}, I.~Y., {Kniazev}, A.~Y., {Kasparova}, A.~V., \& {Sil'chenko}, O.~K.
  2019, \mnras, 483, 2413, \dodoi{10.1093/mnras/sty3268}

\bibitem[{{Kauffmann} {et~al.}(2003){Kauffmann}, {Heckman}, {Tremonti},
  {Brinchmann}, {Charlot}, {White}, {Ridgway}, {Brinkmann}, {Fukugita}, {Hall},
  {Ivezi{\'c}}, {Richards}, \& {Schneider}}]{Kauffmann2003}
{Kauffmann}, G., {Heckman}, T.~M., {Tremonti}, C., {et~al.} 2003, \mnras, 346,
  1055, \dodoi{10.1111/j.1365-2966.2003.07154.x}

\bibitem[{{Kewley} {et~al.}(2001){Kewley}, {Dopita}, {Sutherland}, {Heisler},
  \& {Trevena}}]{Kewley2001}
{Kewley}, L.~J., {Dopita}, M.~A., {Sutherland}, R.~S., {Heisler}, C.~A., \&
  {Trevena}, J. 2001, \apj, 556, 121, \dodoi{10.1086/321545}

\bibitem[{{Khoperskov} {et~al.}(2010){Khoperskov}, {Bizyaev}, {Tiurina}, \&
  {Butenko}}]{Khoperskovetal2010}
{Khoperskov}, A., {Bizyaev}, D., {Tiurina}, N., \& {Butenko}, M. 2010,
  Astronomische Nachrichten, 331, 731, \dodoi{10.1002/asna.200911402}

\bibitem[{{Khoperskov} {et~al.}(2003){Khoperskov}, {Zasov}, \&
  {Tyurina}}]{Khoperskov2003}
{Khoperskov}, A.~V., {Zasov}, A.~V., \& {Tyurina}, N.~V. 2003, Astronomy
  Reports, 47, 357, \dodoi{10.1134/1.1575851}

\bibitem[{{Kroupa}(2001)}]{Kroupa2001}
{Kroupa}, P. 2001, \mnras, 322, 231, \dodoi{10.1046/j.1365-8711.2001.04022.x}

\bibitem[{{Kulier} {et~al.}(2020){Kulier}, {Galaz}, {Padilla}, \&
  {Trayford}}]{Kulier2020}
{Kulier}, A., {Galaz}, G., {Padilla}, N.~D., \& {Trayford}, J.~W. 2020, \mnras,
  496, 3996, \dodoi{10.1093/mnras/staa1798}

\bibitem[{{Kurucz} {et~al.}(1984){Kurucz}, {Furenlid}, {Brault}, \&
  {Testerman}}]{1984sfat.book.....K}
{Kurucz}, R.~L., {Furenlid}, I., {Brault}, J., \& {Testerman}, L. 1984, {Solar
  flux atlas from 296 to 1300 nm}

\bibitem[{{Lelli} {et~al.}(2016){Lelli}, {McGaugh}, \&
  {Schombert}}]{Lelli2016sparc}
{Lelli}, F., {McGaugh}, S.~S., \& {Schombert}, J.~M. 2016, \aj, 152, 157,
  \dodoi{10.3847/0004-6256/152/6/157}

\bibitem[{{Lelli} {et~al.}(2019){Lelli}, {McGaugh}, {Schombert}, {Desmond}, \&
  {Katz}}]{Lellietal2019}
{Lelli}, F., {McGaugh}, S.~S., {Schombert}, J.~M., {Desmond}, H., \& {Katz}, H.
  2019, \mnras, 484, 3267, \dodoi{10.1093/mnras/stz205}

\bibitem[{{Luridiana} {et~al.}(2015){Luridiana}, {Morisset}, \&
  {Shaw}}]{Luridiana2015}
{Luridiana}, V., {Morisset}, C., \& {Shaw}, R.~A. 2015, \aap, 573, A42,
  \dodoi{10.1051/0004-6361/201323152}

\bibitem[{{Lyu} {et~al.}(2019){Lyu}, {Rieke}, \& {Smith}}]{2019ApJ...886...33L}
{Lyu}, J., {Rieke}, G.~H., \& {Smith}, P.~S. 2019, \apj, 886, 33,
  \dodoi{10.3847/1538-4357/ab481d}

\bibitem[{{Mancera Pi{\~n}a} {et~al.}(2021){Mancera Pi{\~n}a}, {Posti},
  {Pezzulli}, {Fraternali}, {Fall}, {Oosterloo}, \& {Adams}}]{Mancera2021}
{Mancera Pi{\~n}a}, P.~E., {Posti}, L., {Pezzulli}, G., {et~al.} 2021, \aap,
  651, L15, \dodoi{10.1051/0004-6361/202141574}

\bibitem[{{Marigo} {et~al.}(2013){Marigo}, {Bressan}, {Nanni}, {Girardi}, \&
  {Pumo}}]{Marigo2013}
{Marigo}, P., {Bressan}, A., {Nanni}, A., {Girardi}, L., \& {Pumo}, M.~L. 2013,
  \mnras, 434, 488, \dodoi{10.1093/mnras/stt1034}

\bibitem[{Masci {et~al.}(2023)Masci, Laher, Rusholme, Shupe, Paladini, Groom,
  Wold, Miller, \& Drake}]{masci2023new}
Masci, F.~J., Laher, R.~R., Rusholme, B., {et~al.} 2023, A New Forced
  Photometry Service for the Zwicky Transient Facility.
\newblock \doarXiv{2305.16279}

\bibitem[{Meisner {et~al.}(2023)Meisner, Caselden, Schlafly, \&
  Kiwy}]{Meisner_2023}
Meisner, A.~M., Caselden, D., Schlafly, E.~F., \& Kiwy, F. 2023, The
  Astronomical Journal, 165, 36, \dodoi{10.3847/1538-3881/aca2ab}

\bibitem[{{Miyazaki} {et~al.}(2018){Miyazaki}, {Komiyama}, {Kawanomoto}, {Doi},
  {Furusawa}, {Hamana}, {Hayashi}, {Ikeda}, {Kamata}, {Karoji}, {Koike},
  {Kurakami}, {Miyama}, {Morokuma}, {Nakata}, {Namikawa}, {Nakaya}, {Nariai},
  {Obuchi}, {Oishi}, {Okada}, {Okura}, {Tait}, {Takata}, {Tanaka}, {Tanaka},
  {Terai}, {Tomono}, {Uraguchi}, {Usuda}, {Utsumi}, {Yamada}, {Yamanoi},
  {Aihara}, {Fujimori}, {Mineo}, {Miyatake}, {Oguri}, {Uchida}, {Tanaka},
  {Yasuda}, {Takada}, {Murayama}, {Nishizawa}, {Sugiyama}, {Chiba}, {Futamase},
  {Wang}, {Chen}, {Ho}, {Liaw}, {Chiu}, {Ho}, {Lai}, {Lee}, {Jeng}, {Iwamura},
  {Armstrong}, {Bickerton}, {Bosch}, {Gunn}, {Lupton}, {Loomis}, {Price},
  {Smith}, {Strauss}, {Turner}, {Suzuki}, {Miyazaki}, {Muramatsu}, {Yamamoto},
  {Endo}, {Ezaki}, {Ito}, {Kawaguchi}, {Sofuku}, {Taniike}, {Akutsu}, {Dojo},
  {Kasumi}, {Matsuda}, {Imoto}, {Miwa}, {Suzuki}, {Takeshi}, \&
  {Yokota}}]{2018PASJ...70S...1M}
{Miyazaki}, S., {Komiyama}, Y., {Kawanomoto}, S., {et~al.} 2018, \pasj, 70, S1,
  \dodoi{10.1093/pasj/psx063}

\bibitem[{{M{\"u}ller} {et~al.}(2018){M{\"u}ller}, {Pawlowski}, {Jerjen}, \&
  {Lelli}}]{2018Sci...359..534M}
{M{\"u}ller}, O., {Pawlowski}, M.~S., {Jerjen}, H., \& {Lelli}, F. 2018,
  Science, 359, 534, \dodoi{10.1126/science.aao1858}

\bibitem[{{M{\"u}ller} {et~al.}(2019){M{\"u}ller}, {Vudragovi{\'c}}, \&
  {B{\'\i}lek}}]{muller2019}
{M{\"u}ller}, O., {Vudragovi{\'c}}, A., \& {B{\'\i}lek}, M. 2019, \aap, 632,
  L13, \dodoi{10.1051/0004-6361/201937077}

\bibitem[{{Osterbrock} \& {Ferland}(2006)}]{Osterbrock2006}
{Osterbrock}, D.~E., \& {Ferland}, G.~J. 2006, {Astrophysics of gaseous nebulae
  and active galactic nuclei}, 2nd edn. (University Science Books)

\bibitem[{{Ostriker} \& {Binney}(1989)}]{1989MNRAS.237..785O}
{Ostriker}, E.~C., \& {Binney}, J.~J. 1989, \mnras, 237, 785,
  \dodoi{10.1093/mnras/237.3.785}

\bibitem[{{Pawlowski} {et~al.}(2024){Pawlowski}, {M{\"u}ller}, {Taibi},
  {J{\'u}lio}, {Kanehisa}, \& {Heesters}}]{Pawlowski2024}
{Pawlowski}, M.~S., {M{\"u}ller}, O., {Taibi}, S., {et~al.} 2024, arXiv
  e-prints, arXiv:2405.06016, \dodoi{10.48550/arXiv.2405.06016}

\bibitem[{{Pe{\~n}arrubia} {et~al.}(2006){Pe{\~n}arrubia}, {McConnachie}, \&
  {Babul}}]{Penarrubia2006}
{Pe{\~n}arrubia}, J., {McConnachie}, A., \& {Babul}, A. 2006, \apjl, 650, L33,
  \dodoi{10.1086/508656}

\bibitem[{{Pfeffer} {et~al.}(2022){Pfeffer}, {Bekki}, {Forbes}, {Couch}, \&
  {Koribalski}}]{2022MNRAS.509..261P}
{Pfeffer}, J.~L., {Bekki}, K., {Forbes}, D.~A., {Couch}, W.~J., \&
  {Koribalski}, B.~S. 2022, \mnras, 509, 261, \dodoi{10.1093/mnras/stab2934}

\bibitem[{{Pickering} {et~al.}(1997){Pickering}, {Impey}, {van Gorkom}, \&
  {Bothun}}]{Pickering1997}
{Pickering}, T.~E., {Impey}, C.~D., {van Gorkom}, J.~H., \& {Bothun}, G.~D.
  1997, \aj, 114, 1858, \dodoi{10.1086/118611}

\bibitem[{{Pilyugin} \& {Grebel}(2016)}]{Pilyugin16}
{Pilyugin}, L.~S., \& {Grebel}, E.~K. 2016, \mnras, 457, 3678,
  \dodoi{10.1093/mnras/stw238}

\bibitem[{{Pilyugin} {et~al.}(2014{\natexlab{a}}){Pilyugin}, {Grebel}, \&
  {Kniazev}}]{Pilyugin2014a}
{Pilyugin}, L.~S., {Grebel}, E.~K., \& {Kniazev}, A.~Y. 2014{\natexlab{a}},
  \aj, 147, 131, \dodoi{10.1088/0004-6256/147/6/131}

\bibitem[{{Pilyugin} {et~al.}(2014{\natexlab{b}}){Pilyugin}, {Grebel},
  {Zinchenko}, \& {Kniazev}}]{Pilyugin2014b}
{Pilyugin}, L.~S., {Grebel}, E.~K., {Zinchenko}, I.~A., \& {Kniazev}, A.~Y.
  2014{\natexlab{b}}, \aj, 148, 134, \dodoi{10.1088/0004-6256/148/6/134}

\bibitem[{{Pohlen} \& {Trujillo}(2006)}]{Pohlen2006}
{Pohlen}, M., \& {Trujillo}, I. 2006, \aap, 454, 759,
  \dodoi{10.1051/0004-6361:20064883}

\bibitem[{{Ponomareva} {et~al.}(2018){Ponomareva}, {Verheijen}, {Papastergis},
  {Bosma}, \& {Peletier}}]{Ponomareva2018}
{Ponomareva}, A.~A., {Verheijen}, M. A.~W., {Papastergis}, E., {Bosma}, A., \&
  {Peletier}, R.~F. 2018, \mnras, 474, 4366, \dodoi{10.1093/mnras/stx3066}

\bibitem[{{Potanin} {et~al.}(2020){Potanin}, {Belinski}, {Dodin},
  {Zheltoukhov}, {Land er}, {Postnov}, {Savvin}, {Tatarnikov}, {Cherepashchuk},
  {Cheryasov}, {Chilingarian}, \& {Shatsky}}]{2020AstL...46..836P}
{Potanin}, S.~A., {Belinski}, A.~A., {Dodin}, A.~V., {et~al.} 2020, Astronomy
  Letters, 46, 836, \dodoi{10.1134/S1063773720120038}

\bibitem[{{Ramya} {et~al.}(2011){Ramya}, {Prabhu}, \& {Das}}]{Ramya2011}
{Ramya}, S., {Prabhu}, T.~P., \& {Das}, M. 2011, \mnras, 418, 789,
  \dodoi{10.1111/j.1365-2966.2011.19530.x}

\bibitem[{{Revaz} \& {Pfenniger}(2004)}]{2004AA...425...67R}
{Revaz}, Y., \& {Pfenniger}, D. 2004, \aap, 425, 67,
  \dodoi{10.1051/0004-6361:20041386}

\bibitem[{{Reyes} {et~al.}(2011){Reyes}, {Mandelbaum}, {Gunn}, {Pizagno}, \&
  {Lackner}}]{Reyes2011}
{Reyes}, R., {Mandelbaum}, R., {Gunn}, J.~E., {Pizagno}, J., \& {Lackner},
  C.~N. 2011, \mnras, 417, 2347, \dodoi{10.1111/j.1365-2966.2011.19415.x}

\bibitem[{{Rodriguez-Gomez} {et~al.}(2017){Rodriguez-Gomez}, {Sales}, {Genel},
  {Pillepich}, {Zjupa}, {Nelson}, {Griffen}, {Torrey}, {Snyder},
  {Vogelsberger}, {Springel}, {Ma}, \& {Hernquist}}]{Rodriguez-Gomez2017}
{Rodriguez-Gomez}, V., {Sales}, L.~V., {Genel}, S., {et~al.} 2017, \mnras, 467,
  3083, \dodoi{10.1093/mnras/stx305}

\bibitem[{{Rosales-Ortega} {et~al.}(2012){Rosales-Ortega}, {S{\'a}nchez},
  {Iglesias-P{\'a}ramo}, {D{\'\i}az}, {V{\'\i}lchez}, {Bland-Hawthorn},
  {Husemann}, \& {Mast}}]{Rosales-Ortega2012}
{Rosales-Ortega}, F.~F., {S{\'a}nchez}, S.~F., {Iglesias-P{\'a}ramo}, J.,
  {et~al.} 2012, \apjl, 756, L31, \dodoi{10.1088/2041-8205/756/2/L31}

\bibitem[{{Ro{\v{s}}kar} {et~al.}(2010){Ro{\v{s}}kar}, {Debattista}, {Brooks},
  {Quinn}, {Brook}, {Governato}, {Dalcanton}, \&
  {Wadsley}}]{2010MNRAS.408..783R}
{Ro{\v{s}}kar}, R., {Debattista}, V.~P., {Brooks}, A.~M., {et~al.} 2010,
  \mnras, 408, 783, \dodoi{10.1111/j.1365-2966.2010.17178.x}

\bibitem[{{Rupke} {et~al.}(2010){Rupke}, {Kewley}, \& {Barnes}}]{rupke2010}
{Rupke}, D. S.~N., {Kewley}, L.~J., \& {Barnes}, J.~E. 2010, \apjl, 710, L156,
  \dodoi{10.1088/2041-8205/710/2/L156}

\bibitem[{{Saburova}(2018)}]{Saburova2018}
{Saburova}, A.~S. 2018, \mnras, 473, 3796, \dodoi{10.1093/mnras/stx2583}

\bibitem[{{Saburova} {et~al.}(2019){Saburova}, {Chilingarian}, {Kasparova},
  {Katkov}, {Fabricant}, \& {Uklein}}]{Saburova2019}
{Saburova}, A.~S., {Chilingarian}, I.~V., {Kasparova}, A.~V., {et~al.} 2019,
  \mnras, 489, 4669, \dodoi{10.1093/mnras/stz2434}

\bibitem[{{Saburova} {et~al.}(2021){Saburova}, {Chilingarian}, {Kasparova},
  {Sil'chenko}, {Grishin}, {Katkov}, \& {Uklein}}]{saburovaetal2021}
---. 2021, \mnras, 503, 830, \dodoi{10.1093/mnras/stab374}

\bibitem[{{Saburova} {et~al.}(2018){Saburova}, {Chilingarian}, {Katkov},
  {Egorov}, {Kasparova}, {Khoperskov}, {Uklein}, \&
  {Vozyakova}}]{Saburovaetal2018}
{Saburova}, A.~S., {Chilingarian}, I.~V., {Katkov}, I.~Y., {et~al.} 2018,
  \mnras, 481, 3534, \dodoi{10.1093/mnras/sty2519}

\bibitem[{{Saburova} {et~al.}(2023){Saburova}, {Chilingarian}, {Kulier},
  {Galaz}, {Grishin}, {Kasparova}, {Toptun}, \& {Katkov}}]{saburova2023}
{Saburova}, A.~S., {Chilingarian}, I.~V., {Kulier}, A., {et~al.} 2023, \mnras,
  520, L85, \dodoi{10.1093/mnrasl/slad005}

\bibitem[{{Saburova} {et~al.}(2017){Saburova}, {Katkov}, {Khoperskov}, {Zasov},
  \& {Uklein}}]{saburova2017}
{Saburova}, A.~S., {Katkov}, I.~Y., {Khoperskov}, S.~A., {Zasov}, A.~V., \&
  {Uklein}, R.~I. 2017, \mnras, 470, 20, \dodoi{10.1093/mnras/stx1200}

\bibitem[{{Savchenko} {et~al.}(2023){Savchenko}, {Poliakov}, {Mosenkov},
  {Smirnov}, {Marchuk}, {Il'in}, {Gontcharov}, {Seguine}, \&
  {Baes}}]{Savchenko2023}
{Savchenko}, S.~S., {Poliakov}, D.~M., {Mosenkov}, A.~V., {et~al.} 2023,
  \mnras, 524, 4729, \dodoi{10.1093/mnras/stad2189}

\bibitem[{{Sil'chenko} {et~al.}(2018){Sil'chenko}, {Kniazev}, \&
  {Chudakova}}]{2018AJ....156..118S}
{Sil'chenko}, O.~K., {Kniazev}, A.~Y., \& {Chudakova}, E.~M. 2018, \aj, 156,
  118, \dodoi{10.3847/1538-3881/aad37b}

\bibitem[{{Sokolovsky} {et~al.}(2017){Sokolovsky}, {Gavras}, {Karampelas},
  {Antipin}, {Bellas-Velidis}, {Benni}, {Bonanos}, {Burdanov}, {Derlopa},
  {Hatzidimitriou}, {Khokhryakova}, {Kolesnikova}, {Korotkiy}, {Lapukhin},
  {Moretti}, {Popov}, {Pouliasis}, {Samus}, {Spetsieri}, {Veselkov}, {Volkov},
  {Yang}, \& {Zubareva}}]{Sokolovsky2017}
{Sokolovsky}, K.~V., {Gavras}, P., {Karampelas}, A., {et~al.} 2017, \mnras,
  464, 274, \dodoi{10.1093/mnras/stw2262}

\bibitem[{{Spitzer}(1942)}]{1942ApJ....95..329S}
{Spitzer}, Lyman, J. 1942, \apj, 95, 329, \dodoi{10.1086/144407}

\bibitem[{{Subramanian} {et~al.}(2016){Subramanian}, {Ramya}, {Das}, {George},
  {Sivarani}, \& {Prabhu}}]{Subramanian2016}
{Subramanian}, S., {Ramya}, S., {Das}, M., {et~al.} 2016, \mnras, 455, 3148,
  \dodoi{10.1093/mnras/stv2500}

\bibitem[{{Tang} {et~al.}(2020){Tang}, {Chen}, {Zhang}, {Lin}, {Chen}, {Gao},
  {Liang}, {Liu}, \& {Kong}}]{2020ApJ...897...79T}
{Tang}, Y., {Chen}, Q., {Zhang}, H.-X., {et~al.} 2020, \apj, 897, 79,
  \dodoi{10.3847/1538-4357/ab98fd}

\bibitem[{{Thilker} {et~al.}(2005){Thilker}, {Bianchi}, {Boissier}, {Gil de
  Paz}, {Madore}, {Martin}, {Meurer}, {Neff}, {Rich}, {Schiminovich},
  {Seibert}, {Wyder}, {Barlow}, {Byun}, {Donas}, {Forster}, {Friedman},
  {Heckman}, {Jelinsky}, {Lee}, {Malina}, {Milliard}, {Morrissey}, {Siegmund},
  {Small}, {Szalay}, \& {Welsh}}]{Thilker2005}
{Thilker}, D.~A., {Bianchi}, L., {Boissier}, S., {et~al.} 2005, \apjl, 619,
  L79, \dodoi{10.1086/425251}

\bibitem[{{Toptun} {et~al.}(2023){Toptun}, {Chilingarian}, {Grishin}, \&
  {Katkov}}]{2023PASP..135h4102T}
{Toptun}, V.~A., {Chilingarian}, I.~V., {Grishin}, K.~A., \& {Katkov}, I.~Y.
  2023, \pasp, 135, 084102, \dodoi{10.1088/1538-3873/aceca0}

\bibitem[{{Tully} \& {Fisher}(1977)}]{1977A&A....54..661T}
{Tully}, R.~B., \& {Fisher}, J.~R. 1977, \aap, 500, 105

\bibitem[{{van der Kruit} \& {Searle}(1981)}]{vanderKruit1981}
{van der Kruit}, P.~C., \& {Searle}, L. 1981, \aap, 95, 105

\bibitem[{{Veilleux} \& {Osterbrock}(1987)}]{Veilleux1987}
{Veilleux}, S., \& {Osterbrock}, D.~E. 1987, \apjs, 63, 295,
  \dodoi{10.1086/191166}

\bibitem[{{Verro} {et~al.}(2022){Verro}, {Trager}, {Peletier}, {Lan{\c{c}}on},
  {Arentsen}, {Chen}, {Coelho}, {Dries}, {Falc{\'o}n-Barroso}, {Gonneau},
  {Lyubenova}, {Martins}, {Prugniel}, {S{\'a}nchez-Bl{\'a}zquez}, \&
  {Vazdekis}}]{Verro2022}
{Verro}, K., {Trager}, S.~C., {Peletier}, R.~F., {et~al.} 2022, \aap, 661, A50,
  \dodoi{10.1051/0004-6361/202142387}

\bibitem[{Virtanen {et~al.}(2020)Virtanen, Gommers, Oliphant, Haberland, Reddy,
  Cournapeau, Burovski, Peterson, Weckesser, Bright, {van der Walt}, Brett,
  Wilson, Millman, Mayorov, Nelson, Jones, Kern, Larson, Carey, Polat, Feng,
  Moore, {VanderPlas}, Laxalde, Perktold, Cimrman, Henriksen, Quintero, Harris,
  Archibald, Ribeiro, Pedregosa, {van Mulbregt}, \& {SciPy 1.0
  Contributors}}]{2020SciPy-NMeth}
Virtanen, P., Gommers, R., Oliphant, T.~E., {et~al.} 2020, Nature Methods, 17,
  261, \dodoi{10.1038/s41592-019-0686-2}

\bibitem[{{Vudragovi{\'c}} {et~al.}(2021){Vudragovi{\'c}}, {Bilek}, {Muller},
  {Samurovi{\'c}}, \& {Jovanovi{\'c}}}]{2021POBeo.100..169V}
{Vudragovi{\'c}}, A., {Bilek}, M., {Muller}, O., {Samurovi{\'c}}, S., \&
  {Jovanovi{\'c}}, M. 2021, in XIX Serbian Astronomical Conference, Vol. 100,
  169--174, \dodoi{10.48550/arXiv.2103.11721}

\bibitem[{{Ward} {et~al.}(2022){Ward}, {Gezari}, {Nugent}, {Bellm}, {Dekany},
  {Drake}, {Duev}, {Graham}, {Kasliwal}, {Kool}, {Masci}, \&
  {Riddle}}]{2022ApJ...936..104W}
{Ward}, C., {Gezari}, S., {Nugent}, P., {et~al.} 2022, \apj, 936, 104,
  \dodoi{10.3847/1538-4357/ac8666}

\bibitem[{{Yoachim} {et~al.}(2012){Yoachim}, {Ro{\v{s}}kar}, \&
  {Debattista}}]{Yoachim2012}
{Yoachim}, P., {Ro{\v{s}}kar}, R., \& {Debattista}, V.~P. 2012, \apj, 752, 97,
  \dodoi{10.1088/0004-637X/752/2/97}

\bibitem[{{Zasov} {et~al.}(2015){Zasov}, {Saburova}, {Katkov}, {Egorov}, \&
  {Afanasiev}}]{Zasov2015}
{Zasov}, A., {Saburova}, A., {Katkov}, I., {Egorov}, O., \& {Afanasiev}, V.
  2015, \mnras, 449, 1605, \dodoi{10.1093/mnras/stv454}

\bibitem[{{Zasov} \& {Khoperskov}(2003)}]{Zasov_Khoperskov2003}
{Zasov}, A.~V., \& {Khoperskov}, A.~V. 2003, Astronomy Letters, 29, 437,
  \dodoi{10.1134/1.1589861}

\bibitem[{{Zasov} {et~al.}(2011){Zasov}, {Khoperskov}, \&
  {Saburova}}]{Zasov2011}
{Zasov}, A.~V., {Khoperskov}, A.~V., \& {Saburova}, A.~S. 2011, Astronomy
  Letters, 37, 374, \dodoi{10.1134/S1063773711050069}

\bibitem[{{Zasov} {et~al.}(2017){Zasov}, {Saburova}, {Khoperskov}, \&
  {Khoperskov}}]{2017PhyU...60....3Z}
{Zasov}, A.~V., {Saburova}, A.~S., {Khoperskov}, A.~V., \& {Khoperskov}, S.~A.
  2017, Physics Uspekhi, 60, 3, \dodoi{10.3367/UFNe.2016.03.037751}

\bibitem[{{Zhu} {et~al.}(2023){Zhu}, {P{\'e}rez-Monta{\~n}o},
  {Rodriguez-Gomez}, {Cervantes Sodi}, {Zjupa}, {Marinacci}, {Vogelsberger}, \&
  {Hernquist}}]{Zhu2023}
{Zhu}, Q., {P{\'e}rez-Monta{\~n}o}, L.~E., {Rodriguez-Gomez}, V., {et~al.}
  2023, \mnras, 523, 3991, \dodoi{10.1093/mnras/stad1655}

\bibitem[{{Zhu} {et~al.}(2018){Zhu}, {Xu}, {Gaspari}, {Rodriguez-Gomez},
  {Nelson}, {Vogelsberger}, {Torrey}, {Pillepich}, {Zjupa}, {Weinberger},
  {Marinacci}, {Pakmor}, {Genel}, {Li}, {Springel}, \&
  {Hernquist}}]{Zhuetal2018}
{Zhu}, Q., {Xu}, D., {Gaspari}, M., {et~al.} 2018, \mnras, 480, L18,
  \dodoi{10.1093/mnrasl/sly111}

\bibitem[{{Zou} {et~al.}(2017){Zou}, {Zhou}, {Fan}, {Zhang}, {Zhou}, {Nie},
  {Peng}, {McGreer}, {Jiang}, {Dey}, {Fan}, {He}, {Jiang}, {Lang}, {Lesser},
  {Ma}, {Mao}, {Schlegel}, \& {Wang}}]{2017PASP..129f4101Z}
{Zou}, H., {Zhou}, X., {Fan}, X., {et~al.} 2017, \pasp, 129, 064101,
  \dodoi{10.1088/1538-3873/aa65ba}

\end{thebibliography}
\bibliographystyle{aasjournal}
\appendix

\section{LSB disk spectra}
\restartappendixnumbering

\begin{figure*}[h]
    \begin{center}
        \includegraphics[clip,trim={23mm 85mm 25mm 80mm},width=0.97\hsize]{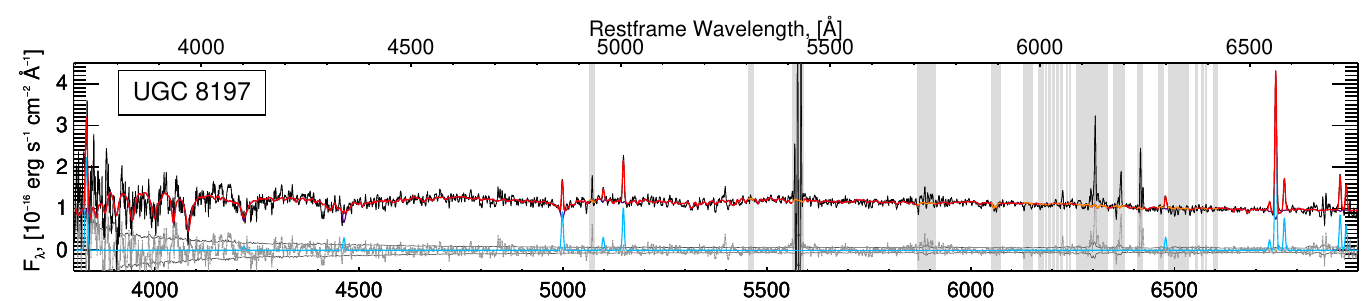}
        \includegraphics[clip,trim={23mm 85mm 25mm 86mm},width=0.97\hsize]{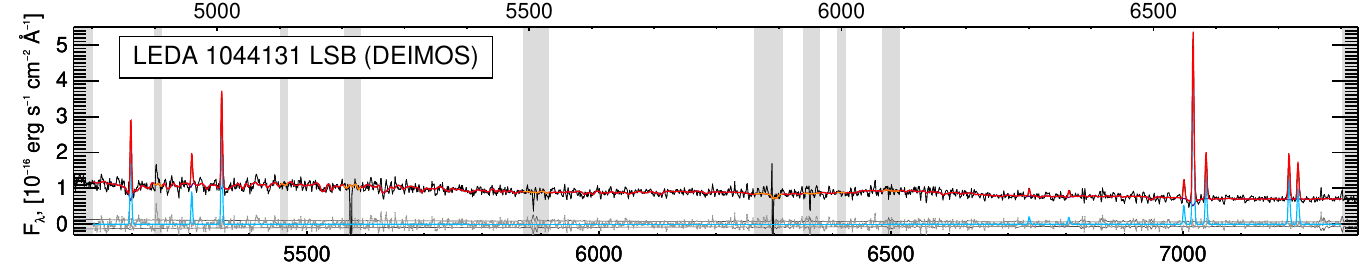}
        \includegraphics[clip,trim={23mm 85mm 25mm 86mm},width=0.97\hsize]{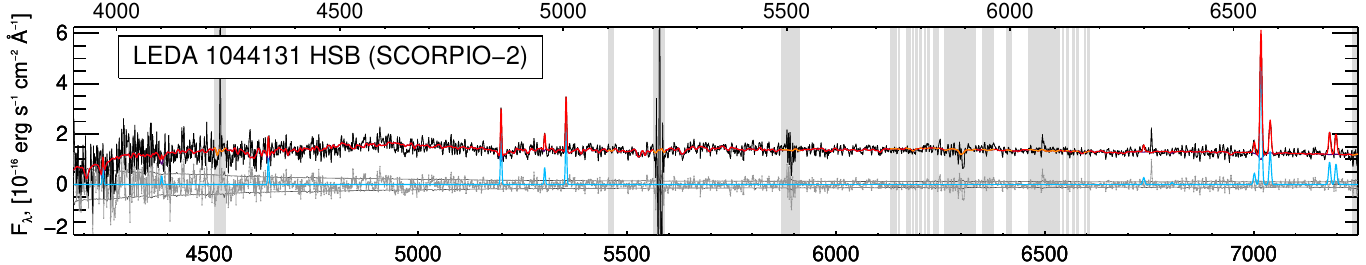}
        \includegraphics[clip,trim={23mm 85mm 25mm 86mm},width=0.97\hsize]{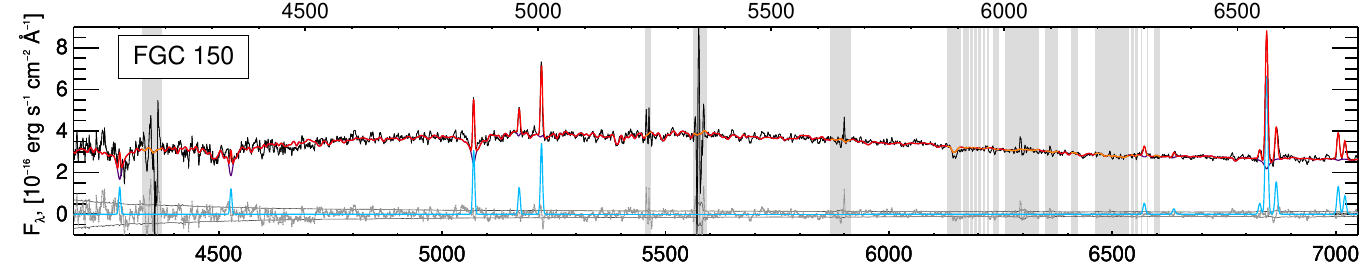}
        \includegraphics[clip,trim={23mm 80mm 25mm 86mm},width=0.97\hsize]{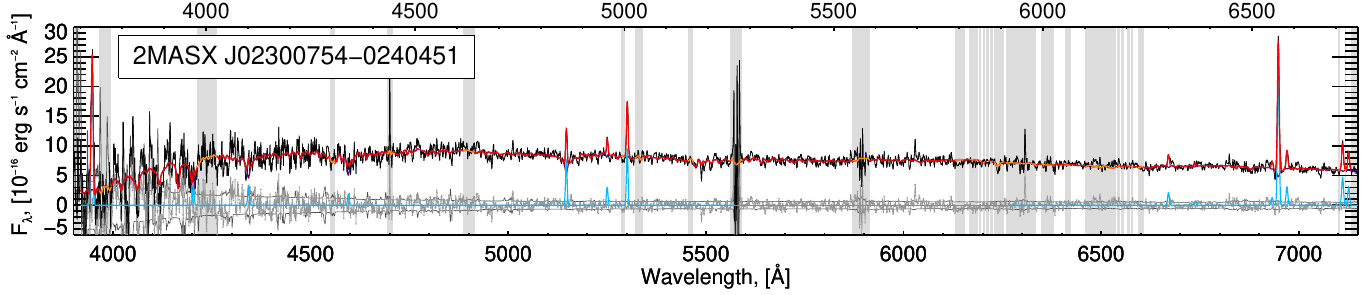}
    \end{center}
    
    \caption{The results of \nb{} full spectrum fitting using XSL SSP models for the spectra of LSB disks (+HSB disk for LEDA~1044131) coadded in the large spatial bins shown by red and yellow regions in Fig.~\ref{fig:spectra}. Black lines show the observed spectra of the LSB disks (flux density); red, cyan, and dark grey lines show the total best-fitting models, pure emission-line components, and fitting residuals respectively; the shaded grey areas denote the masked regions excluded from  fitting.}
    \label{fig:lsb}
\end{figure*}

\end{document}